\documentclass[11pt, preprint]{aastex}
\usepackage{lscape}
\usepackage{graphicx}
\usepackage{longtable}

\newcommand{\dif}{\mathrm{d}}

\begin{document}

\title{Probing Outflows in $z= 1 \sim 2$ Galaxies through \ion{Fe}{2}/\ion{Fe}{2}* Multiplets}

\author{Yuping Tang\altaffilmark{1}, Mauro Giavalisco\altaffilmark{1}, 
        Yicheng Guo\altaffilmark{2}, Jaron Kurk\altaffilmark{3} \\}
 
\altaffiltext{1}{Department of Astronomy, University of Massachusetts, Amherst, MA, 01003, USA, yupingt@astro.umass.edu}
\altaffiltext{2}{UCO/Lick Observatory, Department of Astronomy and Astrophysics, University of California, 
                 Santa Cruz, CA, 95064}
\altaffiltext{3}{Max-Planck-Institut f\"{u}r Extraterrestrial Physik, Gie\ss{}enbachstrasse, 85748, Garching bei München, Germany}

\begin{abstract}

  We report on a study of the $2300-2600\AA$ \ion{Fe}{2}/\ion{Fe}{2}$^*$
  multiplets in the rest--UV spectra of star--forming galaxies at $1.0<z<2.6$
  as probes of galactic--scale outflows. We extracted a mass--limited sample
  of 97 galaxies at $z\sim1.0-2.6$ from ultra--deep spectra obtained during
  the GMASS spetroscopic survey in the GOODS South field with the VLT and
  FORS2.  We obtain robust measures of the rest equivalent width of the
  \ion{Fe}{2}\ absorption lines down to a limit of $W_r>1.5$\AA\ and of the
  \ion{Fe}{2}$^*$ emission lines to $W_r>0.5$\AA. Whenever we can measure the 
  systemic redshift of the galaxies from the [\ion{O}{2}] emission line, 
  we find that both the \ion{Fe}{2}\ and
  \ion{Mg}{2}\ absorption lines are blueshifted, indicative that both species
  trace gaseous outflows. We also find, however, that the \ion{Fe}{2}\ gas has
  generally lower outflow velocity relative to that of \ion{Mg}{2}. We
  investigate the variation of \ion{Fe}{2} line profiles as a function of the
  radiative transfer properties of the lines, and find that transitions with
  higher oscillator strengths are more blueshifted in terms of both line
  centroids and line wings.  We discuss the possibility that
  \ion{Fe}{2}\ lines are suppressed by stellar absorptions. The lower
  velocities of the \ion{Fe}{2}\ lines relative to the \ion{Mg}{2}\ doublet,
  as well as the absence of spatially extended \ion{Fe}{2}$^*$ emission in 2D
  stacked spectra, suggest that most clouds responsible for the
  \ion{Fe}{2}\ absorption lie close ($3 \sim 4$ kpc) to the disks of galaxies.  We
  show that the \ion{Fe}{2}/\ion{Fe}{2}$^*$ multiplets offer unique probes of
  the kinematic structure of galactic outflows. 

\end{abstract}


\section{Introduction}

Outflow winds are observed to be ubiquitous in star--forming galaxies at low
and high redshift (e.g. Tremonti et~al. 2007; Chen et~al. 2010; Weiner
et~al. 2009; Steidel et~al. 2010), and are thought to be an essential physical
process to regulate galaxy formation and evolution. In theoretical models and
simulations, appropriate treatment of the outflows is critical to recover the
observed properties of galaxies over a wide range of cosmic time, mass and
star formation activity: Feedback by outflows powered by the star formation,
although in forms that still remain rather unclear, is traditionally
introduced to prevent excessive star formed from cooling baryons and to
explain the luminosity function of galaxies, especially at the bright and
faint end (White \& Frenk et al. 1991; Katz et al. 1996; Cole et al. 2000).
Outflows might also be a key ingredient responsible for establishing the well
documented relationship between stellar mass and gas metallicity at both low
(Tremonti et al. 2004) and high redshift (Erb et al. 2006), as well as for
enriching the Intergalactic Medium (IGM) (Madau et al. 2001; Adelberger et
al. 2003).

In a cosmological sense, the critical parameter of galactic outflow is the
mass loss rate as a function of the star formation history and halo mass,
which is far from being accurately constrained. Theoretically, how outflow
wind is launched remains poorly understood, especially the relative
contribution by AGN and by star formation in powering cool (i.e. $T\sim 10^4$
K) outflows. Outflow wind is known to be present in post starburst galaxies
and AGN (Rupke 2005, Tremonti et~al. 2007, Cimatti et al. 2013), 
and while Sturm et~al. (2011) have recently reported a positive relationship 
between high velocity ($\sim 1000$ km/s) molecular outflows and AGN activity 
(albeit in a limited sample), Krug et~al. (2010) find small detection rate of 
cool outflows in infrared--faint Seyfert galaxies. Energy input from supernova 
is also expected to be a primary energy source for powering galaxy--scale outflows 
(Chevalier \& Clegg 1985; Heckman et~al. 2002), and theoretical studies 
suggest that radiation pressure could be a important mechanism to drive 
outflow wind (Murray et~al. 2005, 2011; Krumoltz et~al.  2011). 
Observationally, a scaling relationship between
outflow velocity and star formation rate (SFR), $v \propto$ SFR$^{0.3}$, has
also been reported in a number of studies (Martin 2005; Weiner et~al. 2009;
Banerji et~al. 2011).

In the local universe, galactic outflows leave their spectral footprints
across the whole electromagnetic spectrum (Veilleux et~al. 2005 and reference
therein), revealing a variety of gas phases associated with outflow wind, from
plasma to molecular gas. As one moves to distant universe, however,
blueshifted atomic absorption lines arising from cool gas remain almost the
only probes of galactic outflow.  At low and intermediate redshift,
interstellar medium (ISM) absorption lines of \ion{Na}{0}~D $\lambda\lambda
5890,5896$ and \ion{Mg}{2}~$\lambda\lambda 2796,2803$ doublets are widely
studied as probes of cool outflow. These lines are favored because of their
high oscillator strength, moderate contamination from stellar photosphere and
rich abundance in galaxies. The information carried by these absorption lines
is unfortunately limited by the fact that they are usually saturated, thus
line profiles are complicated by covering fraction and fail to provide
accurate estimate of column density. Furthermore, as shown in Prochaska et
al. (2011), photon scattering could substantially modify absorption line
profiles and obscure information of wind structure.  This effect, has been
largely ignored in previous studies.

  Ignoring dust extinction, the propagation of resonant photons in the ISM
  involves two types of processes: resonance scattering and fluorescence.  If
  no fluorescence transition is included, a resonance photon simply undergoes
  random walk in a stationary system. For example, a typical resonant transition 
  in near UV band is \ion{Mg}{2}$2796,2800$.  If a resonance transition is associated
  with fluorescence transition, the random walk of resonance photons could be
  terminated, as resonance photons are re-emitted as optically thin
  fluorescence photons and exit the nebula. In an outflow wind, since absorbing medium has 
  multi velocity components, continuum photons are absorbed in only limited regions of the wind, 
  namely those where the co-moving wavelength of photons are Doppler shifted to local
  resonance wavelength. This means that for a photon of wavelength $\lambda$, 
  absorption is contributed only by parts of wind with velocity around
  $\frac{\lambda-\lambda_0}{\lambda_0}c$, where $\lambda_0$ is the default line
  center of resonance transition and $c$ is light speed. As a conclusion, 
  the absorption and emission transitions associated with scattering and fluorescence 
  can provide invaluable information about the distribution and dynamics of the wind. 

  Recently, a simple model of cool outflow has been developed by Prochaska
  et~al. (2011, hereafter P11), which offers detailed insight into emission
  and absorption line profiles from observation. In this model:

1. The outflow wind is accelerated and it expands with increasing velocity and
decreasing density as it propagate outwards from the inner regions.

2. The profiles of the absorption lines are re-shaped by
re-emitted photons following absorption, up to $50\%$ absorbed photons could
be filled by photons scattered from clouds that are located away from
the line of sight. 

3. Continuum fluorescence of UV photons in outflow wind gives rise to
\ion{Fe}{2}* emission lines, which could efficiently terminate ``random walk''
of resonance scattering.

In this work we investigate the \ion{Fe}{2} multiplets located between $2344$
and $2600$\AA\ as probes of outflow wind in a mass--limited sample of
$\sim100$ star--forming galaxies at $z=1.0-2.6$. After the \ion{Mg}{2}
$2796,2800$ \AA\ doublet, the five \ion{Fe}{2} absorption lines \ion{Fe}{2}
$2344.2$, \ion{Fe}{2} $2374.5$, \ion{Fe}{2} $2382.8$, \ion{Fe}{2} $2586.7$ and
\ion{Fe}{2} $2600.2$ are the most prominent spectral features in the
rest--frame mid UV of star--forming galaxies. The \ion{Fe}{2} absorption lines
are primarily generated in ISM, with minor stellar contamination (Leitherer et
al. 2011). The \ion{Fe}{2} multiplets, with their oscillator strengths
spreading over an order of magnitude, show outstanding potential of
deconstructing the physical structure of wind.  Attempts have been made to
incorporate \ion{Fe}{2} into analysis of cool outflow wind in previous studies
at intermediate redshift (Martin \& Bouch\'{e} 2009; Rubin et al. 2010, 2011,
2012; Kornei et al. 2012, 2013; Erb et al. 2012; Martin et al. 2012, 2013).
These works show that \ion{Fe}{2} lines are common features in the mid--UV
spectra of galaxies with star--formation rates higher than a few
$M_{\odot}/yr$, with equivalent widths (EWs) comparable to that of the \ion{Mg}{2}
doublet. The detection of \ion{Fe}{2}* emission lines predicted by P11 has
been reported by Rubin et al. (2011); Coil et al. (2011); Erb et al. (2012)
and Kornei et al. (2013). More recently, Rubin et al. (2011, 2013) and Martin
et al. (2012) reported redshifted \ion{Fe}{2} and \ion{Mg}{2} absorption line
profiles and suggested that these features are evidence of inflow, further
highlighting the capacity of these lines as probes of the kinematics of cool
gas, i.e. $T\sim 10^4$ K.  All these works have focused on star--forming
  galaxies around $z\sim~1$, with the exception of the sample by Erb et
  al. (2012), which consists of UV color-selected galaxies at $z\sim~1-2$.
  The primary objective of this work is an independent
  examination of the properties of galactic outflows of galaxies from a
  mass-selected sample around $z\sim~1-2$, similar to recent works by Erb et
  al. (2012). In this work we'll include the entire set of 
  \ion{Fe}{2}/\ion{Fe}{2}* and \ion{Mg}{2} transitions into our study. 
  Also, we'll discuss possible effect of stellar absorption on \ion{Fe}{2}* measurements.

This paper is organized as follows: In Section 2 we describe the sample and
the method of averaging spectra. In Section 3 and Section 4 we study the
observational properties of \ion{Fe}{2} absorption lines and \ion{Fe}{2}*
emission lines, respectively, which are followed by discussion and conclusion
in Section 5.  We use cosmological parameters $H_0=70 kms^{-1} Mpc^{-1}$,
$\Omega_M=0.3$, $\Omega_{\Lambda}=0.7$ throughout this work.

\section{The Data}

\subsection {Sample Selection}

The sample of galaxies discussed here is extracted from the Galaxy Mass
Assembly Spectroscopic Survey (GMASS) described by Kurk et al. (2013), a
program of spectroscopic observations of a mid--IR magnitude--limited
($m_{AB}$ of IRAC4.5 $< 23.0$) sample selected from a $6.8' \times 6.8'$ field
in the GOODS-S Field (Giavalisco et al. 2004). The main scientific motivation
of GMASS was to investigate the mass assembly and evolution of galaxies within
the redshift range ($1.3<z<2.6$).  The spectra were obtained at the ESO VLT
Very Large Telescope with the FORS2 instrument using two grisms, a blue one
(300V) covering the spectral range 3600--6000\AA, and a red one (300I) for the
range 6000--10000\AA. A slit width of 1 arcsec yielded a spectral resolution
$500km/s$ with both grisms.  The observation are very deep, with typical
exposure times on target of $>10$ hours for the blue masks and $20-30$ hours
for the red masks.

The main flux--limit selection criterion results in a total of 221 sources in
the footprint of the survey, of which 174 were assigned a slit with either the
blue or red grism.  Spectra were obtained for 102 galaxies in the 300V grism
and 115 with the 300I one.  The targets that satisfy the additional criteria
$B\le 26$, $z-K_s\le 2.3$, and $z_{phot}> 1.4$ have been included for
observations in the blue masks and the 300V grism.  Those with $I\le 26$,
$z-K_s> 2.3$ and $z_{phot}<1.4$ in the red mask with the 300I grism.
Additional information on the observations data reduction can be found in
Kurk et al. (2012). As discussed by Cimatti et al. (2008), Talia et al. (2012) 
and Kurk et al. (2009, 2013) in the targeted redshift range the sample is 
essentially a mass--complete one. 

Since the GMASS survey in entirely contained in the GOODS-S Field (Giavalisco
et al. 2004), the full range of the GOODS deep multi--wavelength photometry is
available, from UV to infrared, including VLT/VIMOS U-band, HST/ACS BViz,
VLT/ISAAC JHK, and Spitzer/IRAC 3.6, 4.5, 5.8, and 8.0 $\mu m$. We use this
panchromatic photometric catalog to measure the integrated parameters of the
stellar populations by fitting the spectral energy distribution (SED) of each
galaxy to the spectral population synthesis models by Chariot \& Bruzual (2009; CB09 henceforth).  
We use the Salpeter IMF with the lower and upper mass limit of $0.1 M_{\odot}$ and $100
M_{\odot}$, respectively. We also used the Calzetti law (Calzetti et al. 1994,
2000) and the recipe of Madau et al. (1995) to model the dust extinction and
the cosmic opacity of IGMs.  The details of the multi-wavelength catalog and
SED-fitting can be found in Guo et al. (2011).  Since the peak of stellar
emission of $z\sim2$ galaxies, redshifted to around the IRAC $3.6 \mu m$, is
covered by our multi-wavelength data, our SED-fitting yields robust mass
estimates, with a typical internal uncertainty of $\sim0.1$ dex, comparable to
the mass uncertainties of SED-fitting in literature.

Unlike the measure of stellar mass, the instantaneous rate of star-formation
(SFR) and dust extinction, parametrized by the E(B-V), are not well
constrained by SED fitting, because they critically depend on the assumption
on the star formation history. We measured both these quantities for each
GMASS galaxy through the slope and luminosity of the rest-frame far UV
continuum. We use the Calzetti law to derive the E(B-V) parameter from the
observed slope of the rest--frame UV continuum in the approximate range
1300--2000 \AA, and derived the unobscured SFR from its dust-corrected
rest-frame UV continuum using the formula by Kennicutt (1998). Compared with
SED fitting, this method is less model-dependent and requires no prior
information on the star-formation history of the galaxies (e.g., Lee et
al. 2010; Maraston et al. 2010; Papovich et al. 2011). Comparing the SFR
measured from the UV continuum to that from the UV+IR luminosity shows that
the UV dust correction method provides a good measurement of SFR for galaxies
with intermediate SFR, namely $10-100 M_{\odot}/yr$ (Wuyts et al. 2011). Although the scatter between 
the $SFR_{UV, corr}$ and $SFR_{UV+IR}$ is $\sim0.35$ dex, the systematic offset is 
essentially negligible (see also Nordon et al. 2010). For galaxies with 
$SFR_{UV,corr}>100 M_{\odot}/yr$, however, both studies find that the UV dust correction 
method tends to overestimate the SFR by a factor of 1.5--2. In our sample, we 
have 16 galaxies ($~16\%$ of our sample) with $SFR_{UV, corr}>100 M_{\odot}/yr$, 
whose SFRs should therefore be used with caution.

Uncertainties are estimated with Monte-Carlo re-sampling of the photometry of
each band of a galaxy, assuming a Gaussian distribution with the mean and
1-sigma deviation equal to the flux and flux uncertainty of the band, and
re-run the SED-fitting and UV SFR 100 times for the galaxy. Then the
uncertainty of the mass, SFR, and E(B-V) is taken to be the standard
deviation of the 100 runs.

\subsection {Redshift Measurment}

To explore the dynamic properties of cold/warm gas responsible for
absorption/emission, such as outflows and inflows, an accurate determination
of the systemic redshift of the galaxies is required. In the rest--frame
wavelength probed by the GMASS spectra considered here and given the available
S/N ratio, the doublet [\ion{O}{2}]$3727$ emission line, unresolved at the
available resolution, is the only useful spectral feature to determine the
systemic redshift. We have first selected the galaxies in our sample which
have [\ion{O}{2}] detected at the $>5\sigma$ level in their spectra, for which
the systemic redshift is measured from fitting a single Gaussian profile to
the line, whose rest wavelength we set to $3727.30$ \AA. This sub
sample contains 43 galaxies, and in the following we will refer to it as
``Sample [\ion{O}{2}]''.  The redshifts of the galaxies in Sample [\ion{O}{2}] are
expected to be the systemic ones. In all cases, the interpretation of the
emission line as [\ion{O}{2}] is confirmed by examining the alignments of UV
absorption lines at shorter wavelengths. For the wavelength/redshift ranges
covered by the GMASS sample, the most prominent UV absorption lines are
FeII/MgII lines, plus \ion{C}{4}$1548/1550 \AA$, \ion{Si}{2}
$1526 \AA$, \ion{Fe}{2} $1608 \AA$ and \ion{Al}{2} $1671 \AA$. We estimate the S/N ratios
of the individual spectra in two continuum bands, [1740-1820 \AA] and [2680-
2760 \AA]; typically, the blue band has S/N=7 and the red band has a
typical S/N=4. 

\begin{figure}
\includegraphics[width=12cm]{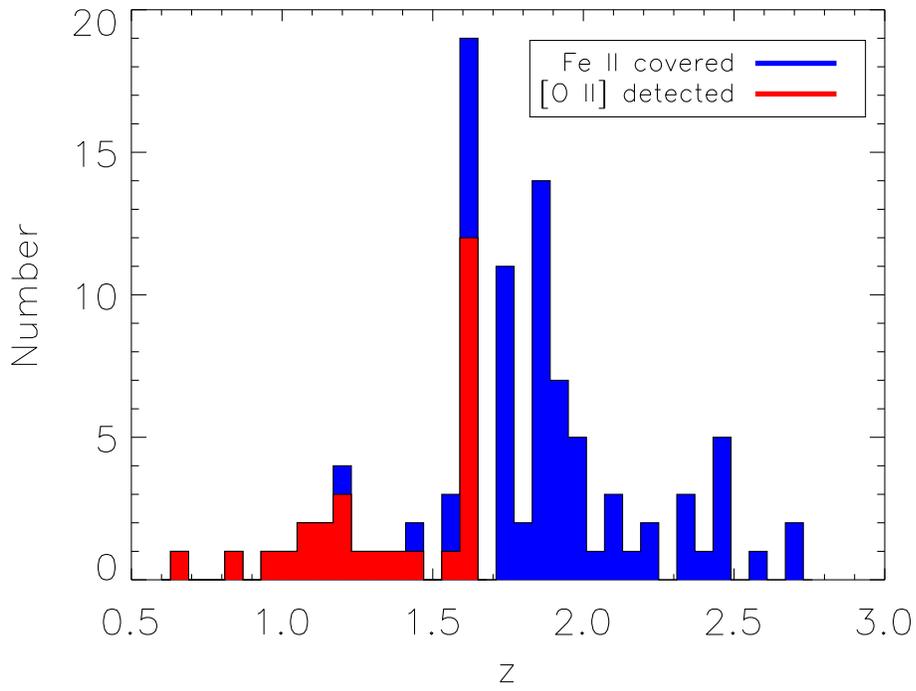}
\caption{Distribution of redshifts of GMASS galaxies. 
Blue: all objects covering $2200-2700\AA$ in rest frame.
Red: objects covering $2200-2700\AA$ and with [O II] $3727$ detection.} \label{fig:z_dis}
\end{figure}

For the remaining galaxies in the sample the redshift of the [\ion{O}{2}]
could not be measured, either because the line is outside of the available
spectral range or because the S/N ratio was poor (typically due to overlapping
OH night sky emission lines). To measure their redshifts we have stacked the
spectra from Sample [\ion{O}{2}] to generate a high S/N template spectrum, and then
we have used the IRAF task XCSAO to cross--correlate this template with the
spectra with no [\ion{O}{2}]. For these galaxies the cross--correlation
determines the redshift primarily from interstellar absorption features, and
since the velocity of winds in the individual galaxies is in general different
from that of the template spectrum (which represents average wind properties),
the redshifts measured in this way deviate from systemic by a random amount of
the order of a few hundred km/s. We have visually inspected the spectra with
cross-correlation redshift, and we have excluded low quality spectra,
i.e. those with less than three identified features. The resulting sample of
cross--correlation redshifts contains 98 galaxies, and we will refer to this
sub sample as ``Sample Abs'' hereafter. Thus, in total, we have redshifts for
141 galaxies, and for 97 of them the spectra cover the \ion{Fe}{2} multiplets
at $2300-2600$ \AA.  These 97 galaxies form the final sample that we have used
in our subsequent analysis. Their redshift distribution is shown in
Figure~\ref{fig:z_dis}. The clear spike shows that a large number of the
galaxies belong to the large structure at $z=1.6$ studied by Kurk et
al. (2009), who conclude that such an over-density of galaxies represents a
proto--cluster.

  The uncertainty in the measure of the redshift of Sample [\ion{O}{2}]
  galaxies is dominated by the uncertainty in the determination of the line
  centroid, which is estimated from the covariance matrices during Gaussian
  fit. The corresponding errors are small, less than 25km/s. For the
    cross--correlated redshifts, we have followed Tonry and Davis(1979) and
    used the XCSAO package to calculate the error as a function of the r
    statistics of the correlation peak (Kurtz \& Mink et al. 1998).  Assuming
    sinusoidal noise with a half width of the sinusoid equal to the half-width
    of the correlation peak, the mean error in the estimation of the peak of
    the correlation function is: $err = \frac{3w}{8(1+r)}$, where $w$ is the
    FWHM of the correlation peak and r is the ratio of the correlation peak
    height to the amplitude of the antisymmetric noise. The effectiveness of
    this assumption has been discussed in detail by Kurtz \& Mink (1998). In
    our case, the value of the uncertainty is $100 \sim 200 km/s$, but we
    also have to consider an additional source of uncertainty due to the
    variation in outflow velocities, which is also in the range $100 \sim
    200km/s$ (Erb et al. 2012).

\begin{figure}
\includegraphics[width=15cm]{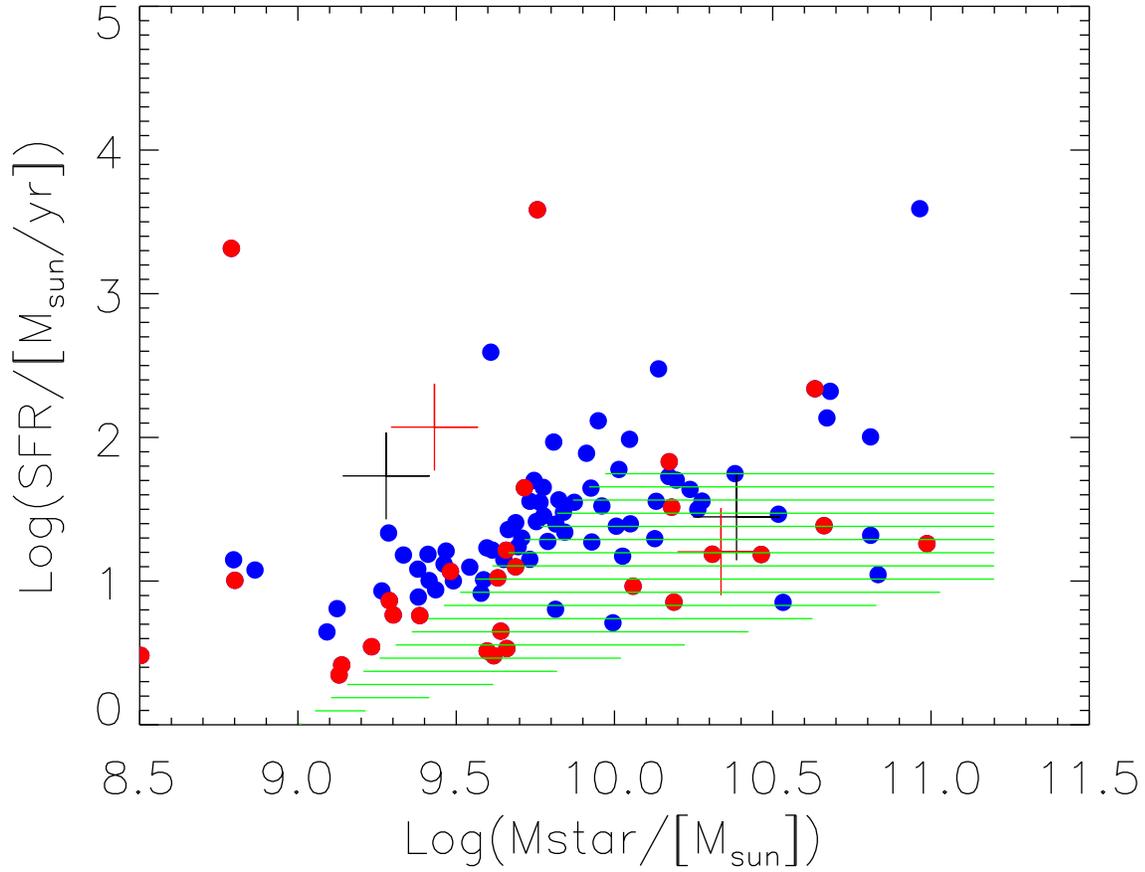}
\caption{$SFR-M_{*}$ diagram. Red dots indicate detection of [\ion{O}{2}] and 
blue dots indicate non-detection of [\ion{O}{2}] . The green shadowed area defines 
the region of DEEP2 sample studied by Kornei et al. (2012) and Kornei et al. (2013). 
The black plus signs represent mean values of the low mass and high 
mass subsample defined in Erb et al. (2012), the red plus signs represent
mean values of their low age and high age subsample.} \label{fig:sfr_m}
\end{figure}

  In Figure~\ref{fig:sfr_m}, we show the SFR-M* diagram for the 97 GMASS
  galaxies in our sample overplotted with the region defined by the DEEP2
  sample studied by Kornei et al. (2012, 2013). In this work SFR and $M_*$ are estimated 
  based on the Salpeter IMF, therefore both are scaled down by a factor of 1.8 in order to
  be compared with previous results based on the Chabrier IMF.
  The mean locations of the low mass and high mass subsamples and low age and high 
  age subsamples studied by Erb et al (2012) are also plotted in the figure for 
  comparison. While the range of the GMASS galaxies overlaps with those of the other studies, for
  given stellar mass the GMASS sample appears to include larger numbers of
  galaxies with larger SFR relative to the DEEP2 ones, which is very likely
  due to the higher redshifts of the former.
  
  The galaxies in the sample by Erb et al. (2012) cover a range of stellar
  mass that is lower than that of the GMASS galaxies, even if the redshifts of
  the two samples are similar. This is very likely the result of the UV
  selection of the former and of the mid--IR selection of the latter. At the
  low mass end the GMASS sample is sparse in the region covered by the Erb et
  al.'s sample.

\subsection{Averaging Galaxy Spectra}
We have averaged the spectra of both Sample [\ion{O}{2}] and Sample Abs
separately, and then combined both. Prior to averaging, the spectra are
shifted back to their rest frame and are linearly interpolated to a uniform
grid with $0.5\AA$/pixel.  The rest frame wavelength coverage of each spectrum
is recorded, after being trimmed of low S/N edges, for further selections of
subsamples with constrained rest-wavelength coverage.  To avoid being biased
towards bright objects, which have higher S/N and higher flux density,
individual spectra are normalized before averaging.  In this study, we focus
on \ion{Fe}{2} and \ion{Mg}{2} lines between $2300\AA$ and $2900\AA$. We
extract three spectral windows from each individual spectrum, $2220-2460\AA$,
$2505-2690\AA$ and $2740-2870\AA$ (band 1, 2 and 3, respectively).  In each
window, the individual spectrum is normalized by the continuum determined by
fitting a straight line to the two spectral regions adjacent to the
  features: [$2220-2305\AA$, $2415-2460\AA$] for band 1, [$2505-2550$,
    $2640-2690$] for band 2 and [$2740-2780$, $2815-2840$] for band 3.
  Finally, we calculate the median for each pixel to derive a
  ``representative'' spectrum of our sample. Throughout this work, we have
  estimated the uncertainty of any measurement based on the ``representative'' or average 
  spectra from the variance of 200 boostrap realizations of the data.

\begin{figure}
\includegraphics[width=15cm]{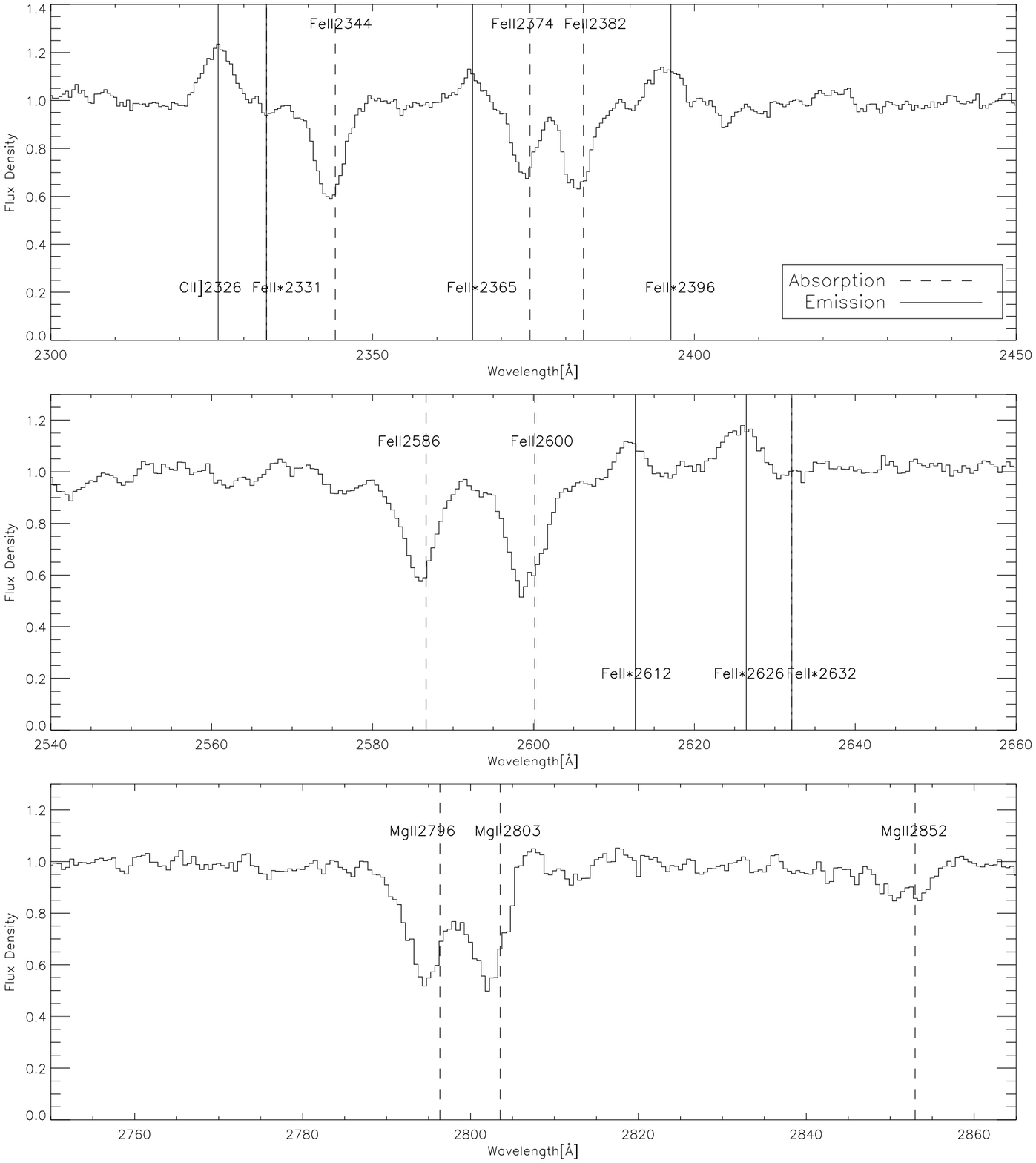}
\caption{Median spectrum of 83 GMASS galaxies with rest frame $2200-2900\AA$ coverage. 
Absorption Lines are marked by dash lines and emission lines are marked by solid lines. 
\ion{Fe}{2}* $2632$ is undetected in this median spectrum.} \label{fig:avg_spec}
\end{figure}

Figure~\ref{fig:avg_spec} shows the extracted windows from the average spectra
around the \ion{Fe}{2} multiplets, and around the $2800\AA$ \ion{Mg}{2}
doublet. Blueshifted \ion{Fe}{2} and \ion{Mg}{2} are clearly present in both
samples, with velocity of maximum absorption in the range $\sim100-200$ km/s. The absorption
lines are fairly separated, with slight blending for
\ion{Fe}{2} $2374,2382$ ($\Delta v=1050 km/s$) and \ion{Mg}{2}$2796,2803$ ($\Delta v=768 km/s$).
This is not simply due to the relatively low
resolution of our spectra ($\sim450$ km/s), but must be due to the broad blue
wings of the absorption lines characterizing outflow kinematics.

Four emission features are clearly present around the wavelengths of
\ion{Fe}{2} absorption lines at $2365$, $2396$, $2612$ and $2626$ \AA, as
marked in Figure~\ref{fig:avg_spec}. The detection of these emission features
in the spectra of star--forming galaxies has only been recently reported
(Rubin et al. 2011; P11; Coil et al. 2011). As suggested by these authors,
emission features around \ion{Fe}{2} absorption lines could simply be
explained as \ion{Fe}{2}* fluorescence emissions, which are downward
transitions to excited fine structure levels from the same upper levels of the
observed \ion{Fe}{2} absorptions.  The high S/N achieved by averaging our
spectra together offers an ideal chance to study the origin of these emission
features, and in Section 4 we examine whether indeed photon scattering could
be the primary excitation mechanism for \ion{Fe}{2}* emissions.

\section{\ion{Fe}{2} Absorptions in GMASS Galaxies}

\subsection{Absorption and Continuum Fluorescence}
\begin{figure}
\includegraphics[width=15cm]{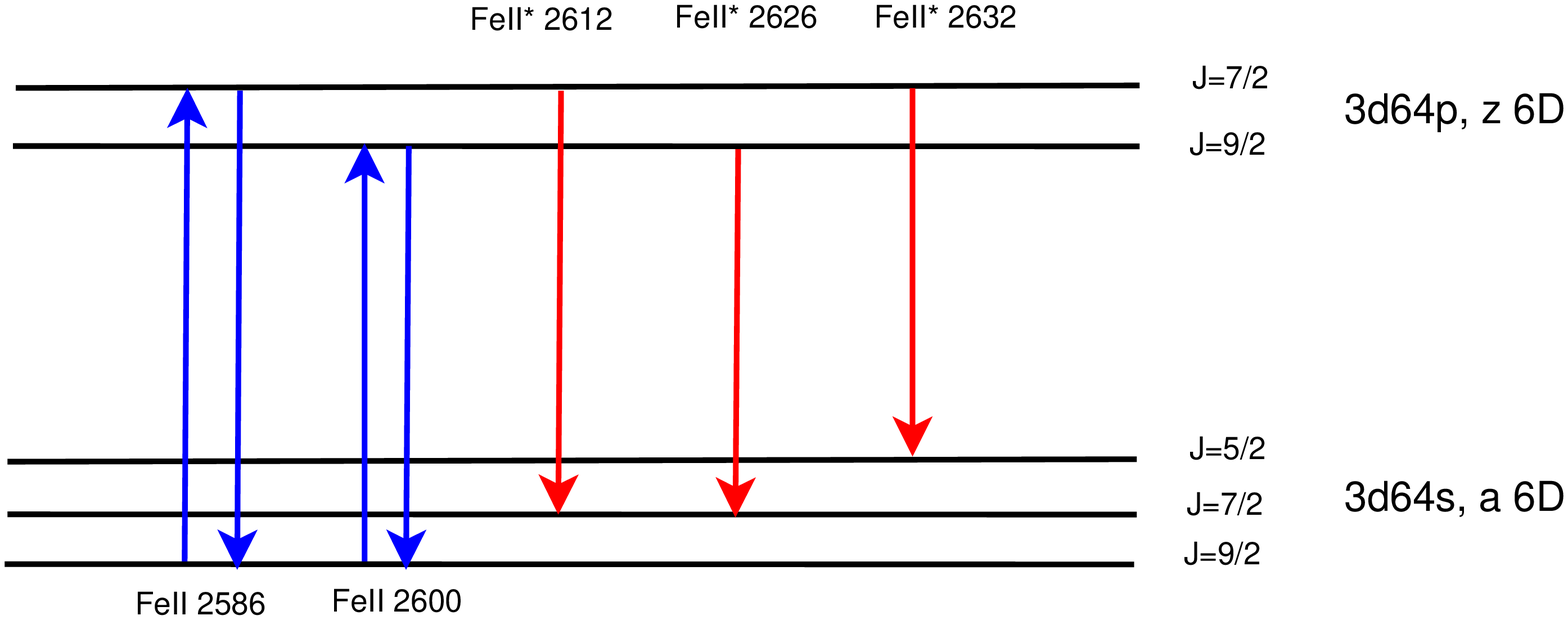}
\includegraphics[width=15cm]{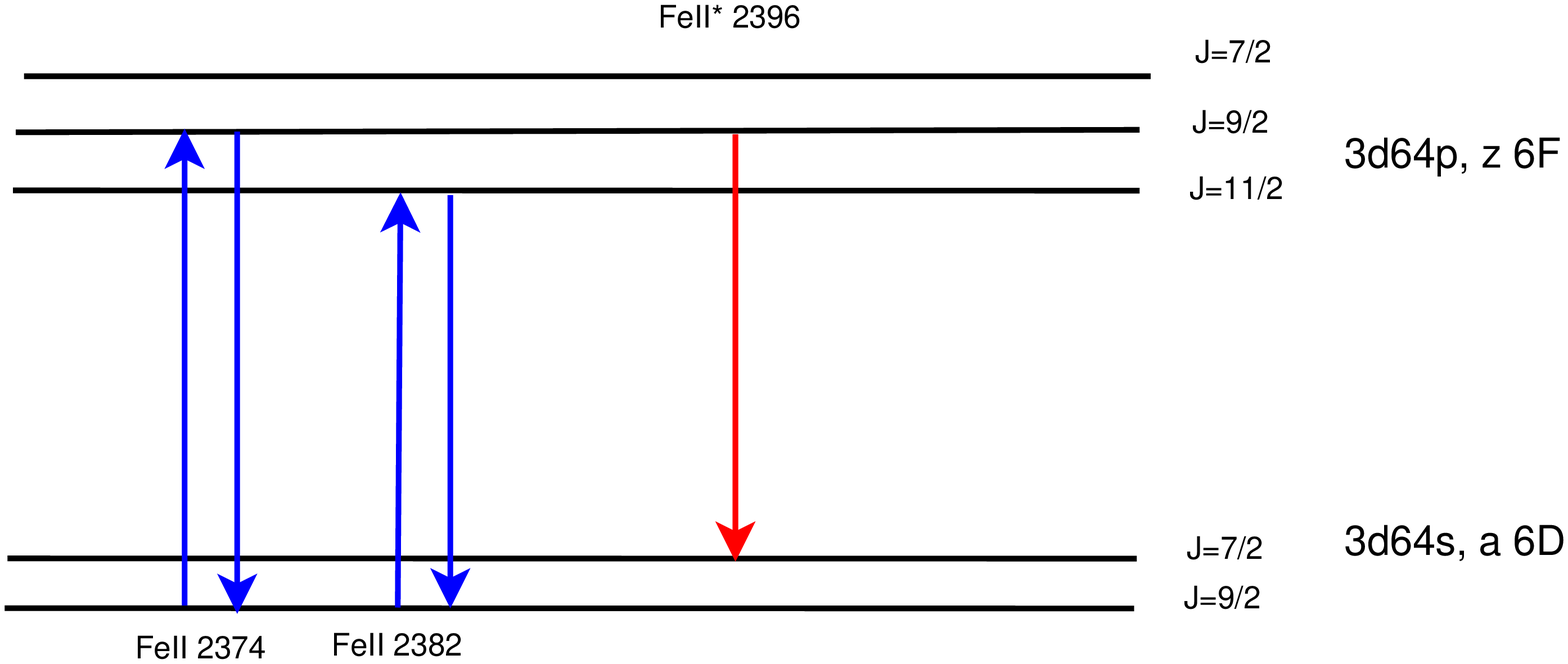}
\includegraphics[width=15cm]{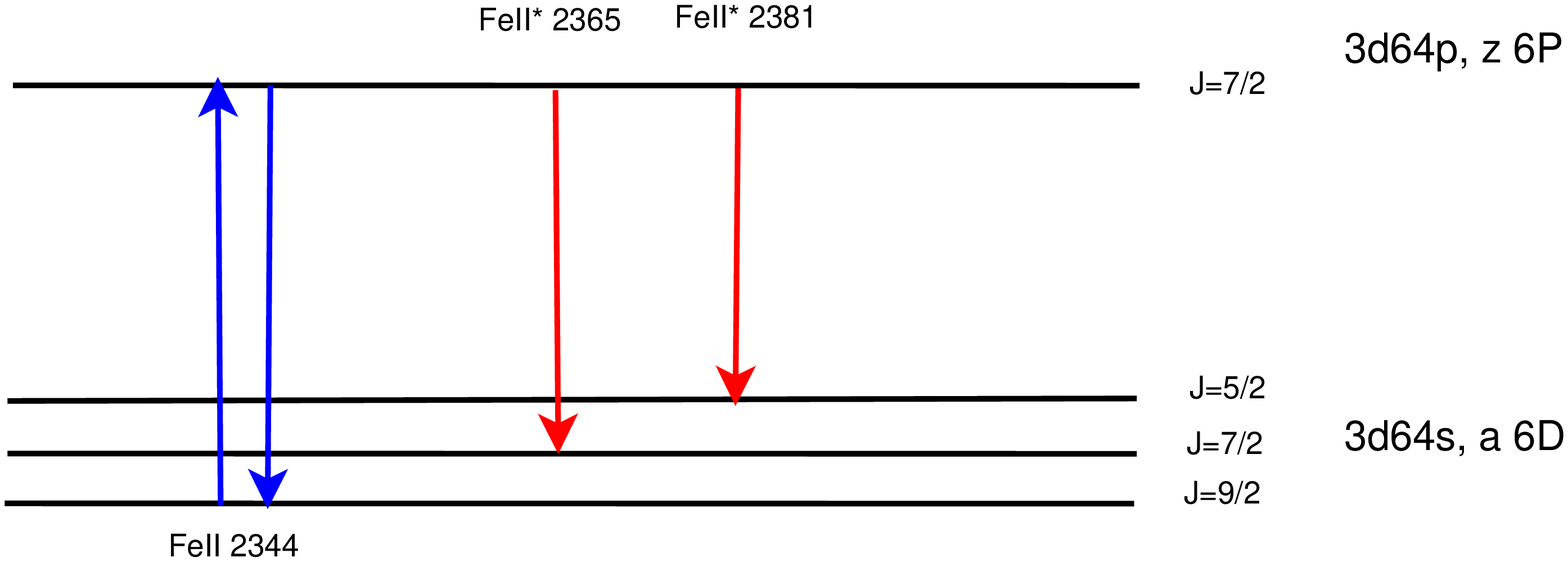}
\caption{Energy level diagram of \ion{Fe}{2}/\ion{Fe}{2}* transitions. 
Resonance lines are shown in blue and fluorescence transitions are
shown in Red.} \label{fig:level_dia}
\end{figure}

Similarly to Figure 1 in P11, we show the energy level diagrams of \ion{Fe}{2}
and associated \ion{Fe}{2}* lines in Figure~\ref{fig:level_dia}. We use the
atomic data from Morton (2003).  The five observed \ion{Fe}{2} resonance lines
in absorption are associated with six downward permitted transitions to
excited fine structure levels (also called fluorescence transitions),
respectively. These transitions are :\ion{Fe}{2} $2344.2$ $\rightarrow$
\ion{Fe}{2}* $2365.6$/\ion{Fe}{2}* $2381.5$; \ion{Fe}{2} $2374.5$
$\rightarrow$ \ion{Fe}{2}* $2396.4$; \ion{Fe}{2} $2586.7$ $\rightarrow$
\ion{Fe}{2}* $2612.7$/\ion{Fe}{2}* $2632.1$; \ion{Fe}{2} $2600.2$
$\rightarrow$ \ion{Fe}{2}* $2626.5$.  Note that for \ion{Fe}{2} $2382.8$, no
fluorescence downward transition is permitted.  The transition \ion{Fe}{2}*
$2381$ is blended with \ion{Fe}{2} $2382$, and the \ion{Fe}{2}* $2632$ is
undetected in our average spectra. The null-detection of \ion{Fe}{2}* $2632$
could pose a potential challenge to the picture shown above, and we will come
back to this point in Section 4.2.

Unless lower excited levels are significantly populated, the cold gas should
be optically thin to the \ion{Fe}{2}* emission, and once these photons are
produced they can escape the interstellar medium or outflow winds. This
suggests that the \ion{Fe}{2}* lines should be observed at systemic velocity,
since in a symmetric wind, the \ion{Fe}{2}* lines from the advancing and
receding parts of the wind are both received by the observer (Rubin et
al. 2011). 

\subsection{Classification of \ion{Fe}{2} Absorption Lines}

In the following sections we discuss the dependence of the EW(\ion{Fe}{2}) and EW(\ion{Mg}{2}) 
on the location of \ion{Fe}{2} absorbing medium and the wind kinematics and discuss the
implications. We start by discussing the profile of the \ion{Fe}{2}
absorption lines.
  
Two parameters control the propagation of resonance photons through the ISM,
namely the lower level oscillator strength of that transition, $f_l$, which
determines how frequently a photon is absorbed, and the probability of
fluorescence, which measures the likelihood that the random walk of resonant
photons in the nebula is terminated by an absorption event. The probability of
fluorescence is defined as:
    
\begin{equation}
P_{fluo}(Fe II_{i})=\frac{\sum_{j}A_{ul}(Fe II*_{j})}{\sum_{j}{A_{ul}(Fe
    II*_j)}+{A_{ul}(Fe II_i)}}, 
\end{equation}  

where $A_{ul}(Fe II*_{j})$ and $A_{ul}(Fe II_i)$ are the Einstein coefficient
for spontaneous emission of fluorescence transition \ion{Fe}{2}*$_{j}$ and
resonance transition \ion{Fe}{2}$_j$, respectively. Thus, $P_{fluo}(Fe
II_{i})$ is simply the possibility for a resonance photon \ion{Fe}{2}$_i$ to
be re-emitted as a \ion{Fe}{2}* photon after an absorption event.

\begin{figure}
\includegraphics[width=15cm]{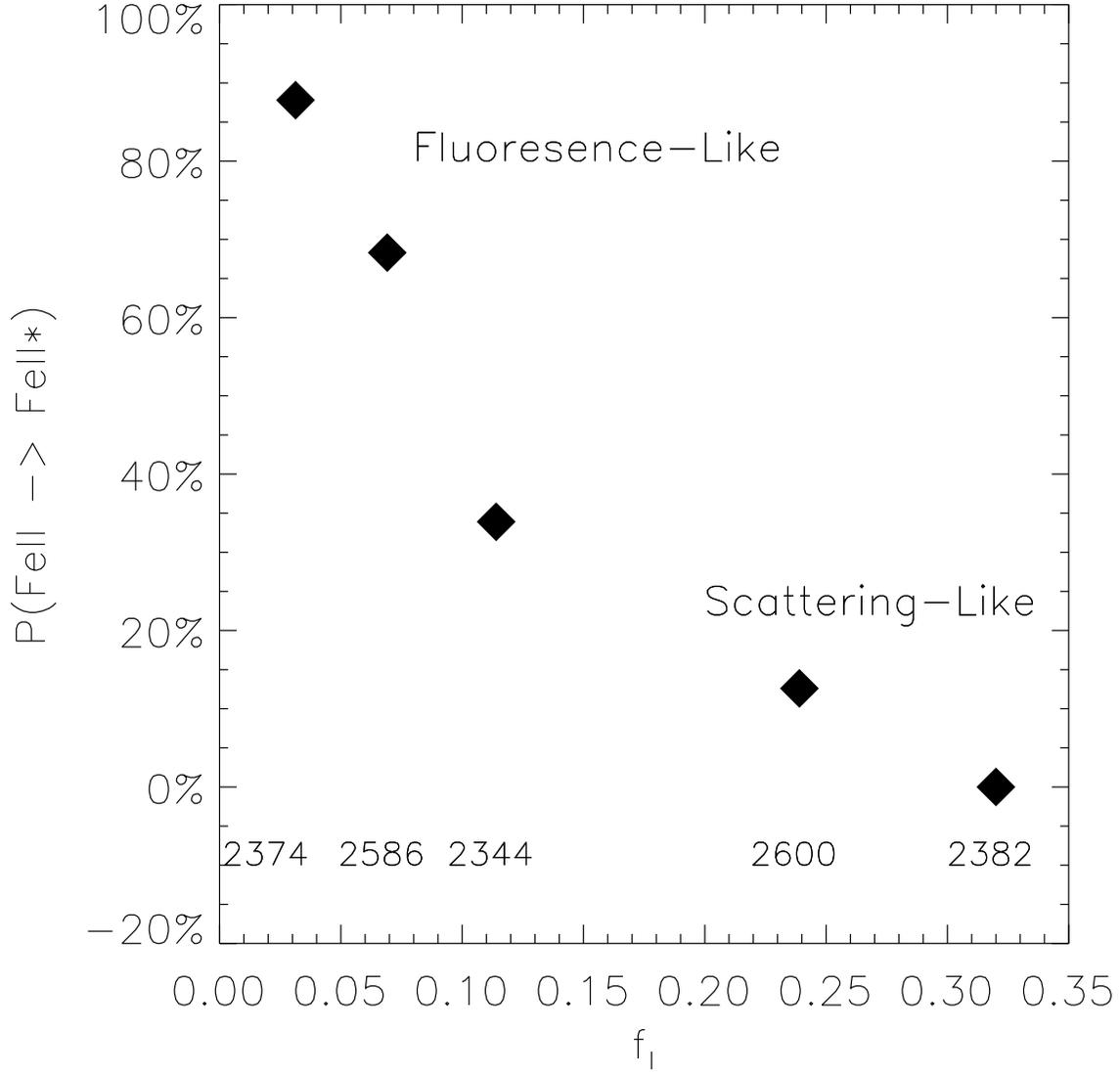}
\caption{Fluorescence Probability from \ion{Fe}{2} $\rightarrow$ \ion{Fe}{2}* 
plotted as a function of line oscillator strength.
$P_{fluo}(Fe II_{i})=\frac{\sum_{j}A_{ul}(Fe II*_{j})}{\sum_{j}{A_{ul}(Fe II*_j)}+{A_{ul}(Fe II_i)}}$.} \label{fig:os_conv}
\end{figure}

$P_{fluo}$ and $f_l$ of \ion{Fe}{2} transitions are listed in
Table~\ref{tab:abs}.  For the \ion{Fe}{2} absorption lines, line profiles
are entirely controlled by $f_l$ and $P_{fluo}$. 
Before discussing how this is achieved, 
we first point out that there exists a monotonic decreasing 
relationship between $f_l$ and $P_{fluo}$ for the five \ion{Fe}{2} 
transitions of concern in this study, which is
shown in Figure~\ref{fig:os_conv}. The \ion{Fe}{2} transitions with higher
oscillator strength always have lower $P_{fluo}$, and this relationship 
can be approximated as $f_l \times P_{fluo} \approx 0.03$.

\begin{figure}
\includegraphics[width=8.4cm]{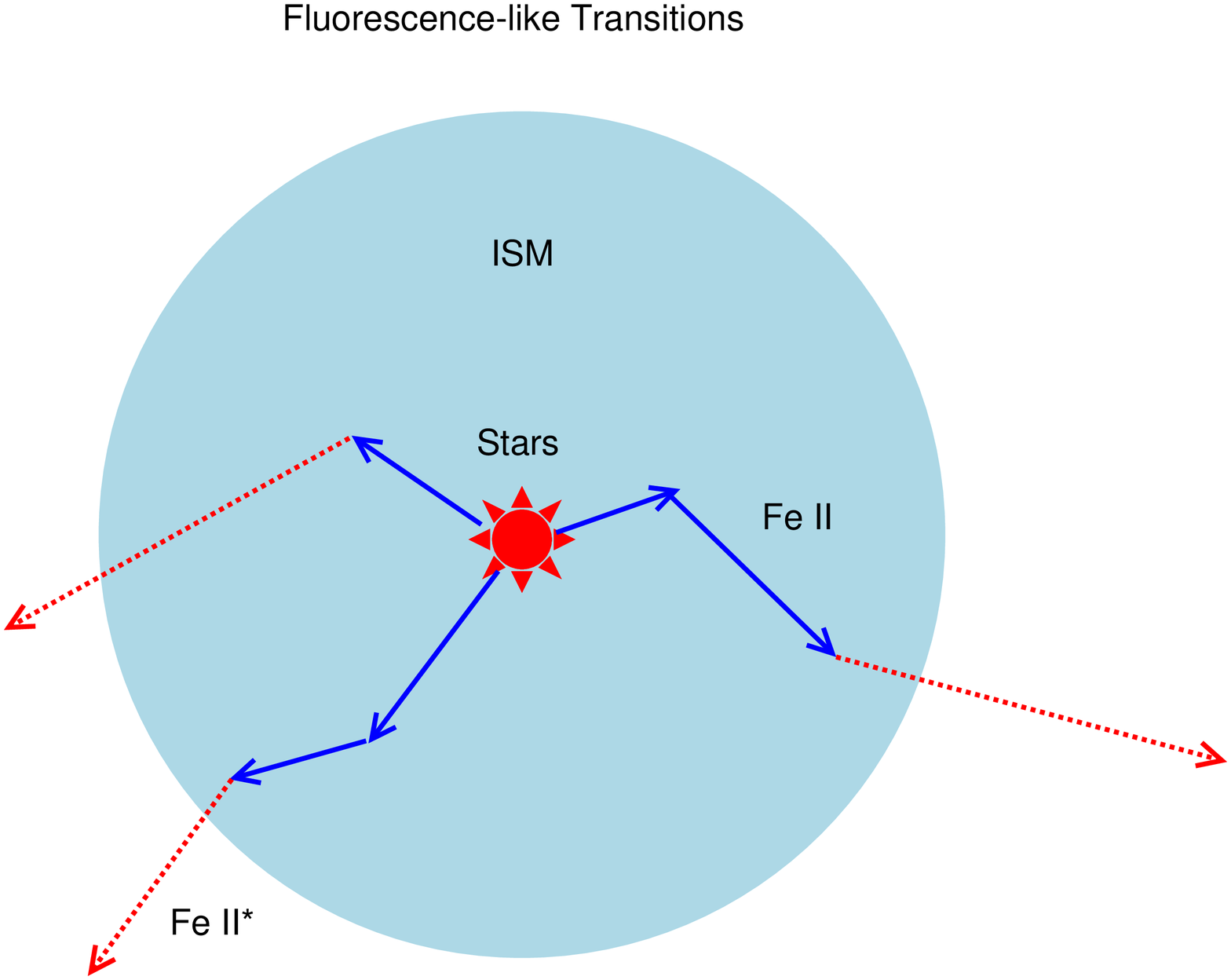}
\includegraphics[width=6.7cm]{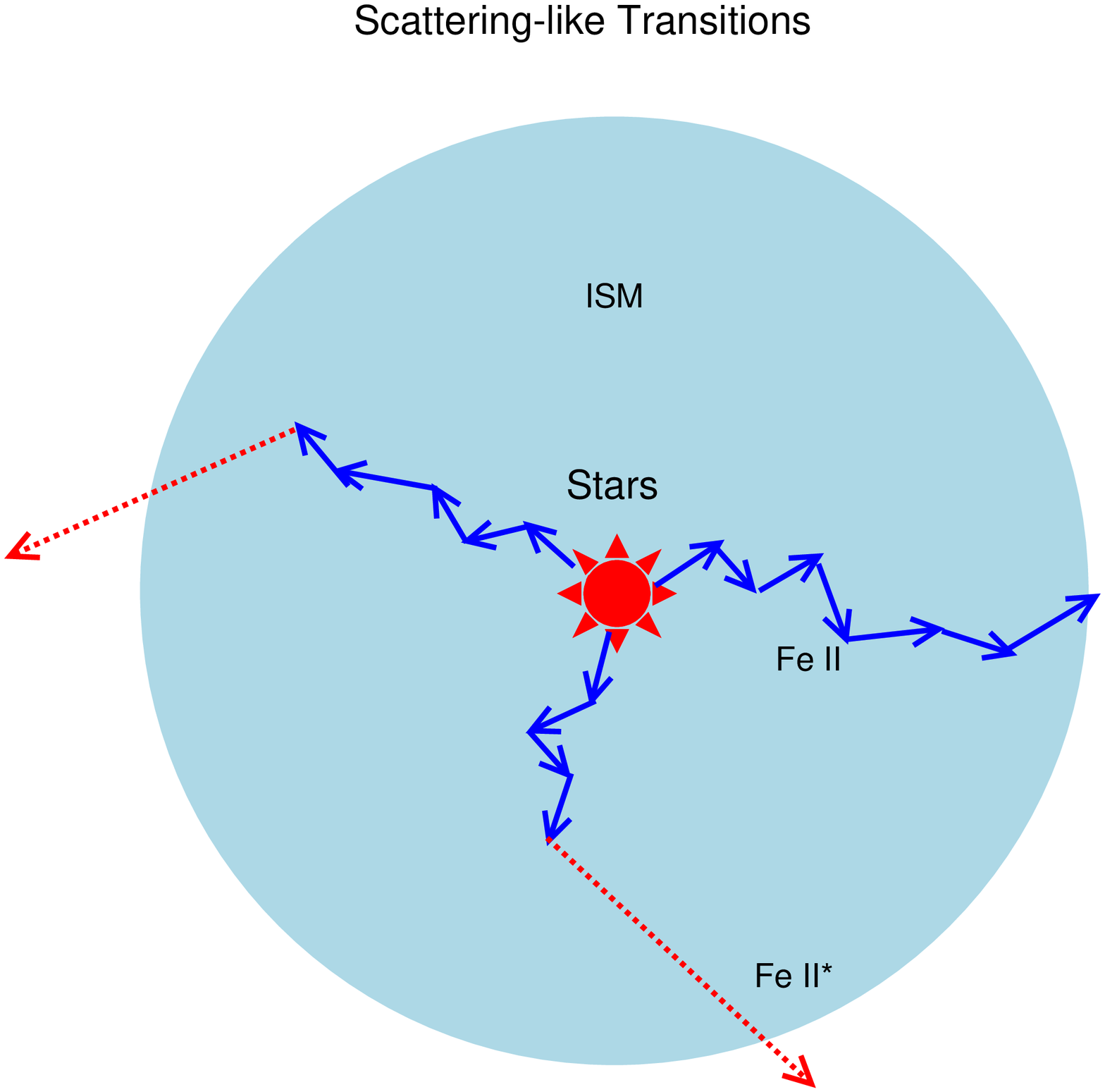}
\caption{Illustration of two different types of radiative transfer: 
fluorescence-like and scattering-like.} \label{fig:propagation}
\end{figure}

Figure~\ref{fig:os_conv} shows that, \ion{Fe}{2} transitions can be placed on a
monotonic sequence: Transitions on the left-top corner have low $f_l$ and high
$P_{fluo}$, resonance photons of these transitions are less absorbed, each
absorption is likely followed by a re-emission of \ion{Fe}{2}* photon. We
refer to them hereafter as ``fluorescence-like'' transitions.  At the opposite
end, transitions on the right--bottom corner have high $f_l$ and low $P_{fluo}$ and
thus resonance photons of these transitions are more frequently absorbed but
less frequently or never give rise to fluorescence transitions. We refer to
them hereafter as ``scattering-like'' transitions.  In an homogeneous,
dustless ISM, the propagation of these two types of resonance photons is
illustrated by Figure~\ref{fig:propagation}. We note here again that 
Figure~\ref{fig:propagation} only represents a special case in a static ISM. 
The propagation of photons in an outflow wind further depends on velocity gradient in the flowing fluid, 
as continuum photons are absorbed in regions those where the co-moving wavelength of photons are Doppler 
shifted to local resonance wavelength. 

\subsection{Equivalent widths of Absorption Lines}

To measure the EWs of \ion{Fe}{2} and \ion{Mg}{2}
absorption lines, we combined Sample [\ion{O}{2}] and Sample Abs to achieve higher S/N
ratio. We used only the 83 spectra with common coverage the rest frame
spectral range $2200-2900$ \AA. We fit a Gaussian profile to each absorption 
line, since several lines are slightly blended (\ion{Fe}{2} $2586/2600$ 
and \ion{Mg}{2}$2796/2803$) and can not be
measured by direct integration of each line's spectrum. Although the intrinsic
line profile should be asymmetric due to the outflow, our spectral resolution
is too low to observe this effect, and the absorption line profiles show small
deviation from the Gaussian one. The bootstrap resamplings of the spectra show
that internal errors of our measures are $\approx 10$\% or less. The results 
are listed in Table~\ref{tab:abs}.

The oscillator strengths of the five \ion{Fe}{2} lines range over one order of
magnitude, and including the \ion{Mg}{2} doublet further extended this range
by a factor of 2. In contrast, their EWs span a relatively
narrow range from $1.5-2.2$.  The line with the lowest EW, \ion{Fe}{2} $2374$,
also has the lowest oscillator strength $f_l=0.0313$. This has been
interpreted in previous work as evidence that the lines are saturated. 
However, as shown in P11, Fig 5, in an expanding wind model the presence of scattered 
emission could mimic the effect of partial covering of the continuum source. 
This is because a) in an outflow wind, transitions with higher
$f_l$ are associated with stronger redshifted P-cygni emission features from
back-scattered photons, which are likely blended with absorption lines in our
low resolution, average spectrum. We will show evidence of this effect in
Section 3.7; b) absorption lines with high $f_l$ could be ``re-filled'' by
scattered photons. According to the $P_{fluo}-f_l$ diagram, low $f_l$ lines
are efficiently converted to fluorescence photons instead of being resonantly
scattered.
 
It is unclear to us whether these effects can reproduce the narrow range of
EWs that we observed, but it is important to keep in mind that
that contributions from unsaturated absorption lines should not be ignored.

\subsection{Modeling Line Profiles}

\begin{figure}
\includegraphics[width=15cm]{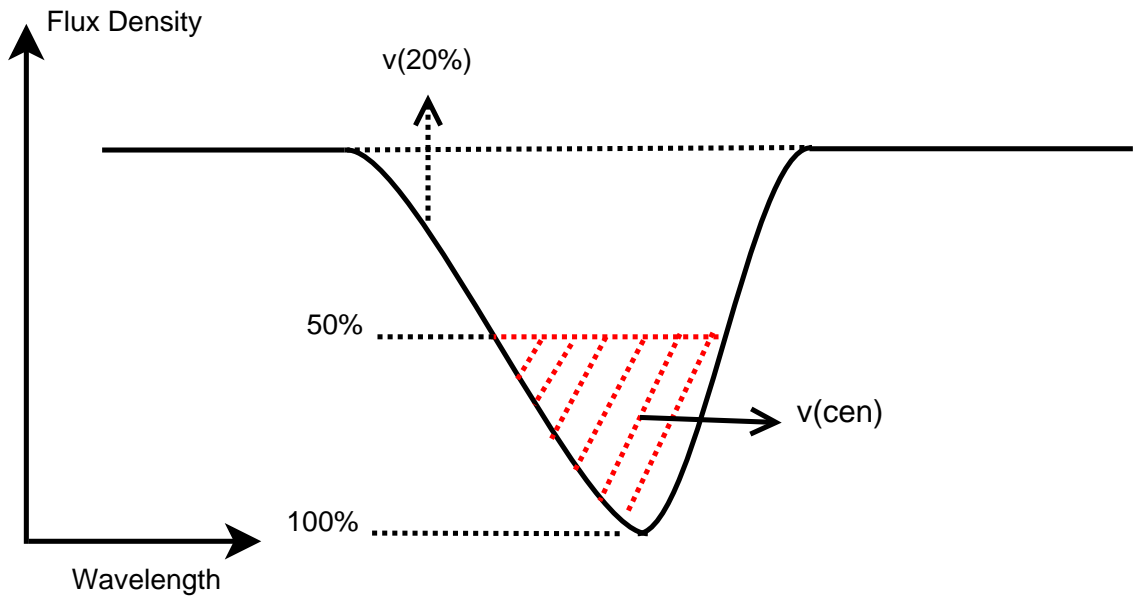}
\caption{Illustration of how we quantify line profile: $v_{20\%}$ and $v_{cen}$.} \label{fig:measure_v}
\end{figure}

In this work, we model line profiles by means of a two parameter
representation.  As illustrated by Figure~\ref{fig:measure_v}, first we
measure the extent of blue line wing by $v_{20\%}$, the velocity where
$20\%$ of the continuum is absorbed.  We then calculate the velocity
centroid of absorption as:

\begin{equation}
\lambda_{cen}=\frac{\int_{\lambda_{(50\%blue)}}^{\lambda_{(50\%red)}}
  \tau(\lambda) \cdot \lambda \dif \lambda}
       {\int_{\lambda_{(50\%blue)}}^{\lambda_{(50\%red)}} \tau(\lambda) \dif
         \lambda }
\end{equation}

\begin{equation}
v_{cen}=c \cdot \frac{\lambda_{cen} - \lambda_{rest}}{\lambda_{rest}}, 
\end{equation}  

where $\lambda_{(50\%blue)}$ and $\lambda_{(50\%red)}$ are the wavelengths at
which the depth of absorption trough decrease to $50\%$ of the peak depth on
the blue side and the red side, respectively.  The wavelength centroid
$\lambda_{cen}$ is calculated between $\lambda_{(50\%blue)}$ and
$\lambda_{(50\%blue)}$, weighted by apparent optical depth
$\tau(\lambda)$. $\tau(\lambda)=\log(1-\frac{F(\lambda)}{F_c(\lambda)})$,
where $1-\frac{F(\lambda)}{F_c(\lambda)}$ is the depth of absorption trough at
wavelength $\lambda$.

In the following sections, we discuss how $v_{20\%}$ and $v_{cen}$ depend on
locations of \ion{Fe}{2} in $P_{fluo}-{f_l}$ diagram.

\subsection{Dependence of $v_{20\%}$ on Line Properties}

Consider an expanding wind with increasing velocity and decreasing density 
moving away from the galactic center. To first order, $v_{20\%}$ increases
with $f_l$, since lines with high $f_l$ can trace down to low density regions
at large radii and hence higher velocities. The five \ion{Fe}{2} absorption
lines considered here span a factor of 10 in $f_l$, suggesting a substantial
spread in the extent of line wing. On the other hand, 
at a fixed oscillator strength, low $P_{fluo}$ lines
suffer more from scattered refilling and thus should be less extended in
velocity to the blue. Therefore, to a certain degree, 
due to the inverse $P_{fluo}-f_l$ relationship, the increasing of 
the absorption EW at higher oscillator strengths is counter-blanced by the
increasing of the scattered refilling.  However, it is easy to perceive that the 
scattered refilling is still a secondary effect in shaping line profiles, 
since any scattering must be initiated by a resonance absorption.

\subsection{Dependence of $v_{cen}$ on Line Properties}

If the re-shaping of absorption lines by re-emitted photons is ignored, 
$v_{cen}$ simply traces the highest density region of the wind, which should be constant for
different \ion{Fe}{2} lines. However, P11 shows that the peak velocity of
absorption lines could be severely blended with nearby, redshifted P-cygni
emissions, which is built from photons scattered by receding part of
wind. This effect could be even more significant in our low resolution data.
``Scattering-like'' transitions are associated with strong P-cygni emissions,
as their corresponding resonance photons are more scattered and less
fluoresced. For these transitions, the P-cygni emission is blended with absorption
line and ``pushes'' the line centroid $v_{cen}$ toward shorter wavelengths. At
the opposite, ``Fluorescence-like'' transitions suffer less contamination from
P-cygni emission, as their corresponding photons have long mean free path and
high $P_{fluo}$.  Therefore $v_{cen}$ can also be positively correlated with
$f_l$. This effect has been clearly demonstrated in P11, Figure 23.

\subsection{Profiles of \ion{Fe}{2} and \ion{Mg}{2} Absorption Lines in Average Spectrum}

Similar to previous analyses, we combined Sample [\ion{O}{2}] and Sample Abs and
selected galaxies with wavelength coverage $2200-2700$ \AA. The
average spectra are first smoothed using a boxcar of 3 pixels. 
Note that, although Sample Abs is characterized by large uncertainties 
in systematic redshifts, which are estimated from cross-correlation, 
since our primary interest is relative differences in line profiles, 
as long as our comparison is carried out on the same set of spectra, 
the uncertainty of the absolute systemic redshift, which shifts all 
lines by an equal amount, is canceled out.

\begin{figure}
\includegraphics[width=17cm]{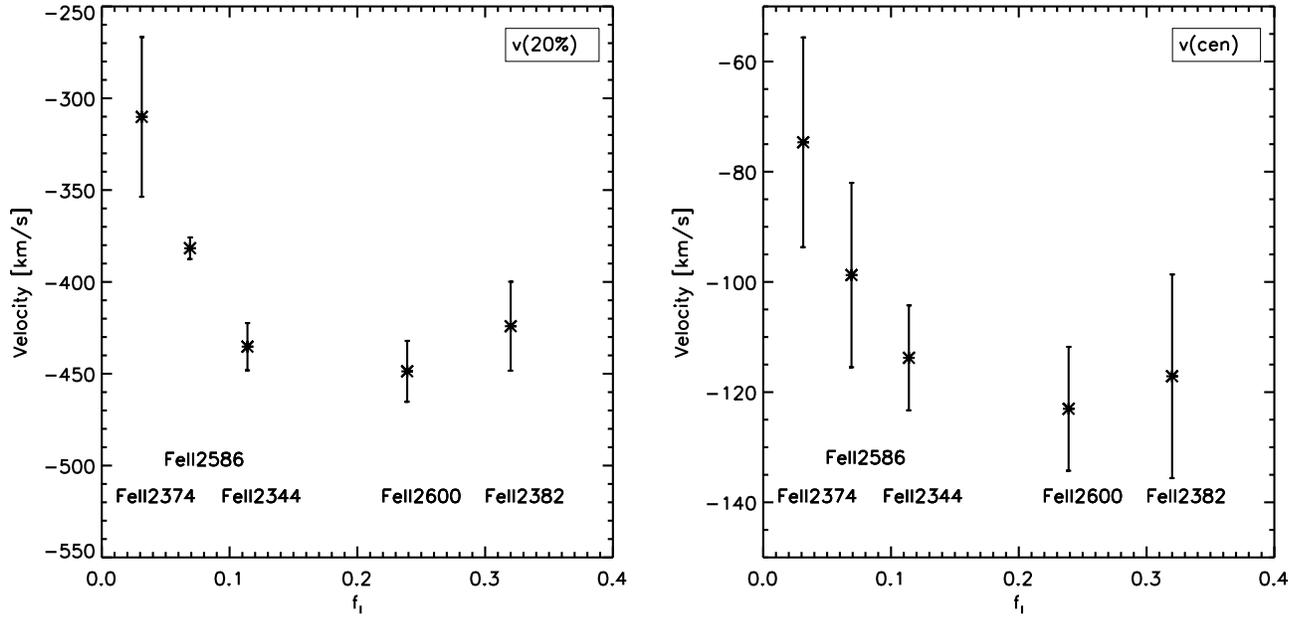}
\caption{Variation of $v_{20\%}$/$v_{cen}$ as a function of line oscillator strength. For this measurement, 
\ion{Fe}{2} $2382$ is removed of contribution from \ion{Fe}{2}* $2381$ emission.} \label{fig:fe_v}
\end{figure}

At our spectral resolution, the \ion{Fe}{2} $2382$ absorption line is blended
with the \ion{Fe}{2}* $2381$ emission.  Since \ion{Fe}{2}* $2381$ and
\ion{Fe}{2}* $2365$ share same upper level, we scale \ion{Fe}{2}* $2365$ to
remove \ion{Fe}{2}* $2381$ component from \ion{Fe}{2} $2382$. To do this, we
first select a spectral window, $-700-600$ km/s wide, around \ion{Fe}{2}*
$2365$, then we scale the emission line profile in this window by the ratio of
Einstein A Coefficient $A_{2381}/A_{2365}$, and subtract it from the spectra. 
Without this correction, $v_{cen}$ of \ion{Fe}{2} $2382$
would be lower by $10-20$ km/s. We plot $v_{cen}$ and $v_{20\%}$ of \ion{Fe}{2} 
in Figure~\ref{fig:fe_v} as a function of $f_l$.  For lines of low $f_l$, 
such as \ion{Fe}{2} $2374$, \ion{Fe}{2} $2586$, FeII $2374$, the 
relationship between $v_{cen}$ and $v_{20\%}$ with $f_l$ is in agreement 
with the discussion above---assuming an
accelerating, smooth wind, both increase with $f_l$. This trend starts to
flatten at high velocities.

\begin{figure}
\includegraphics[width=14cm]{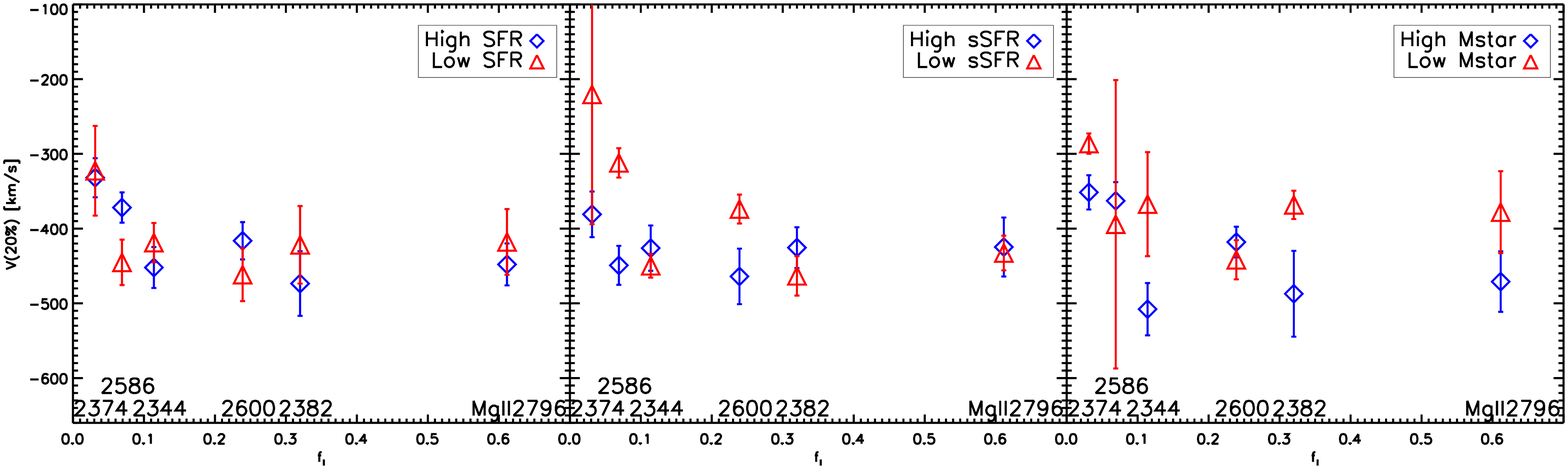}

\includegraphics[width=14cm]{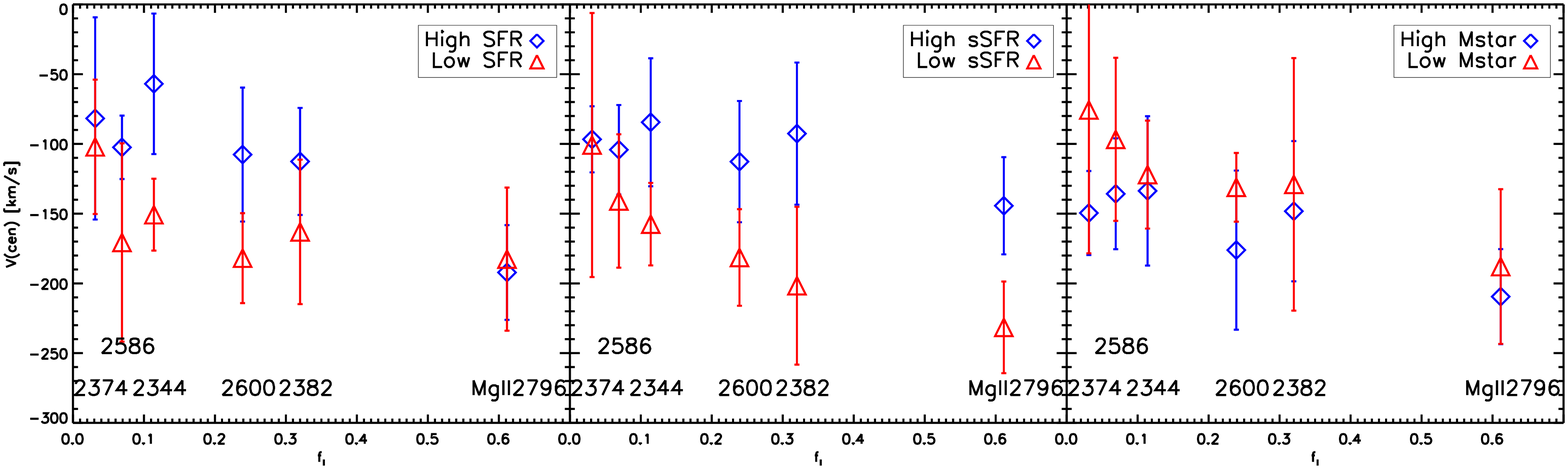}
\caption{Comparing $v_{20\%}$ (top panel) and $v_{cen}$ (lower panel) 
of subsamples with high/low SFR, sSFR and stellar mass,
only galaxies with [\ion{O}{2}] 3727 detection are included.
Red triangles represent low SFR/sSFR/stellar mass, blue diamonds represent high 
SFR/sSFR/stellar mass.} \label{fig:ems_div}
\end{figure}

To further explore kinematic properties of \ion{Fe}{2}, we selected galaxies
from Sample [O II] that have common coverage of the spectral range
$2200-2900$ \AA. We have then divided these 28 galaxies, for which accurate
system redshift is available, into two subsamples by their median
star--formation rate (SFR), $18.95 M_{\odot}yr^{-1}$, specific star formation
rate (sSFR), $-8.62 yr^{-1}$, and stellar mass, $10^{9.9} M_{\odot}$, respectively, 
and have averaged them. We compare the values of $v_{cen}$ and $v_{20\%}$ of each pair 
of subsamples in Figure~\ref{fig:ems_div}, $v_{20\%}$ and
$v_{cen}$ of \ion{Mg}{2} 2796 are also plotted here. Note that since 
\ion{Mg}{2}$2803$ is blended with \ion{Mg}{2}$2796$, $v_{20\%}$ of
\ion{Mg}{2}$2803$ is not measurable in average spectra.

Similar to terminal velocity, $v_{20\%}$ strongly depends on the
high--velocity components of ourflowing fluids. 
Several previous studies have found that $v_{terminal}$ scales
roughly with the star--formation rate as SFR$^{0.3}$ for \ion{Mg}{2} doublet
(Martin et al. 2005; Weiner et al. 2009; Banerji et al. 2011), suggesting that
outflow wind in more intensively star--forming galaxies are accelerated to
higher speed. Using our subsample of 28 galaxies, however, we find that
$v_{20\%}$ has a stronger dependence on stellar mass and on specific star--formation rate, 
as systematic offsets between high M$_*$/sSFR and low M$_*$/sSFR 
samples are higher than $1\sigma$ uncertainty of $v_{20\%}$, 
this might suggest that gravity also plays an important in the kinematics of the wind.
On the other hand, $v_{cen}$ is possiblly related to sSFR, 
as indicated by systematic offsets between subsamples.
Note that unlike $v_{20\%}$, $v_{cen}$ increases with
decreasing sSFR. 

\begin{figure}
\includegraphics[width=14cm]{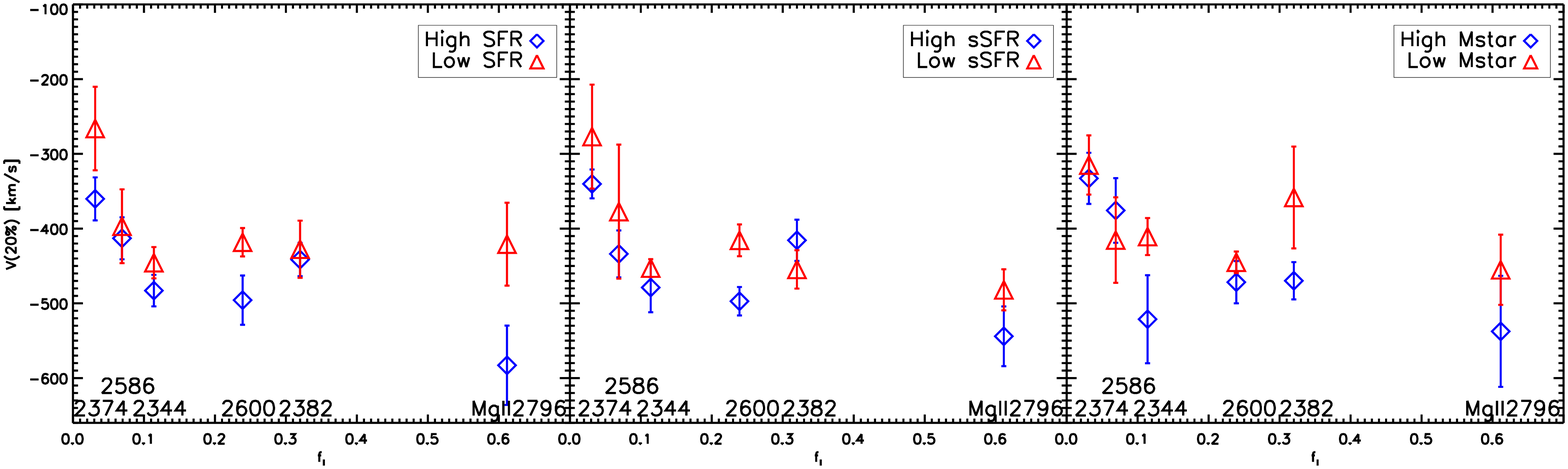}

\includegraphics[width=14cm]{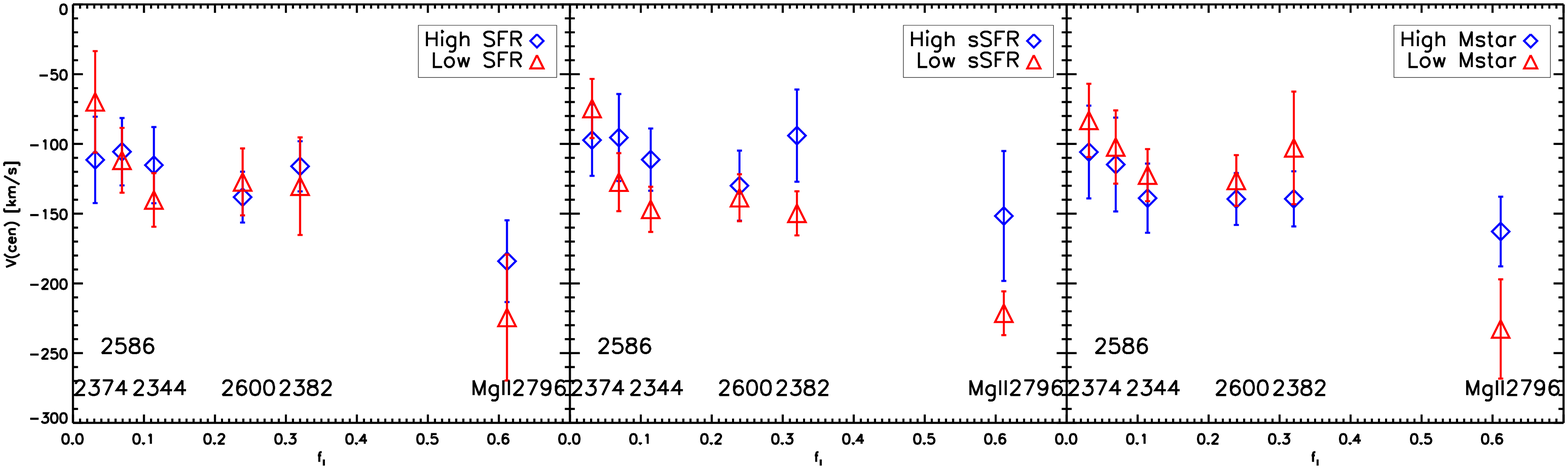}
\caption{Similar to Figure~\ref{fig:ems_div}, but include galaxies with redshifts 
determined from cross-correlation.} \label{fig:femg_div}
\end{figure}

The velocities measured in the average spectra of Sample 
[\ion{O}{2}] have large uncertainties due to the limited sample size. 
Therefore we have repeated the above analysis
including another 55 galaxies from Sample Abs with coverage of the
$2200-2900\AA$ spectral range. The results are shown in Figure~\ref{fig:femg_div}.
Distributions of $v_{20\%}$ and $v_{cen}$ are in agreement 
with Figure~\ref{fig:ems_div}. Note that, since galaxies in Sample Abs do not have
accurate systemic redshift estimates, only relative differences in $v_{20\%}$ and $v_{cen}$
measured from the same set of spectra are meaningful, vertical offsets 
between cross-correlated subsamples could be potentially dominated by 
transitions other than \ion{Fe}{2} and \ion{Mg}{2}.

We also find that, there is a trend of increasing EW(\ion{Mg}{2}) and EW(\ion{Fe}{2}) 
in massive and high--SFR galaxies observed in GMASS galaxies, 
and this is consistent with previous studies (Erb et al. 2012, Kornei et al. 2012, 2013, Martin et al.2012). 
To compare our results with that derived by Erb et al. (2012), which are based on a sample with
similar redshift range, we use exactly the same criteria by Erb et al. (2012) to split galaxies 
into subsamples by their M$_*$, SFR, E(B-V) and age. 
The sample of Erb et al. (2012) is selected based on the rest UV colors 
of the galaxies (the so--called ``BM'' color criteria) and $R$ magnitude. 
In their study, composite spectra are constructed based 
on galaxies properties of SFR, M*, Age and E(B-V). 
The trend between the profiles of absorption lines in the
wind and the sSFR has not been directly studied by Erb et al. (2012). However,
their comparison based on stellar age shows that EW(\ion{Mg}{2}) is enhanced in 
old and massive galaxies, which also exhibit a mean sSFR 0.8 dex lower 
than the young and low--mass galaxies. 

\begin{figure}

\includegraphics[width=17cm]{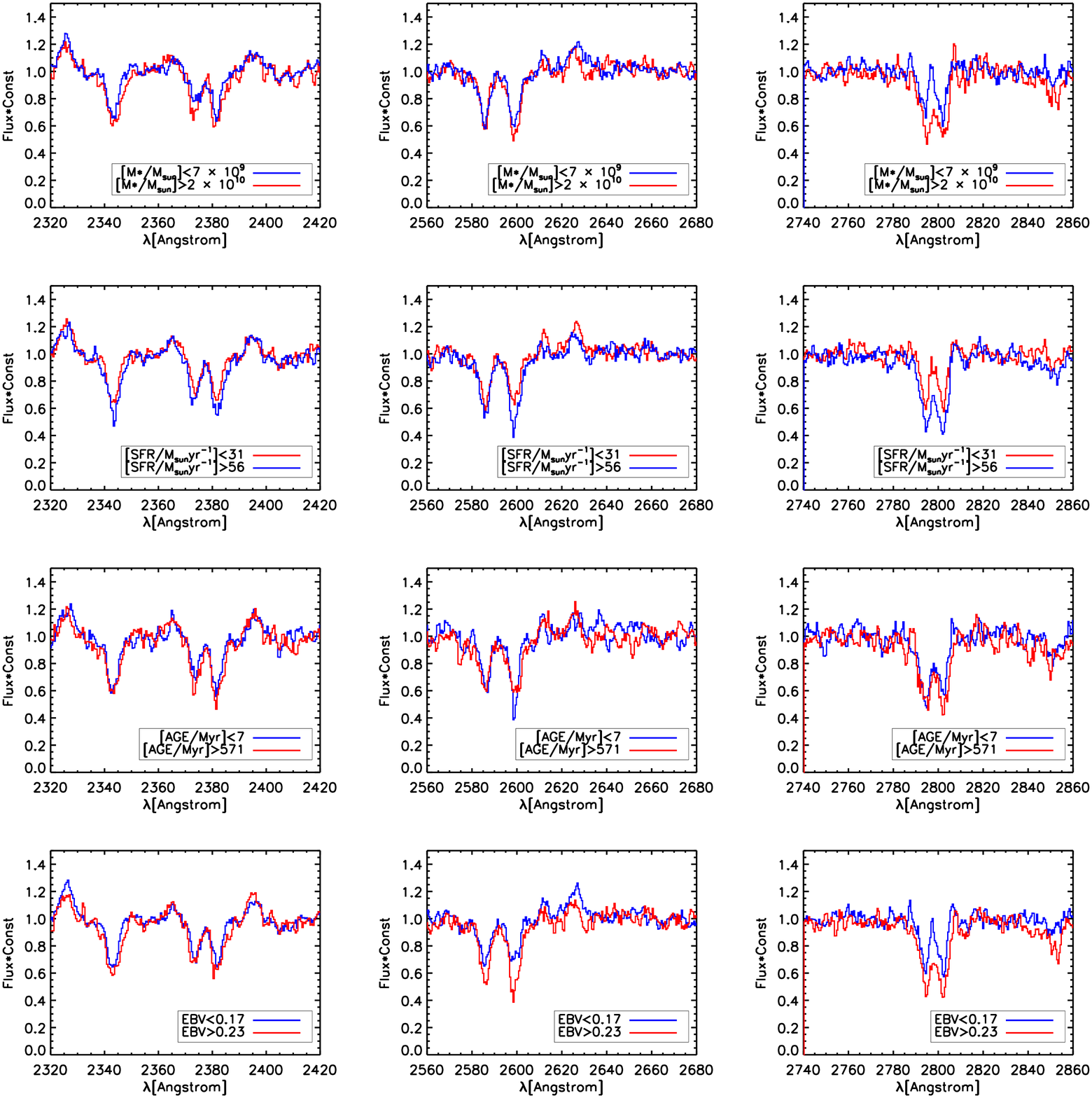}

\caption{The galaxies are split into subsamples by $M_*$, SFR, age and E$(B-V)$, 
using the same criteria used by Erb et al. (2012). The composite spectra of each pair of 
subsamples with low/high $M_*$, SFR, age and E$(B-V)$ are plotted 
together for comparison.} \label{fig:comp_erb}
\end{figure}

The composite spectra of each pair of subsamples are plotted in comparison 
in Figure~\ref{fig:comp_erb}. The mean SFR/$M_*$/Age/E(B-V) of each
subsample are listed in Table~\ref{tab:mean_sub} and the EWs of \ion{Fe}{2} and \ion{Mg}{2} are listed in 
Table~\ref{tab:ew_sub}. Again, since Erb et al. use the Chabrier IMF, we scale up 
their criteria of SFR and M* by a factor of 1.8 since our measurements are based on the Salpeter IMF. 
The typical S/N of composite spectra is $17-27/pixel$.
The general trends of \ion{Mg}{2} are increasing EWs in massive/high SFR galaxies. 
Also, MgII absorption is slightly weaker in the young-age subsample than in the old-age subsample, both in agreement
with Erb et al. (2012). On the other hand, EWs of the \ion{Fe}{2} absorptions show smaller variation
across subsamples. All these trends are similar to that found by Erb et al. (2012) 
and Kornei et al. (2012). 

\section {\ion{Fe}{2}* Emissions in GMASS Galaxies}

\subsection{Relative Line Strengths of \ion{Fe}{2}*}

Since emission features around \ion{Fe}{2} are not resolved in our low
resolution data, the possibility can not be ruled out that the \ion{Fe}{2}*
emission features are blended with other \ion{Fe}{2} transitions. The energy
level configuration of Fe$^{+}$ is among the most complex among ion species in
astrophysical environments, as Fe$^{+}$ has 6 valence electrons in its
outermost shell and exhibit substantial fine structure splittings.  If Fe$^+$
is excited by mechanisms other than continuum fluorescence ---which is true in
AGN broad line regions--- the emission features could be broad and could be
associated with thousands of transitions (Vestergaard \& Wilkes 2001, Sigut \&
Pradhan 2003).  In this work, we are not trying to explore other possible
excitation mechanisms (collision excitation, recombination, etc.), and we
focus only on continuum fluorescence and examine whether this simple
excitation mechanism offers a satisfactory explanation for the observed
\ion{Fe}{2}* emissions in $z=1.0-2.6$ galaxies.

Assuming a flat incident continuum between $2300-2700\AA$, 
the EW of an \ion{Fe}{2}* emission line is proportional to the 
fraction of the incident resonance photons being eventually fluoresced 
through that particular \ion{Fe}{2}* transition. For a transition 
\ion{Fe}{2}*$_{i}$, we define the total conversion fraction as

\begin{equation}
F(Fe II*_{i})=\frac{N_{fluo}(Fe II*_{i})}{N_{resonance}(Fe II_{i})}, 
\end{equation}

where $N_{resonance}(Fe II_{i})$ is the number of the incident resonance
photons associated with the $Fe II*_{i}$ transition, and $N_{fluo}(Fe
II*_{i})$ is the number of the \ion{Fe}{2}$*_{i}$ photons being eventually
produced.  For a single absorption event, $F(Fe II*_{i})$=$P_{abs, single}(Fe
II*_{i}) \times P_{fluo, single}(Fe II*_{i})$, where $P_{abs, single}$ is the
probability of absorption and $P_{fluo, single}$ has been defined by Eq(1).
For multiple absorption events, the $i_{th}$ absorption event and the
${i+1}_{th}$ absorption event are not independent, and $F(Fe II*_{i})$ depends 
on the specific process of radiative transfer.

This problem could be simplified in two extreme cases: the optically thick
limit and the optically thin limit. In the optically thick limit, after
multiple absorption and re-mission, all resonance photons are eventually
fluoresced. If, for instance, \ion{Fe}{2}*$_{i}$ is the only permitted
fluorescence transition from its upper level, $F(Fe II*_{i})=100\%$.
Otherwise, decaying through other fluorescence transition must be accounted.
Without loss of generality, the total conversion fraction for \ion{Fe}{2}* in
the optically thick limit is:

\begin{equation}
F_{thick}(Fe II*_{i})=\frac{A_{ul}(Fe II*_{i})}{\sum_{j}{A_{ul}(Fe II*_{j})}}, 
\end{equation}  

where $A_{ul}(Fe II*_{i})$ is the Einstein coefficient for spontaneous
de-excitation to excited fine structure levels and $\sum_j{A_{ul}(Fe II*_{j})}$ 
is the sum of $A_{ul}$ coefficients for all permitted Fe II*
transitions decaying from the same upper level as $Fe II*_{i}$.

In the optically thin limit, the total conversion fraction of \ion{Fe}{2}*$_i$
is

\begin{equation}
F_{thin}(Fe II*_i) = P_{abs, single}(Fe II*_{i}) \times P_{fluo, single}(Fe
II*_{i}).
\end{equation}  

Here $P_{abs, single} \propto f_l(Fe II_i)$, where $f_l(Fe II_i)$ is the
oscillator strength of \ion{Fe}{2}$_i$. 

\begin{figure}
\includegraphics[width=14cm]{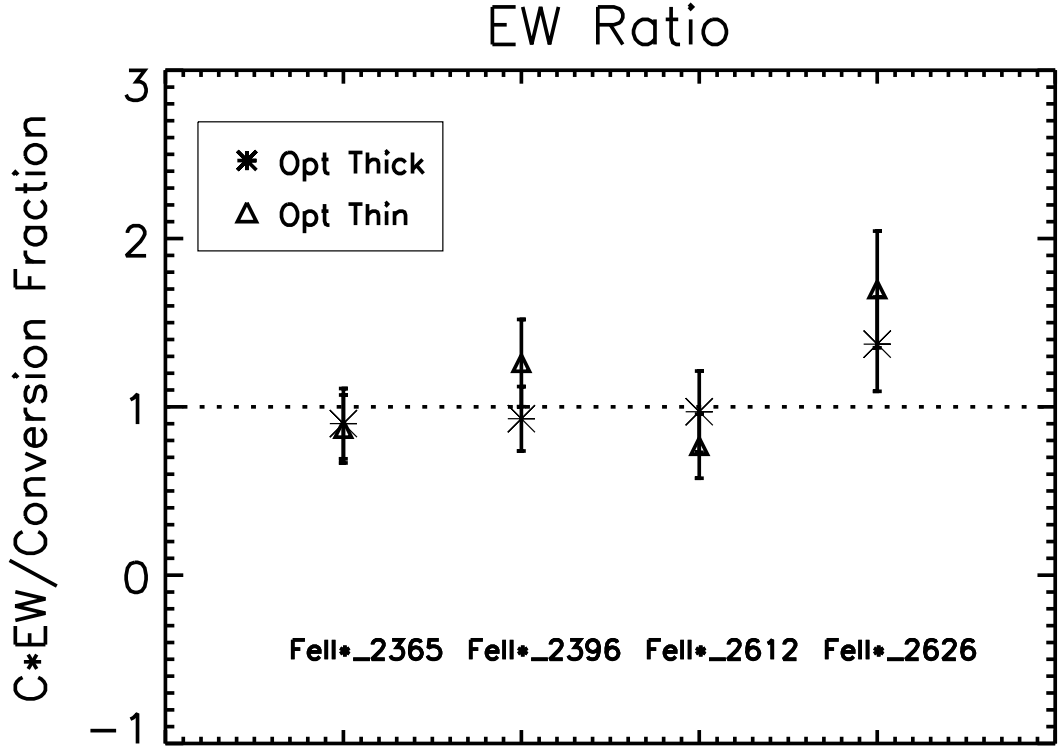}
\caption{Comparing EWs of \ion{Fe}{2}* lines. The EW of each \ion{Fe}{2}* line is divided 
by its conversion fraction from \ion{Fe}{2} $\rightarrow$ \ion{Fe}{2}*, the latter is proportional to
$f_{thick}(Fe II*_{i})=\frac{A_{ul}(Fe II*_{i})}{\sum_{j}{A_{ul}(Fe II*_j)}}$ in optically 
thick limit (stars) and is proportional to $f_l(Fe II_j) \times P_{single}(Fe II*_{i})$
in optically thin limit (open triangles), where
$ P_{single}(Fe II*_{i})=\frac{A_{ul}(Fe II*_{i})}{\sum_{j}{A_{ul}(Fe II*_j)}+{A_{ul}(Fe II_i)}}$, see text for details.}
 \label{fig:ew_ratio}
\end{figure}

In Figure~\ref{fig:ew_ratio} we show the ratios $\frac{EW(Fe II*_{i})}{F_{thin}(Fe II*_{i})}$
and $\frac{EW(Fe II*_{i})}{F_{thick}(Fe II*_{i})}$ for all \ion{Fe}{2}* transitions. 
The ratios are re-normalized such that the average ratio in each 
case equals 1 ($AVG(\frac{EW}{F_{thin}})=1$ and $AVG(\frac{EW}{F_{thick}})=1$). 
To measure EW(\ion{Fe}{2}, we averaged all
spectra which have common coverage of the range $2200-2700$ \AA. The EWs are
measured by integrating the line profile within specified wavelength ranges,
as listed in Table~\ref{tab:ems}.  To avoid contamination from the neighboring
\ion{Fe}{2} absorption lines, each \ion{Fe}{2} absorption line is fitted by a
Gaussian profile and subtracted from the original spectra prior to the
integration. For a flat continuum, the ratio between EW and conversion
fraction $\frac{EW(Fe II*_i)}{F(Fe II*_i)}$ should be a constant.

  The relative conversion fractions are
  $F(2365):F(2396):F(2612):F(2626)=0.66:1.0:0.66:1.0$ in the optically thick
  limit, and $F(2365):F(2396):F(2612):F(2626)=0.26:0.27:0.31:0.30$ in the
  optically thin limit. In the optically thin limit, the similar conversion
  fractions among \ion{Fe}{2}* lines result from the inverse relationship
  between $P_{single}(Fe II*_j)$ and $f_l(Fe II*_j)$ as illustrated by
  Fig~\ref{fig:os_conv}, $P_{single}(Fe II*_j)*f_l(Fe II*_j)\approx0.03$,
which basically states that lines that are more absorbed are less
fluoresced. This effect leads to a roughly constant line ratio with varying
optical depth. We find that in P11, all except two models predict the line ratio
of \ion{Fe}{2}*2612 and 2626 that fall between the two limits discussed
here. These two exceptions have extreme physical conditions, namely a bipolar
wind with small opening angle ($45^{\circ}$) and sharp edges, and a model
which simply ignores resonant trapping. Since the relative strengths of
\ion{Fe}{2}* lines are nearly constant within a wide range of column density,
they are poor diagnostics of density gradients.  Nevertheless, since
this test shows that the ratios of \ion{Fe}{2}* lines agree
with the scenario of continuum fluorescence, no other mechanism is really
necessarily required to explain their excitation mechanism.

There is, however, an exception to this picture. As will be discussed in the
next section, the \ion{Fe}{2}* transition F(2632) is not detected in the 
average spectrum of the whole sample. And this null detection is not simply a 
result of low S/N ratio.

\subsection{Null Detection of \ion{Fe}{2}* 2632}

\begin{figure}
\includegraphics[width=17cm]{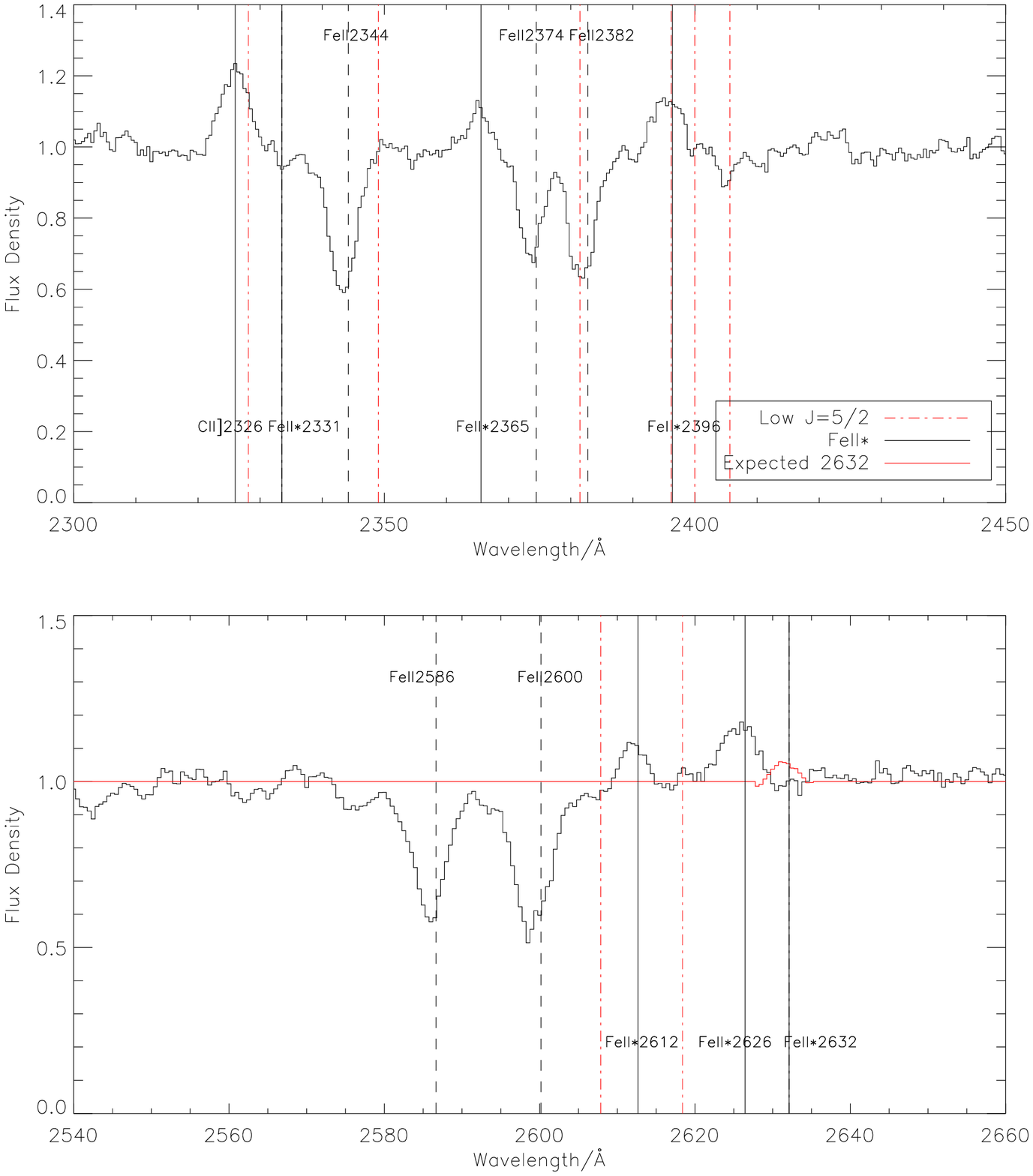}
\caption{Same as Figure~\ref{fig:avg_spec}, with estimated \ion{Fe}{2}* $2632$ emission 
being shown in solid red. Transitions with lower level J=5/2 are marked by
red dash-dot line.} \label{fig:show2632}
\end{figure}

Considering the large EWs of \ion{Fe}{2} resonance lines ($|EW| >
1.3\AA$), one might expect that all \ion{Fe}{2}* lines associated with
resonance transitions should be present in the high S/N average spectrum,
except \ion{Fe}{2}* $2381$, which is blended with \ion{Fe}{2} $2382$. The
emission line \ion{Fe}{2}* $2632$, however, is undetected in our average
spectra. Since \ion{Fe}{2}* $2632$ decays from the same upper level as
\ion{Fe}{2}* 2612, the expected strength of \ion{Fe}{2}* 2632 can be estimated
by scaling \ion{Fe}{2}* $2612$ with the ratio of their Einstein A
Coefficients. To do this, we have extracted the profile of \ion{Fe}{2}* 2612
from a velocity window $[-550-400]$ km/s and multiplied it by
$A_{ul}(2632)/A_{ul}(2612)$.  The expected profile is shown in
Figure~\ref{fig:show2632}, where we see that the 3-sigma detection limit
of the EW of the 2632 emission is $0.1 \AA$ while the expected EW estimated
for the 2612 emission is $0.2 \AA$. 

One possibility for the lack of detection is that the lower level of
\ion{Fe}{2}* $2632$, $J=5/2$, is heavily populated through non-radiative
excitations. If this is true, we should also detect other absorption lines
arising from the same lower level. All transitions with lower level $J=5/2$
and $f_l$ $>0.02$ are marked in Figure~\ref{fig:show2632}. No isolated
absorption line corresponding to these transitions is clearly detected.
Especially, \ion{Fe}{2}* $2405$ and \ion{Fe}{2}* $2400$ have $f_l$ $2.7$ and
$1.4$ times as that of \ion{Fe}{2}* $2632$ and so one should expect obvious
absorption features at these wavelengths if $J=5/2$ is significantly occupied.
We conclude, therefore, that it is unlikely that the null detection is the
result of heavily populated lower level.

\begin{figure}
\includegraphics[width=17cm]{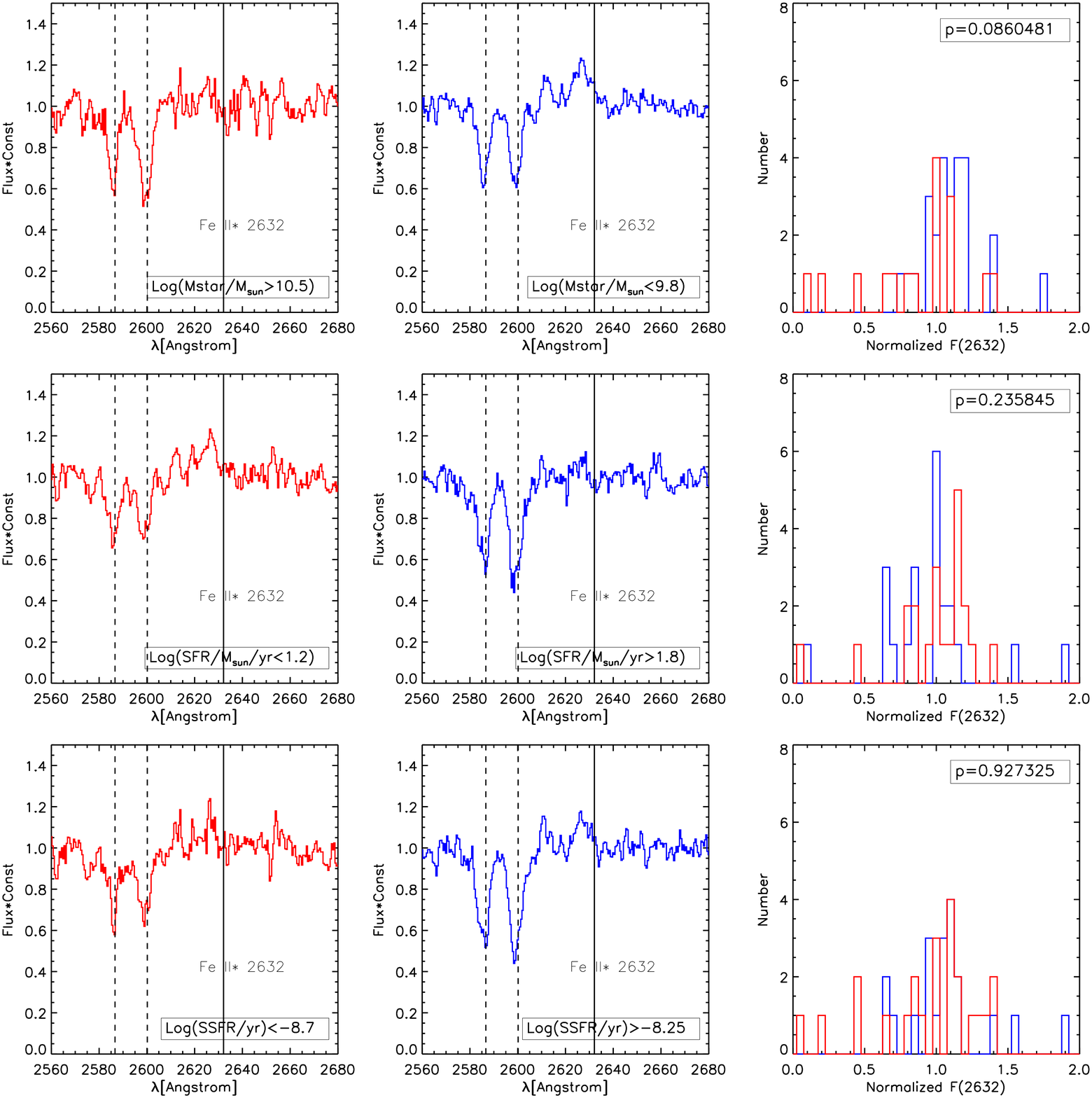}
\caption{Comparing \ion{Fe}{2}* 2632 in subsamples split by M$_*$/SFR/sSFR. 
Left Panels show subsamples with high stellar mass, low SFR and low sSFR, from
top to bottom, respectively. Right panels show low stellar mass, high SFR and high sSFR.
We perform a K-S test on each pair of subsamples to examine
if there is a significant segregation of F(2632). We see a possible separation of \ion{Fe}{2}* 2632 only in 
M*-split subsamples, with a significance level of $8.6\%$ rejecting the null hypothesis that the two 
distributions are drawn from the same parent sample.} \label{fig:com2632}
\end{figure}

\ion{Fe}{2}* $2632$ emission feature could also be obscured by underlying
absorption features in stellar continuum.  In passive galaxies, the most
prominent stellar feature around \ion{Fe}{2}* $2632$ is the B2640 continuum
break (Spinrad 1997, Cimatti et al. 2004), which primarily reflects UV1 close
spaced \ion{Fe}{2} resonance multiplets on the shorter wavelength side of
$2640\AA$ produced in photosphere of late F and G type stars. Stellar absorption 
is actually found to be significant for passive galaxies in the GMASS sample (Cimatti
et al. 2008). To examine whether \ion{Fe}{2}* $2632$ could be obscured by the 
absorption features in our average spectrum, we compare in Figure~\ref{fig:com2632} the
average spectra of galaxies in subsamples with the highest half and lowest
half of stellar mass, SFR, and sSFR. In the last column of Figure~\ref{fig:com2632}, 
the distributions of single-pixel continuum-normalized flux densities measured in individual spectra 
at $2632\AA$ are also plotted for comparison. We perform a K-S test on each pair of subsamples to examine
if there is a significant segregation of F(2632). We see a possible separation of \ion{Fe}{2}* 2632 only in 
M*-split subsamples, with a significance level of $8.6\%$ rejecting the null hypothesis that the two 
distributions are drawn from the same parent sample. In SFR and sSFR divided subsamples, \ion{Fe}{2}* 2632
show no statistically significant separation. 

\begin{figure}
\includegraphics[width=12cm]{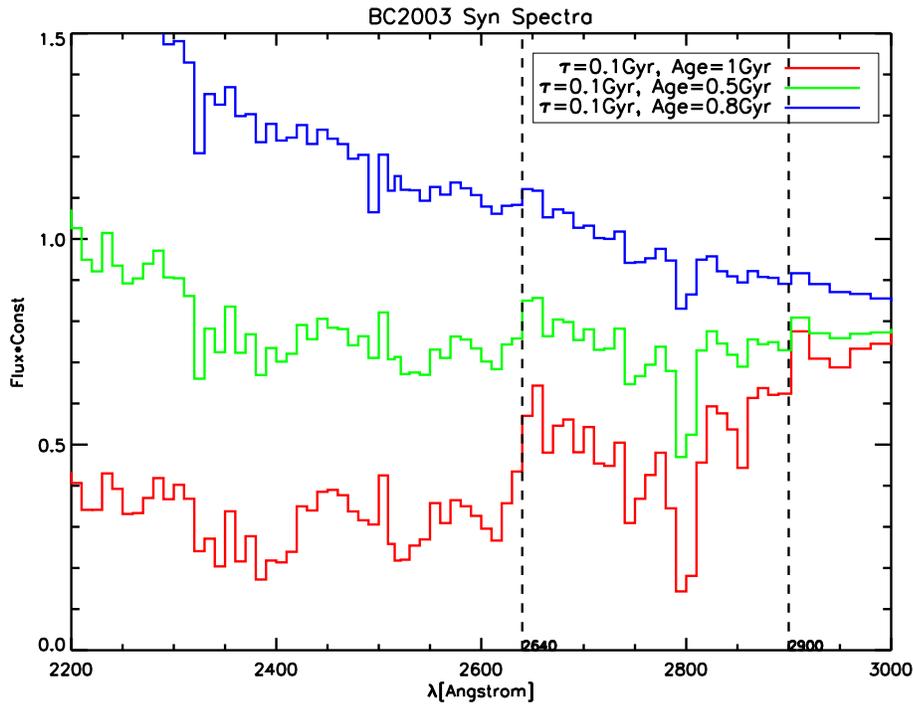}
\caption{Synthesized spectra generated from BC03 model. An exponential decaying star-formation history 
with an e-folding time of 0.1 Gyr is assumed here. The spectra are shown for three different 
ages, $0.3, 0.5$ and $1 Gyr$. The stellar \ion{Fe}{2} absorptions and 2632 break are 
less prominent in younger galaxies.} \label{fig:bc03_spec}
\end{figure}

The intensity of the B2640 break should be related to the location of the 
galaxy in Color-Magnitude Diagram (Spinrad et al. 1997). For a subsample of
passive GMASS galaxies, Cimatti et al. (2008) estimate a decrease of $59\%$
in the continuum level across the $2640$ \AA\ break. Starforming galaxies,
however, should have a less pronounced break. Nevertheless, 
it cannot be ruled out that stellar absorption responsible for 
this discontinuity is potentially sufficient to suppress the FeII* $2632$ emission. 
We show the synthesized spectra generated from the BC03 templates in Figure~\ref{fig:bc03_spec}.  
For a galaxy with Age=0.3-0.5 Gyr, the continuum discontinuity is between a few 
to 10 percent. This is comparable to the expected intensity of \ion{Fe}{2}* 2632 estimated from \ion{Fe}{2}* 2612, 
which has a peak flux density lower than $10\%$ of the continuum level, 
as shown in Figure~\ref{fig:show2632}.
  
\begin{figure}
\includegraphics[width=17cm]{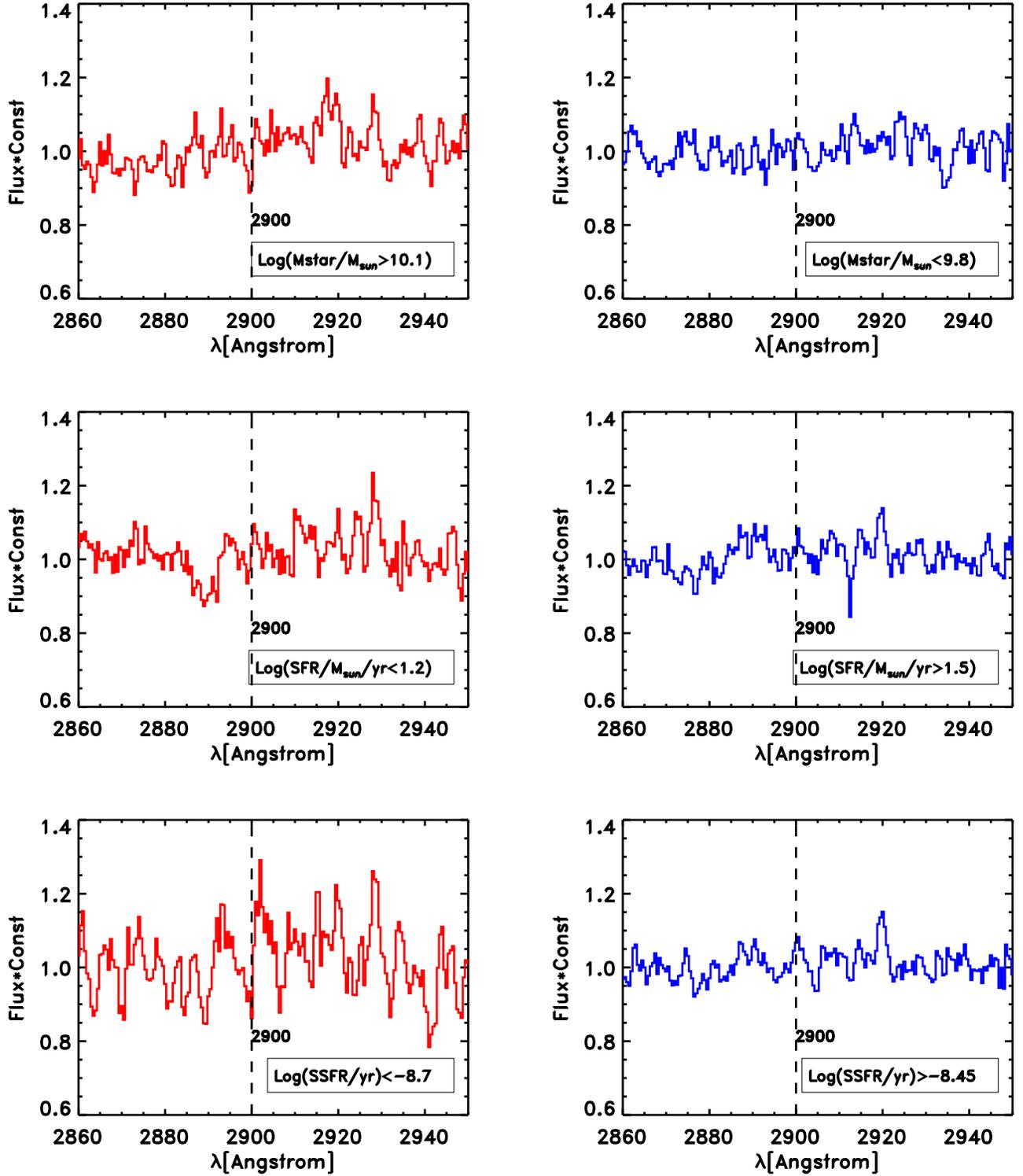}
\caption{Comparing B2900 of subsamples with low/high M$_*$/SFR/sSFR. 
Left Panels show subsamples with high stellar mass, low SFR and low sSFR, from
top to bottom, respectively. Right panels show low stellar mass, high SFR and high sSFR.} \label{fig:com2900}
\end{figure}

\begin{figure}
\includegraphics[width=17cm]{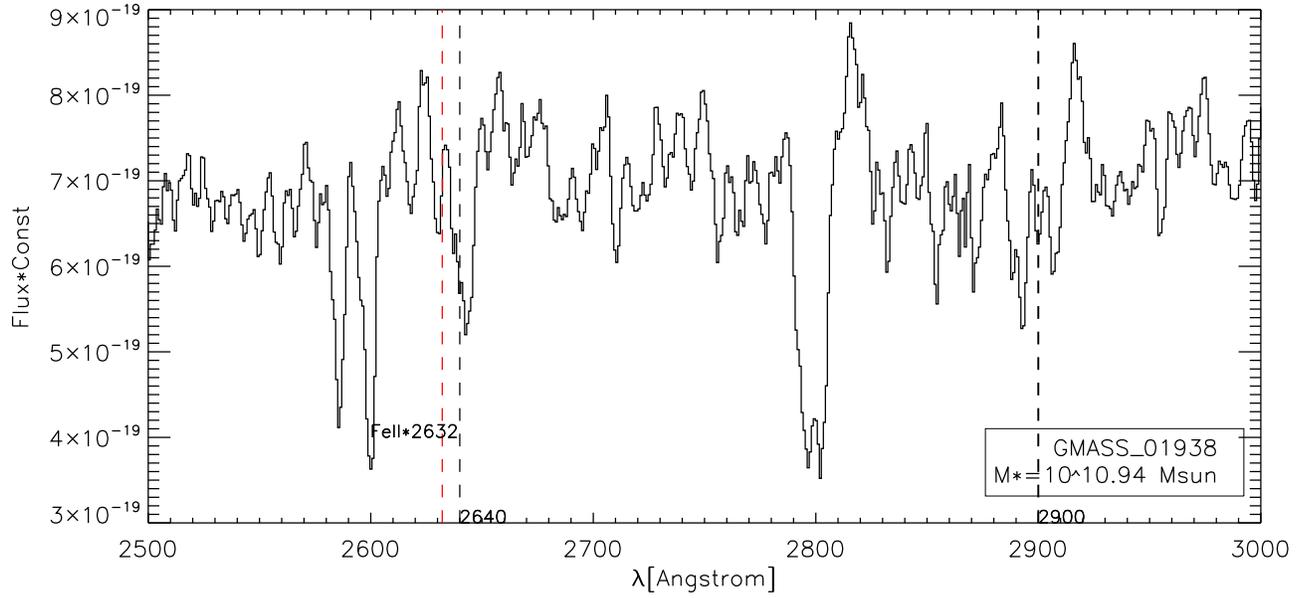}
\caption{The spectrum of GMASS01938, which has a stellar mass of $10^{10.94} M_{\odot}$. 
Absorption features blueward of $2640 \AA$ and $2900 \AA$ are possibly present.} \label{fig:G01938}
\end{figure}

  The S/N ratios and the spectral resolution of the GMASS data do not allow
  us to decompose the stellar absorption and the FeII* $2632$ emission.  One way
  to further test this scenario is to examine the B2900 break, another
  continuum break caused by similar metal absorptions in F and G-type
  stars. (i.e. FeII, FeI etc. Heap et al. 1998). In Figure~\ref{fig:com2900},
  we compare the strengths of B2900 break in the average spectra of the lowest
  half and the highest half sSFR, SFR and stellar mass subsamples. In order to retain
  this feature of continuum break, we simply normalized each individual 
  spectrum by its median value over the wavelength range $2820-2970 \AA$ prior to coaddition, 
  instead of normalizing each spectrum by its continuum. The high
  stellar mass and the low sSFR spectra do show signs of absorption blueward of
  $2900$ \AA, which is absent in the average spectrum of the bluest
  galaxies. We also show a high S/N spectrum of one massive galaxy,
  GMASS-01938, in Figure~\ref{fig:G01938}.  In this single case, absorption
  features are possibly present blueward of $2640\AA$ and $2900\AA$.

  Another important question is whether stellar absorption features could
  contaminate other \ion{Fe}{2}/\ion{Fe}{2}* lines.  As shown in Spinrad et
  al. (1997) and Cimatti et al. (2008), the troughs of stellar-originated
  resonance FeII absorptions in passive galaxies are shallower than the
  B2640 and B2900 discontinuities. In our co-added spectra, if the absence of
  the \ion{Fe}{2}*2632 emission were due to stellar absorption lines blueward of
  B2640, these stellar absorptions should have a integrated intensity roughly
  equal to EW(\ion{Fe}{2}* 2632). By scaling EW(\ion{Fe}{2}* 2612), we estimate
  that EW(\ion{Fe}{2}*2632)$\approx 0.2$ \AA. The EWs of other less-significant
  stellar resonance absorption features are unlikely to be higher than this.
  It is still possible, however, that \ion{Fe}{2}*2612 and \ion{Fe}{2}*2626
  are also suppressed by stellar absorptions. We do see signs of this
  possibility in our result, as will be discussed later in Section 4.4.
  

\subsection{Stacking 2D Spectral Images}

To explore the spatial extent of \ion{Fe}{2}* emission we have stacked the 2D
spectral images of individual galaxies.  Similar to 1D spectral stacking, each
image is interpolated to a uniform grid with $0.5\AA$/pixel in the dispersion
direction after being shifted to the rest frame.  We cut out three windows
  around \ion{Fe}{2}* emissions from the 2D spectrum of each individual
  source, namely: $2350-2450$, $2550-2650$ and $2765-2865$\AA.  For each
  cut-out, pixels are averaged along the dispersion axis to produce a 1D
  distribution of surface brightness in the spatial direction, The peak of
  this 1D surface brightness distribution is identified through a 4th order
  polynomial fitting. Each cut-out is then regridded by linear interpolation
  to be centered at the fitted peak. We assume that the curvature of object trace
  is negligible within each narrow band. The fitted peak of the
  bluest band $2350-2450\AA$ and that of the reddest band $2765-2865\AA$ are
  typically offset by less than 1 pixel ($0.126''$), in no case more than 2
  pixels, indicating a slope of up to 0.005 pixel/$\AA$. For each regridded
  cut-out, a continuum intensity is calculated by summing up all pixels within
  +/0.5'' (4 pixels) from the center. The 2D cut-out is normalized by this
  continuum intensity.  We use mean stack to create the final composite 2D
  spectra.

\begin{figure}
\includegraphics[width=12cm]{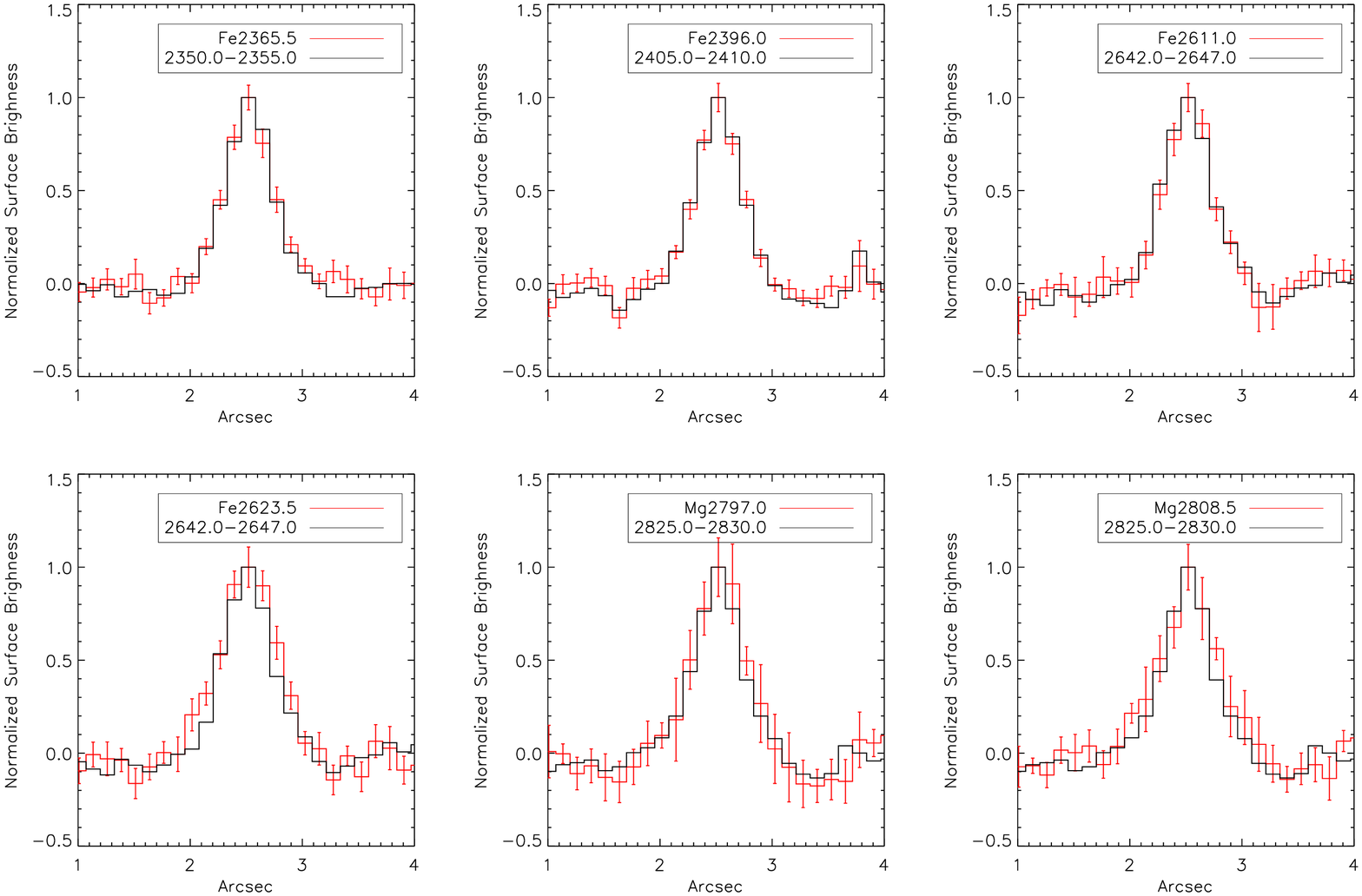}
\includegraphics[width=12cm]{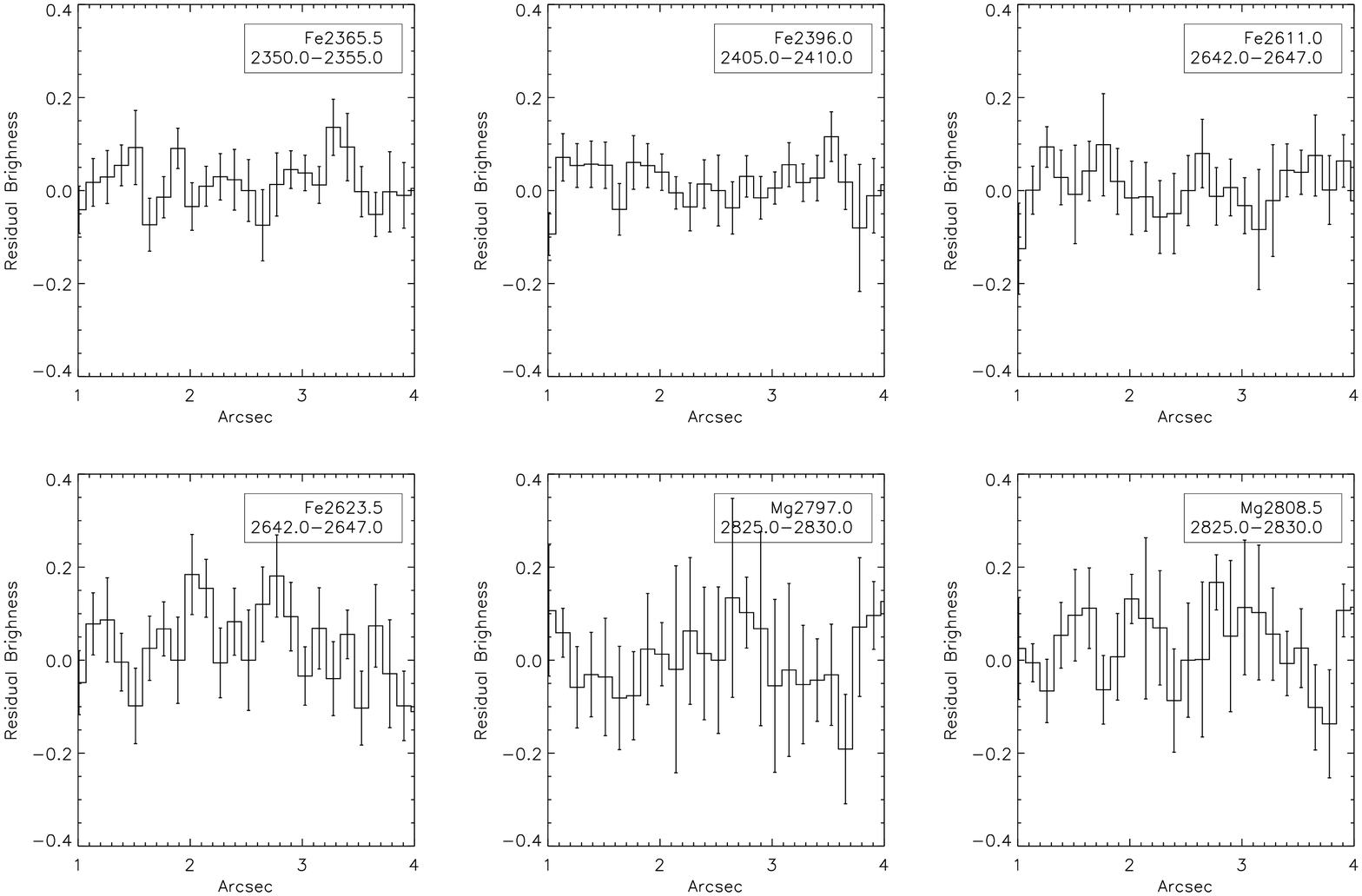}
\caption{Top:Spatial distributions of \ion{Fe}{2}* and MgII emissions (red) 
in stacked 2D spectral images comparing to that of continuum band (black). 
The final stacked image has a pixel scale of $0.5\AA$ along wavelength direction
and a pixel of $0.126''$ along spatial direction. For each line, the plot corresponds to
the wavelength with most extended distribution in a velocity range of approximately [$-500-500$] km/s
around line center. Bottom: Residual signals of \ion{Fe}{2}* and MgII emissions 
with continuum emission being subtracted.} 

\label{fig:rdis}
\end{figure}

Although in our average spectrum \ion{Fe}{2}* lines peak approximately at
systemic velocity, the velocities of spatially extended components of 
the \ion{Fe}{2}* emission lines are necessarily systemic. 
To probe the most extended emission, we calculate FWHMs in spatial 
direction within a $-4$---$4\AA$ window around line center, pixel 
by pixel, and identify the wavelength with the highest FWHM, 
$\lambda_{ext}$.  The spatial distribution of each
\ion{Fe}{2}* line at $\lambda_{ext}$ is overplotted with the average spatial
distribution of a nearby continuum band in
Figure~\ref{fig:rdis}.  The uncertainties of each pixel are
estimated from 200 bootstrap realsamplings of the 2D spectral images. No
extended emission could be detected for any \ion{Fe}{2}* lines.  At the
$1''$ spatial resolution of our data, this indicates that a large fraction
of the \ion{Fe}{2}* emission lines are within the central 4 kpc from the
galactic center. Erb et al. (2012) report only marginal detections of extended
\ion{Fe}{2}* lines in stacked 2D spectral images, in agreement with our
finding that the excess of extended \ion{Fe}{2}* line is weak.

The compact \ion{Fe}{2}* emission could be naturally explained by the low
$f_l$ of resonance \ion{Fe}{2} transitions. In P11, it is shown that the
surface brightness of \ion{Fe}{2}* line decreases by more than a order of
magnitude to 4 kpc for a rapidly decreasing $r^{-2}$ density profile. This is
also consistent with our previous result that \ion{Fe}{2} resonance lines
exhibit lower velocities than \ion{Mg}{2} lines. For an accelerating wind,
this could indicate that the \ion{Fe}{2} absorption lines arise from an inner
region. Recent studies (Rubin et al. 2011, Erb et al. 2012, Kornei et
al. 2013, and Martin et al. 2013) show that in contrast with strong
redshifted P-cygni emission lines usually seen for the \ion{Mg}{2} doublet,
\ion{Fe}{2} transitions rarely present such feature, including \ion{Fe}{2} $2382$,
which, like \ion{Mg}{2}, only involves pure scattering. A potential explanation for
this absence of FeII scattered emission is that the \ion{Fe}{2} absorptions and 
the \ion{Mg}{2} absorptions originate from distinguishable regions of the ISM/wind.
Such segregation could be caused by an evolution in
ionization states of the Fe ions, as discussed by Martin et al. (2013),
Fe irons are mainly in Fe$^{+3}$ in low column density regions,
while Mg ions could remain as Mg$^{++}$, since \ion{Mg}{3} requires a
much higher ionization potential.

Since the individual spectra are normalized by their continuum intensities
before stacking, we cannot easily define a detection limit in terms of
absolute surface brightness. Instead, we provide an upper limit relative to
the continuum surface brightness in the co-added 2-D spectra. The 1-D
continuum surface brightness is estimated from the mean of the central 9
pixels in the 1-D surface brightness distribution.  
The 3-sigma detection limit for \ion{Fe}{2}* 2626 (usually the brightest
\ion{Fe}{2}* line) is $30\%$ of the continuum surface brightness 
over the entire profile.

\subsection{Dependence of \ion{Fe}{2}* Emission on Galaxy Properties}

To further examine whether the intensities of \ion{Fe}{2}* emission lines have any
dependence on galaxy properties, such as stellar mass, SFR and sSFR, we split
our 97 spectra into subsamples by their stellar population properties,
and construct a median spectrum from each subsample.
The intensity of the \ion{Fe}{2}* emissions is quantified by their EWs, which are 
measured by integrating pixels within specified velocity ranges listed in
Table~\ref{tab:ems}. We average the spectra in the same way described in
previous sections.

\begin{figure}
\includegraphics[width=7cm]{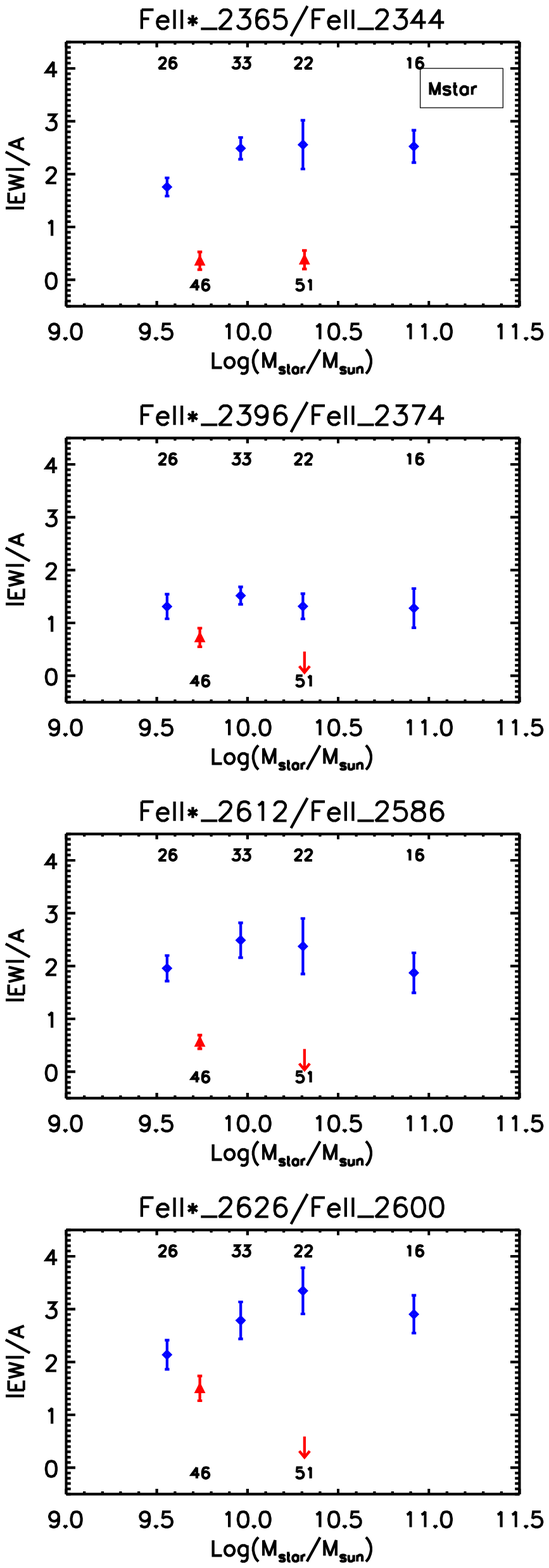}
\includegraphics[width=9cm]{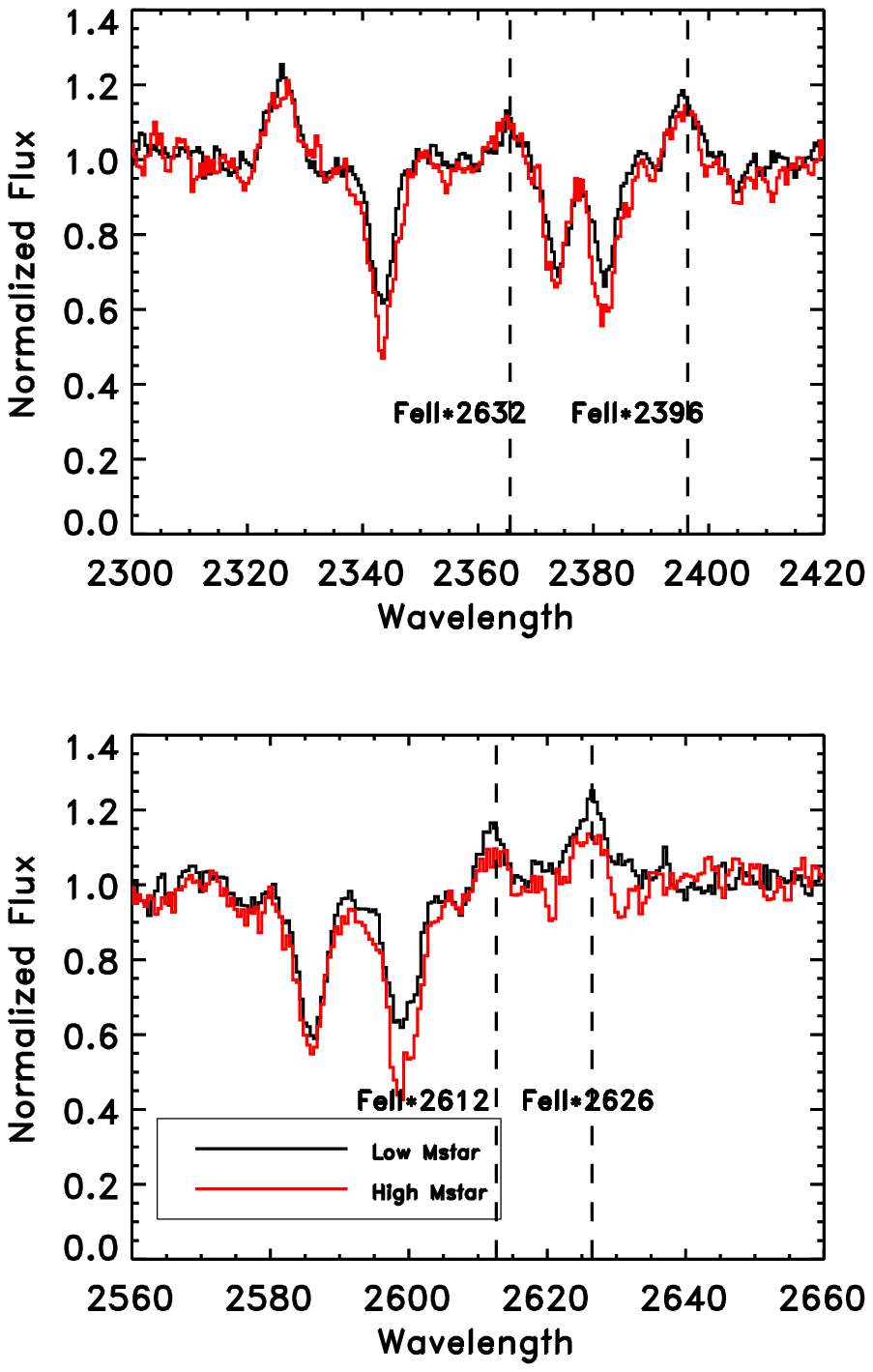}

\caption{Left: EWs of \ion{Fe}{2} (blue) and \ion{Fe}{2}* lines (red) plotted as 
         functions of M$_*$, numbers of objects in each subsample are shown at bottom and top.
         Right: Comparing co-added spectra of low M$_*$ (black) and that of high 
         M$_*$ (red).} \label{fig:ew_mst}
\end{figure}

\begin{figure}
\includegraphics[width=7cm]{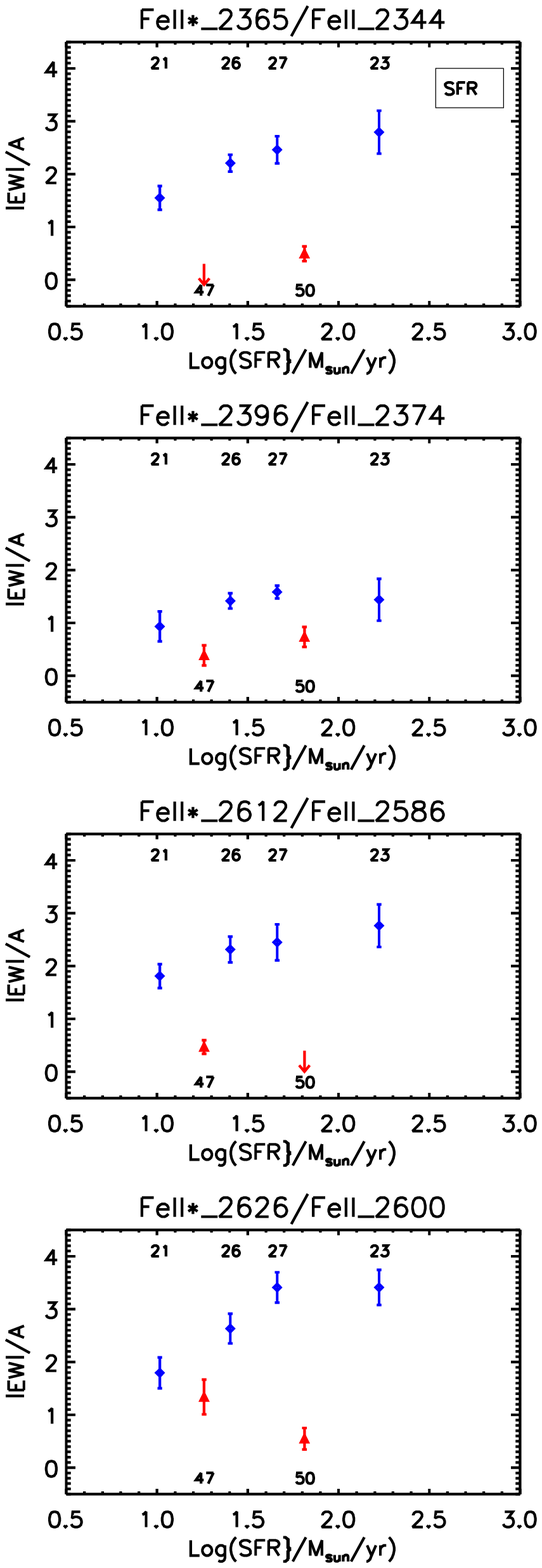}
\includegraphics[width=9cm]{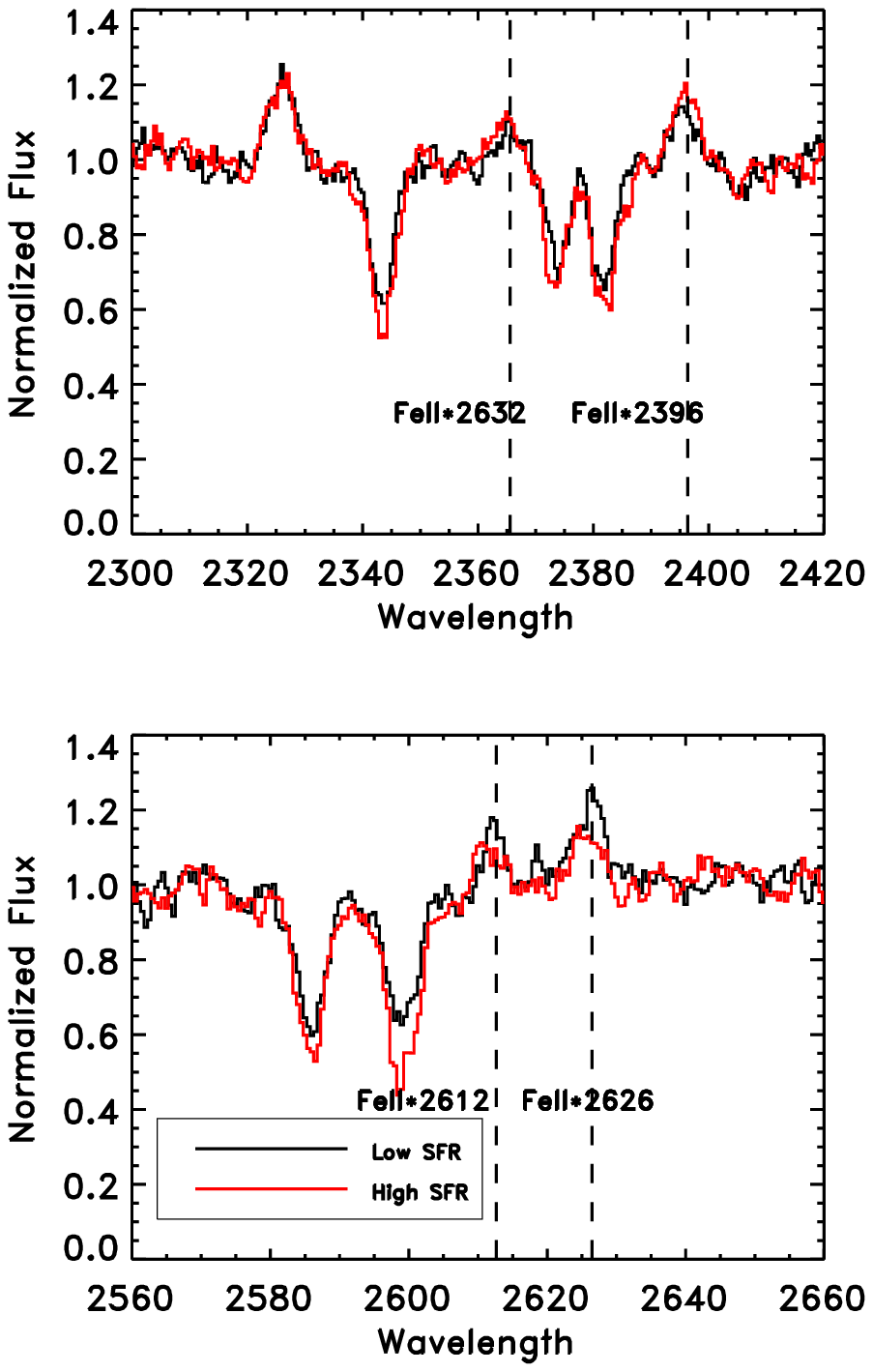}

\caption{Left: EWs of \ion{Fe}{2} (blue) and \ion{Fe}{2}* lines (red) plotted as functions of SFR, 
         numbers of objects in each subsample are shown at bottom and top.
         Right: Comparing co-added spectra of low SFR (black) and that of high SFR (red).} \label{fig:ew_sfr}
\end{figure}

\begin{figure}
\includegraphics[width=7cm]{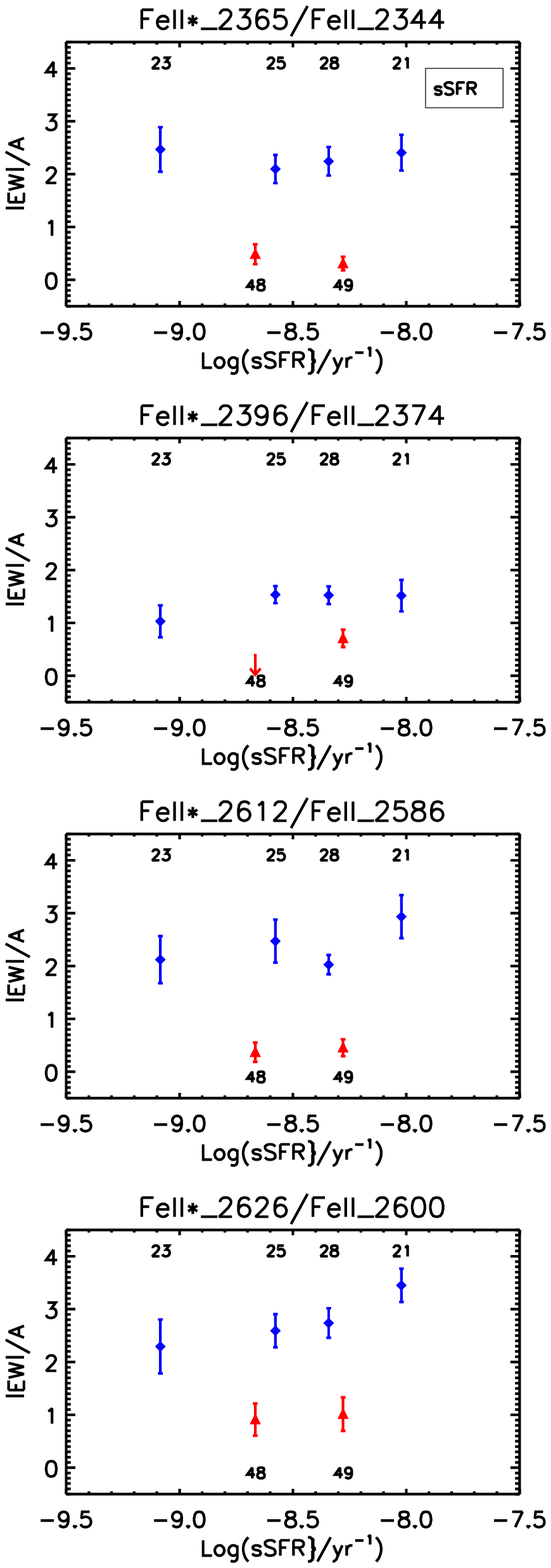}
\includegraphics[width=9cm]{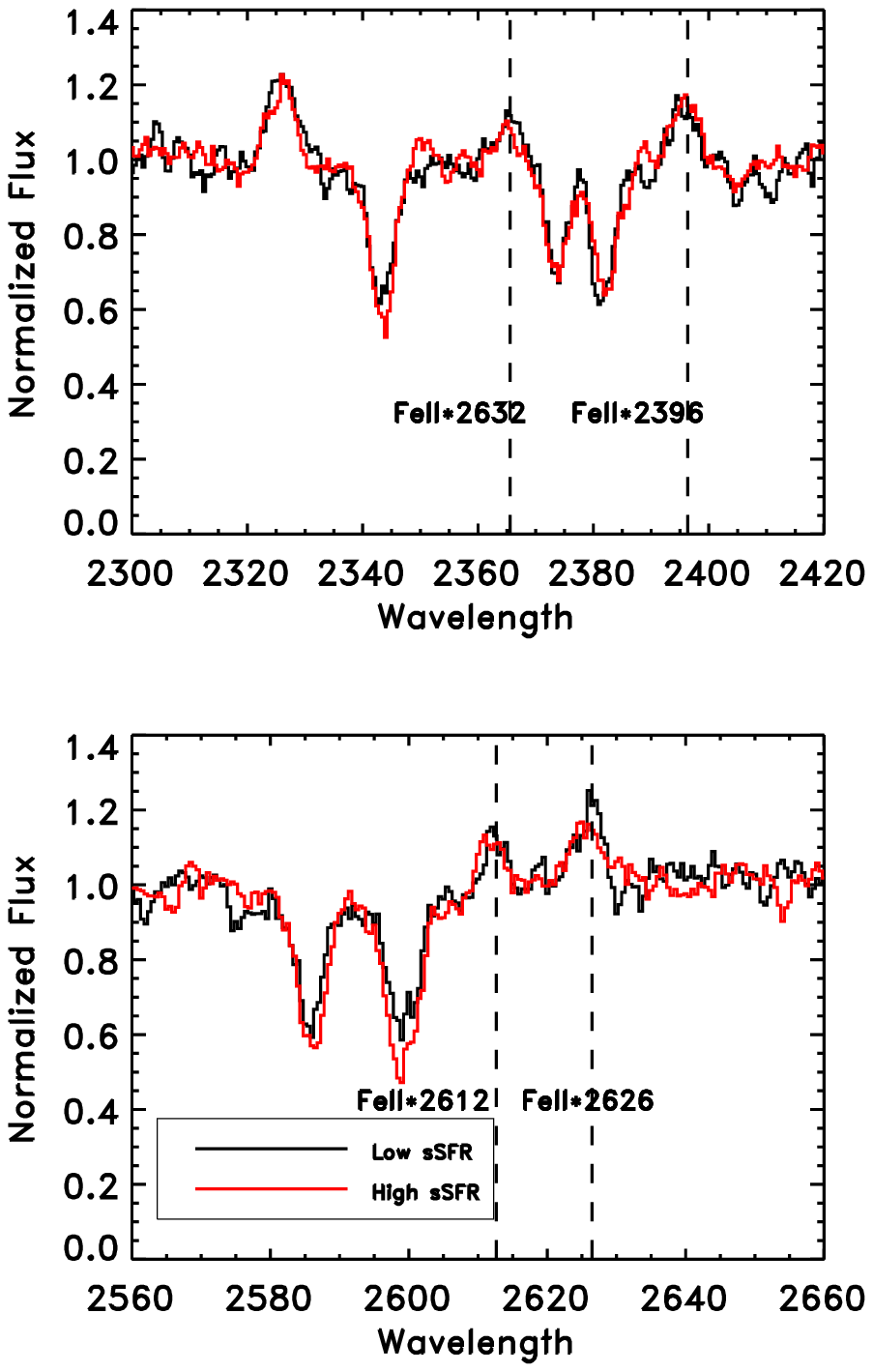}

\caption{Left: EWs of \ion{Fe}{2} (blue) and \ion{Fe}{2}* lines (red) plotted as functions of sSFR, 
         numbers of objects in each subsample are shown at bottom and top.
         Right: Comparing co-added spectra of lower sSFR (black) and that of higher sSFR (red).} \label{fig:ew_ssfr}
\end{figure}

\begin{figure}
\includegraphics[width=7cm]{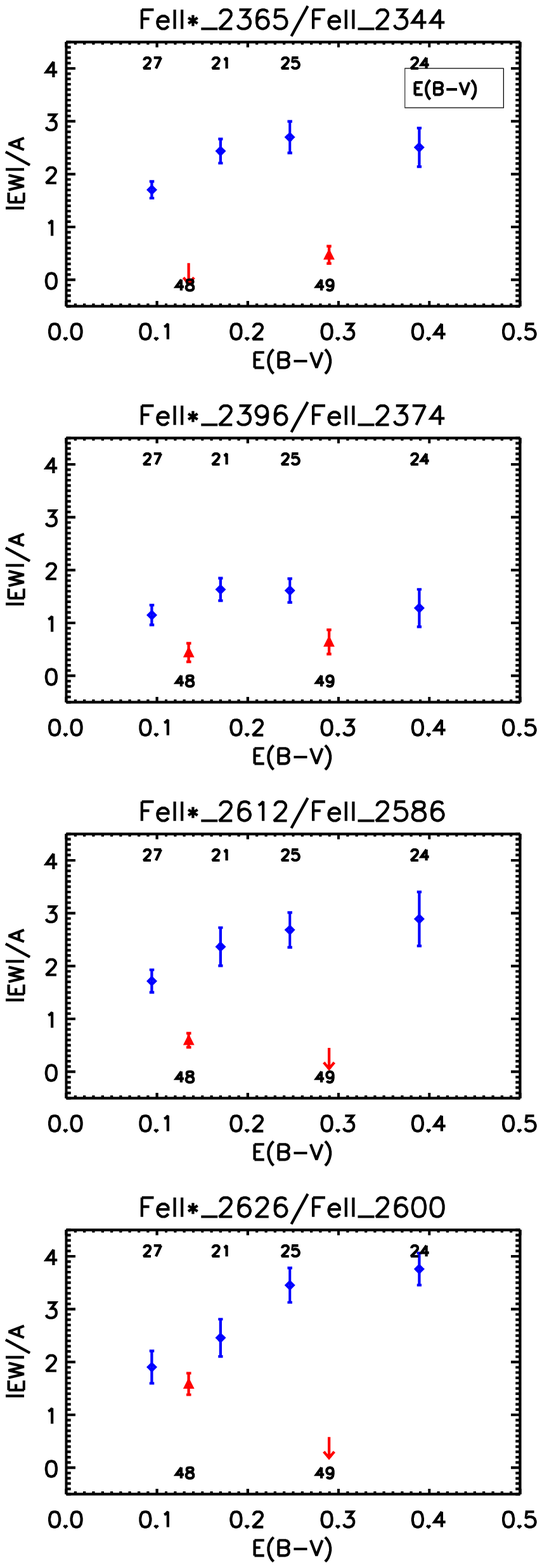}
\includegraphics[width=9cm]{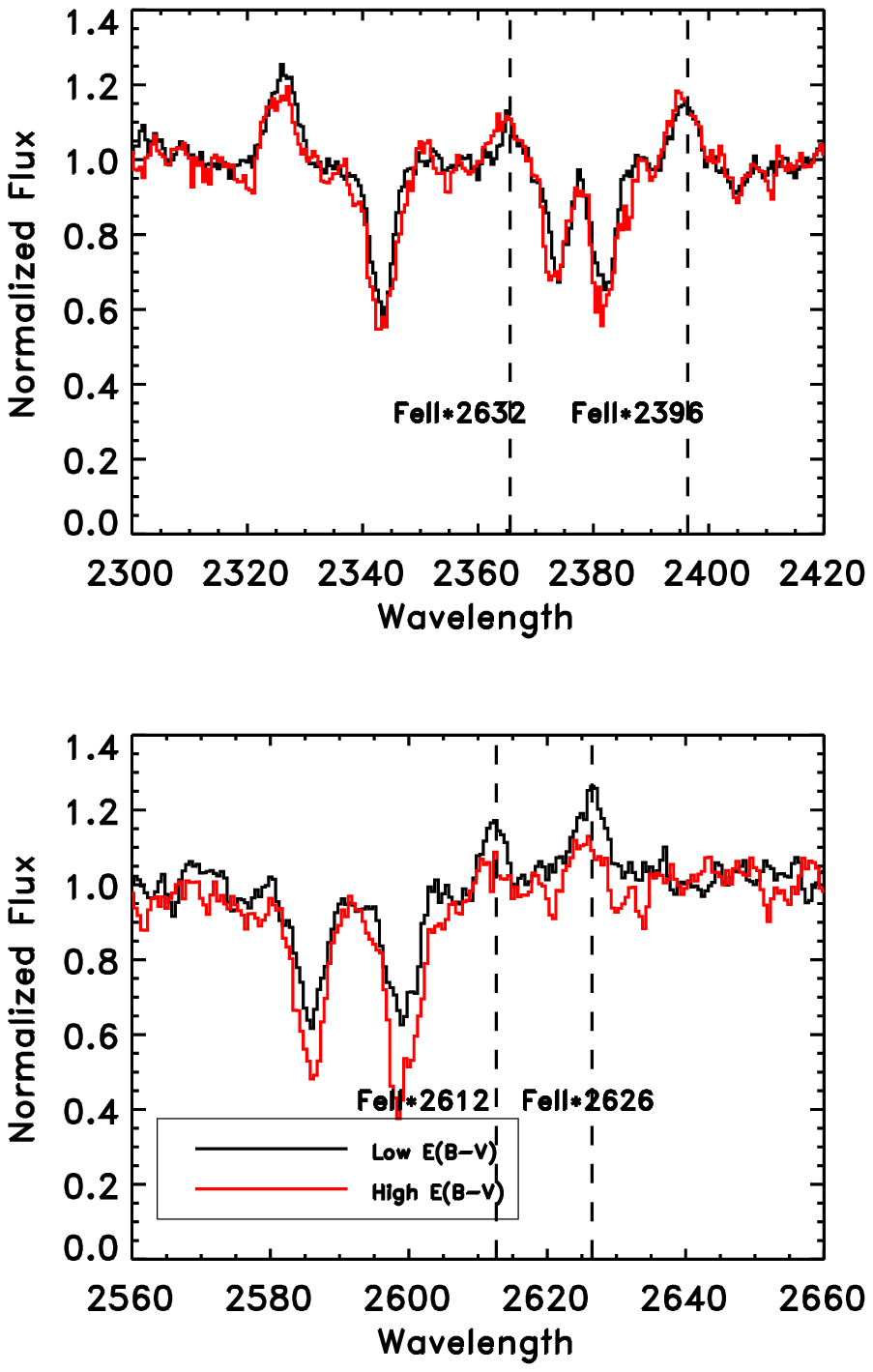}

\caption{Left: EWs of \ion{Fe}{2} (blue) and \ion{Fe}{2}* lines (red) plotted as functions of E(B-V), 
         numbers of objects in each subsample are shown at bottom and top.
         Right: Comparing co-added spectra of lower E(B-V) (black) and that of higher E(B-V) (red).} \label{fig:ew_ebv}
\end{figure}

  Specifically, the whole GMASS sample is split into 4 subsamples by each galaxy property 
  (the stellar mass, SFR, sSFR and E(B-V)). 
  We have chosen the subsamples to avoid spikes in the parameter distribution, and to assign
  each subsample with a roughly equal number of objects. For measuring the \ion{Fe}{2}* emissions, 
  which have smaller EWs and lower S/N ratios compared to the \ion{Fe}{2} absorptions,
  the 4 subsamples are merged into 2. The EWs of the \ion{Fe}{2} absorptions and 
  the \ion{Fe}{2}* emissions are plotted as a function of the median stellar mass/SFR/sSFR/E(B-V) 
  in Figure~\ref{fig:ew_mst} through Figure~\ref{fig:ew_ebv} (for
    the \ion{Fe}{2}* emission lines the upper limit is $2\sigma$). In general,
    the dependence of both EW(\ion{Fe}{2}) and EW(\ion{Fe}{2}*) on galaxy
    properties is weak. For galaxies with increasing star formation activity,
    the strengths of \ion{Fe}{2}* remains constant over more than one order of
    magnitude in SFR and sSFR.  Coil et al. (2011), who report \ion{Fe}{2}*
    detection in K+A galaxies with $|EW|> 1\AA$, reach similar conclusions.

The EW of the \ion{Fe}{2}* transitions, whenever detected, is
$\sim10\%-50\%$ of that of the associated \ion{Fe}{2}\ absorption.  For an ideal
homologous, dustless wind, the conservation of the number of photons implies
EW(em)$\approx$EW(abs). Since \ion{Fe}{2}\ emission is not present in our
average spectra, EW(\ion{Fe}{2}*) should be approximately equal to
EW(\ion{Fe}{2}). A number of factors potentially capable of reducing the
strength of emission lines have been discussed in P11, including bipolar morphology
of the wind, dust extinction and slit losses.  The null-detection of extended
\ion{Fe}{2}* casts doubt on slit losses as a reason for the attenuated
emission. Dust extinction preferentially suppresses the red side of \ion{Fe}{2}* 
emission, as photons scattered from the receding part of the wind travel
longer paths to reach the observer.  The flux-weighted line centroids, listed in 
Table~\ref{tab:v_ems}, are consistently blueshifted in all \ion{Fe}{2}* lines, 
although no blueshift velocity is significant above the $2\sigma$ level. 
This result is in agreement with Erb et al. (2012).  On the other hand, 
high-z star-forming galaxies commonly have blueshifted centroids of absorption lines and 
this high frequency of outflowing absorbing material (Martin et al. 2012, Rubin et al. 2013)
has been interpreted as an evidence of large opening angles of the winds and 
probably argues against simplified models with sharp-edged bipolar morphology. 
However, one should keep in mind that our flux-limited sample
could be biased toward ``face-on'' objects such that we are facing a direction
where the ISM column density is low, and the wind can most easily propagate
out. When viewed from an edge-on direction, the strength of the \ion{Fe}{2} 
absorptions is expected to be reduced relative to the \ion{Fe}{2}* emissions, due to 
less covering fraction and scattered filling.

EW(\ion{Fe}{2}* 2612) and EW(\ion{Fe}{2}* 2626) appear to decrease
with increasing SFR, E(B-V) and stellar mass.  FeII*(2365) and
FeII*(2396), however, show opposite trends, that is, increasing EWs with
increasing SFR, E(B-V) and possibly stellar mass.  This discrepancy among
\ion{Fe}{2}* lines has already been noted by Erb et al. (2012).  Its origin is
not understood. Erb et al. (2012) propose that the decreasing trends of the
FeII*(2626) with increasing SFR, E(B-V) and stellar mass are mainly driven by
slit loss.  This is supported by their results that \ion{Fe}{2}* emissions in
massive objects are more extended than in low mass objects. Kornei et
al. (2013) instead find that the strengths of the \ion{Fe}{2}* emissions are
primarily modulated by dust extinction, however, they use average equivalent width
of \ion{Fe}{2}* 2396 and \ion{Fe}{2}* 2626 as a measure of the emission
line strength, there is the possibility that this lack
of correlation with galaxy properties other than dust extinction might be 
a result that the \ion{Fe}{2}* 2396 and the \ion{Fe}{2}* 2626
have different dependences on galaxy properties.

Trends of FeII*(2365/2396) seen both here and in Erb et al. (2012) are not 
statistically significant and require larger sample to be confirmed or rejected.  
However, if these trends are real, we should question the 
validity of the explanations given above for the
variation of EWs(\ion{Fe}{2}*).  This discrepancy is hardly explained in terms
of different geometrical factors or radiative transfer processes of the
\ion{Fe}{2}\ and \ion{Fe}{2}* transitions.  Geometrical factors (i.e. viewing
angle, slit loss etc.) should affect all \ion{Fe}{2}* lines in a similar
way. On the other hand, note that in Figure~\ref{fig:os_conv}, the
\ion{Fe}{2}* 2612/\ion{Fe}{2} 2586 couple has a $P_{fluo}$ that is higher than that of 
the \ion{Fe}{2}* 2396/\ion{Fe}{2} 2374 couple and lower than that of the
\ion{Fe}{2}* 2365/\ion{Fe}{2} 2344 couple. The radiative transfer properties
of \ion{Fe}{2}* 2612 should lie between \ion{Fe}{2}* 2365 and
\ion{Fe}{2}* 2396, whereas \ion{Fe}{2}* 2612 differs from both lines in
terms of its scaling relationship with galaxy properties. It is true that dust
extinction is likely observed in our co-added spectra, and it could cause
varying degrees of suppression among the \ion{Fe}{2}* lines, but simply
dust extinction could not offer a solution to the observed opposite trends
between transitions with similar scattering and fluorescence frequencies.

A candidate explanation for this complexity is that the \ion{Fe}{2}* 2612 and
the \ion{Fe}{2}* 2626 features are suppressed by overlapped absorption. In
Sedtion 4.2, we've shown that this could be a result of stellar absorption. Since
we are using co-added spectra, \ion{Fe}{2}* emitters are not necessarily the
same objects that exhibit strong stellar absorption.  In
Figure~\ref{fig:ew_mst} through Figure~\ref{fig:ew_ebv}, we also provide a
comparison of co-added spectra for each galaxy property of interest. We do see signs of
absorption around \ion{Fe}{2}* 2626 in subsamples with high $M_*$ and high
E(B-V), and similar features is probably also shown in Erb et al. (2012)
(Figure 10, top and bottom panel). With the S/N ratio and the resolution
of the GMASS spectra a conclusive argument is not possible and shall be
deferred to future improved observations. 
   
\section{Conclusion}

We have discussed the power of interstellar medium \ion{Fe}{2}/\ion{Fe}{2}*
multiplets as tracers of galactic outflow wind, based on a deep, rest
near-infrared flux limited (i.e. stellar mass selected) sample of $97$
star--forming galaxies at $1.0 \lesssim z \lesssim 2.6$ selected from the
GMASS redshift survey. We average our data to create high S/N average spectra,
which enables the identification of faint features, not otherwise visible in
the individual spectra. Because the GMASS survey is contained in the GOODS
South field, a full complement of panchromatic photometry is available to
reliably estimate the integrated properties of the stellar populations of the
galaxies such as star formation rate, stellar mass and dust extinction. These
allow us to explore trends between the line profile and strength of the
\ion{Fe}{2}\ and \ion{Fe}{2}* features as a function of the properties of the
galaxies. In particular:

  (a) We have studied the dependence of line profiles and EWs
  of \ion{Fe}{2}/\ion{Fe}{2}*/\ion{Mg}{2} transitions on galaxy properties. 
  In general, the dependencies of blueshifted velocities and EWs
  of these transitions on SFR/M$_*$/Age/E(B-V) are similar to what have been 
  observed in previous studies on $z=1 \sim 2$ star-forming galaxies.
  We have also confirmed that fluorescence is a plausible excitation mechanism of
  the \ion{Fe}{2}* lines, which are commonly observed in the spectra of
  star--forming galaxies at high redshift, such as those studied here. While
  other excitation mechanisms for this lines are possible, and have not been
  addressed here, the relative strengths of \ion{Fe}{2}* are
  in agreement with the prediction from continuum fluorescence.

 (b) We show that the intensity of \ion{Fe}{2}* $2612$, \ion{Fe}{2}* $2626$
\ion{Fe}{2}* $2632$ is possibly suppressed by underlying stellar
continua. This provides a potential explanation to the opposite trends between
\ion{Fe}{2}* $2612/2626$ and \ion{Fe}{2}* $2365/2396$ with the integrated
properties of the galaxies (SFR, E(B-V) and possibly stellar mass), as well as
the absence of \ion{Fe}{2}* $2632$ in the co-added spectra.

  (c) By stacking 2D spectral images, we find that the region where the
  \ion{Fe}{2}* emission is produced is compact and close to the galactic
  disks, as no extended emission of \ion{Fe}{2}* is detected beyond $3 \sim 4$ kpc in
  radius. Although no evidence is found that \ion{Mg}{2} emissions are more extended than
  \ion{Fe}{2}* in this study, the lower blueshifted velocities of \ion{Fe}{2} lines
  relative to \ion{Mg}{2} doublet, as well as the scarcity of
  scattered emission redward of \ion{Fe}{2} 2382 in \ion{Mg}{2} emitting systems, 
  argue against that \ion{Fe}{2} absorption/\ion{Fe}{2}* transitions 
  probe a region of outflow wind that is spatially and dynamically identical  
  with that probed by \ion{Mg}{2}.	

This study confirms the unique diagnostic power of
\ion{Fe}{2}/\ion{Fe}{2}* multiplets for probing the structure and kinematics
of galactic outflows at $1.0 \lesssim z \lesssim 2.6$.  Our results are
generally consistent with the simple model of outflow developed by Prochaska
et al. (2011).  Some important questions remains unsolved, such as whether the
flattening of \ion{Fe}{2} velocity distribution is associated with a phase
change of wind, and if so, how.  The answer to these questions might be
directly linked to the kinematic profile of cool outflow wind, which is a
crucial factor in constraining mass flow rate of wind.

\clearpage

\section{References}

\noindent Adelberger K. L., Steidel C. C., Shapley A. E., Pettini M., 2003, ApJ, 584, 45

\noindent Banerji M., Chapman S. C., Smail I., Alaghband-Zadeh S., Swinbank A. M., Dunlop J. S., Ivison R. J., Blain A. W., 2011, arXiv1108.0420

\noindent Calzetti D. 1997, AJ, 113, 162

\noindent Calzetti D., Armus L., Bohlin R. C., Kinney A. L., Koornneef J., Storchi-Bergmann T., 2000, ApJ, 533, 682

\noindent Chen Y. M., Tremonti C. A., Heckman T. M., Kauffmann G., Weiner B. J., Brinchmann J., Wang J., 2010, AJ, 140, 445

\noindent Chevalier R. A., Clegg, A. W., 1985, Natur, 317, 44

\noindent Cimatti A., Daddi E., Renzini A., Cassata P., Vanzella E., Pozzetti L., Cristiani S., Fontana A., Rodighiero G., Mignoli M., Zamorani G., 2004, Natur, 430, 184

\noindent Cimatti A., et al., 2008, A\&A, 482, 21

\noindent Cimatti A., et al., 2013, ApJ, 779, 13

\noindent Coil A. L., Weiner B. J., Holz D. E., Cooper M. C., Yan R., Aird J., 2011, arXiv11040681

\noindent Cole S., Lacey C. G., Baugh C. M., Frenk C. S., 2000, MNRAS, 319, 168

\noindent Erb D. K., Steidel C. C., Shapley A. E., Pettini M., Reddy N. A., Adelberger K. L., 2006, ApJ, 646, 107

\noindent Erb, D. K., Quider A. M., Henry A. L., Martin C. L., 2012, ApJ, 759, 26

\noindent Heckman T. M., Norman C. A., Strickland D. K., Sembach, K. R., 2002, ApJ, 577, 691

\noindent Katz, N., Weinberg D. H., Hernquist L., 1996, ApJS, 105, 19

\noindent Kornei K. A., Shapley A. E., Martin C. L., Coil, A. L., Lotz J. M., Schiminovich D., Bundy K., Noeske K. G., 2012, ApJ, 758, 135

\noindent Krug H. B., Rupke D. S. N., Veilleux S., 2010, ApJ, 708, 1145

\noindent Kurk J., et al., 2009, A\&A, 504, 331

\noindent Kurk J., et al., 2013, A\&A, 549, 63star

\noindent Leitherer C., Tremonti C. A., Heckman T. M., Calzetti D., 2011, AJ, 141, 37

\noindent Madau P., Ferrara A., Rees M. J., 2001, ApJ, 555, 92

\noindent Martin, C. L. 2005, ApJ, 621, 227

\noindent Martin C. L. \& Bouch\'{e}, N., 2009, ApJ, 703, 1394

\noindent Martin C. L., Shapley A. E., Coil A. L., Kornei K. A., Bundy K., Weiner B. J., Noeske K. G., Schiminovich D., 2012, ApJ, 760, 127

\noindent Morton D. C., 2003, ApJS, 149, 205

\noindent Murray N., Quataert E., Thompson T. A., 2005, ApJ, 618, 569

\noindent Murray N., M\'{e}nard B., Thompson T. A., 2011, ApJ, 735, 66

\noindent Prochaska J. X., Kasen D., Rubin K., 2011, ApJ, 734, 24

\noindent Reddy et a. 2011, arXiv:1107.2653

\noindent Rubin K. H. R., Prochaska J. X., M\'{e}nard B., Murray N., Kasen D., Koo D. C., Phillips A. C., 2011, ApJ, 728, 55

\noindent Rubin K. H. R.; Prochaska J. X., Koo D. C., Phillips A. C., 2011, arXiv1110.0837

\noindent Rubin K. H. R., Weiner B. J., Koo D. C., Martin C. L., Prochaska J. X., Coil A. L., Newman J. A., 2010, ApJ, 719, 1503

\noindent Rupke D. S., Veilleux S., Sanders D. B., 2005, ApJ, 632, 751

\noindent Sigut T. A. A. \& Pradhan A. K, 2003, ApJS, 145, 15

\noindent Sobolev V. V., 1960, `Moving envelopes of stars', Cambridge: Harvard University Press

\noindent Spinrad H., Dey A., Stern D., Dunlop J., Peacock J., Jimenez R., Windhorst R., 1997, ApJ, 484, 581

\noindent Steidel C. C., Erb D. K., Shapley A. E., Pettini M., Reddy N., Bogosavljevi\'{c} M., Rudie G. C., Rakic O., 2010, ApJ, 717, 289

\noindent Sturm, E. et al., 2011, ApJ, 733, 16

\noindent Talia, M. et al., 2012, A\&A, 539, 61

\noindent Tremonti C. A. et al., 2004, ApJ, 613, 898

\noindent Tremonti C. A., Moustakas J., Diamond-Stanic A. M., 2007, ApJ, 663, 77

\noindent Veilleux S., Cecil G., Bland-Hawthorn J., 2005, ARA\&A, 43, 769767

\noindent Vestergaard M. \& Wilkes B. J., 2001, ApJS, 134, 1

\noindent Weiner B. J., et al., 2009, ApJ, 692, 187

\noindent White Simon D. M., Frenk C. S.  1991, ApJ, 379, 52W

\clearpage

\clearpage

\begingroup
\setlength{\LTleft}{-20cm plus -1fill}
\setlength{\LTright}{\LTleft}
\begin{longtable}{ccccccc}
\caption[cap in list]{Global Properties of Galaxies} \label{tab:id} \\ \hline
GMASS ID &  Alpha (J2000)  & Delta (J2000) & z & SFR & $M_*$ & E(B-V)   \\          
                &                        &                      &    & $M_{\odot}/yr$ & $M_{\odot}$ & \\
\hline
\endfirsthead

\caption[]{(continued)} \\ \hline
GMASS ID &  Alpha (J2000)  & Delta (J2000) & z & SFR & $M_*$ & E(B-V)   \\          
                &                        &                      &    & $M_{\odot}/yr$ & $M_{\odot}$ & \\ 
\hline
\endhead

\hline

\multicolumn{7}{c}{[O II] 3727 detected}  \\

\hline
         428   &       53.076870   &         -27.799203   &          1.0800600   &  $10.35\pm0.56$  &  $9.64\pm0.06$  &  $0.12\pm0.01$\\
         773   &       53.067734   &         -27.784247   &          1.2193600   &  $4.00\pm0.43$  &  $9.39\pm0.02$  &  $0.03\pm0.01$\\
         774   &       53.068581   &         -27.783980   &          1.2179500   &  $13.17\pm0.73$  &  $9.54\pm0.02$  &  $0.12\pm0.01$\\
         793   &       53.191565   &         -27.782669   &          1.2943200   &  $29.51\pm2.19$  &  $9.91\pm0.02$  &  $0.21\pm0.01$\\
         795   &       53.178228   &         -27.783079   &          1.1194100   &  $10.47\pm0.65$  &  $9.56\pm0.04$  &  $0.11\pm0.01$\\
         983   &       53.065793   &         -27.774927   &          1.0212000   &  $18.21\pm0.75$  &  $9.06\pm0.27$  &  $0.15\pm0.01$\\
        1084   &       53.165465   &         -27.769767   &          1.5496700   &  $32.80\pm5.92$  &  $11.24\pm0.04$  &  $0.29\pm0.02$\\
        1227   &       53.088793   &         -27.765170   &          1.2231000   &  $5.47\pm0.38$  &  $8.76\pm0.19$  &  $0.07\pm0.01$\\
        1315   &       53.163594   &         -27.758956   &          1.0954100   &  $392.40\pm42.24$  &  $10.89\pm0.11$  &  $0.62\pm0.01$\\
        1567   &       53.100101   &         -27.751124   &          1.1103900   &  $6.29\pm0.55$  &  $9.49\pm0.05$  &  $0.13\pm0.01$\\
        1585   &       53.160802   &         -27.749997   &         0.97986000   &  $5.87\pm0.55$  &  $9.85\pm0.13$  &  $0.12\pm0.01$\\
        1592   &       53.107569   &         -27.750008   &         0.83184000   &  $3718.82\pm2427.36$  &  $9.04\pm0.04$  &  $1.09\pm0.07$\\
        1652   &       53.173198   &         -27.747988   &          1.3540000   &  $4.69\pm0.44$  &  $9.39\pm0.12$  &  $0.09\pm0.01$\\
        1920   &       53.138204   &         -27.737523   &         0.66454300   &  $6918.20\pm52335.70$  &  $10.01\pm0.03$  &  $1.52\pm0.34$\\
        2135   &       53.160803   &         -27.710286   &          1.2457500   &  $5.45\pm0.69$  &  $9.87\pm0.14$  &  $0.13\pm0.01$\\
        2484   &       53.157004   &         -27.705376   &          1.4362800   &  $16.59\pm1.60$  &  $10.31\pm0.10$  &  $0.11\pm0.01$\\
        1380   &       53.105219   &         -27.758084   &          1.6118400   &  $18.95\pm1.79$  &  $9.89\pm0.02$  &  $0.18\pm0.01$\\
        1399   &       53.173503   &         -27.757124   &          1.6133400   &  $8.07\pm0.84$  &  $9.90\pm0.11$  &  $0.08\pm0.01$\\
        1808   &       53.109013   &         -27.742556   &          1.6085300   &  $20.99\pm1.60$  &  $9.74\pm0.01$  &  $0.20\pm0.01$\\
        1979   &       53.102703   &         -27.735466   &          1.6123500   &  $58.86\pm2.70$  &  $10.44\pm0.01$  &  $0.19\pm0.01$\\
        2081   &       53.124437   &         -27.731926   &          1.6018900   &  $6.11\pm0.43$  &  $9.91\pm0.02$  &  $-0.00\pm0.01$\\
        2142   &       53.098082   &         -27.713721   &          1.6105300   &  $22.62\pm1.82$  &  $9.94\pm0.07$  &  $0.20\pm0.01$\\
        2180   &       53.123146   &         -27.715535   &          1.6088500   &  $121.85\pm7.73$  &  $10.43\pm0.01$  &  $0.36\pm0.01$\\
        2251   &       53.122828   &         -27.722800   &          1.6106400   &  $43.75\pm3.28$  &  $10.92\pm0.03$  &  $0.28\pm0.01$\\
        2368   &       53.071309   &         -27.728343   &          1.6131600   &  $12.86\pm1.29$  &  $10.44\pm0.11$  &  $0.13\pm0.01$\\
        2454   &       53.120374   &         -27.717612   &          1.6021500   &  $27.57\pm3.81$  &  $10.72\pm0.01$  &  $0.17\pm0.02$\\
        2540   &       53.126371   &         -27.711249   &          1.6133000   &  $80.29\pm5.86$  &  $9.97\pm0.03$  &  $0.23\pm0.01$\\
        1495   &       53.147337   &         -27.753514   &          1.6125900   &  $27.69\pm2.51$  &  $10.56\pm0.06$  &  $0.19\pm0.01$\\

\hline 

\multicolumn{7}{c}{Cross Correlation} \\

\hline

          90   &       53.142081   &         -27.819914   &          1.9043547   &  $107.77\pm17.07$  &  $10.27\pm0.07$  &  $0.25\pm0.02$\\
         118   &       53.131508   &         -27.814948   &          1.8858988   &  $52.56\pm19.73$  &  $10.77\pm0.06$  &  $0.30\pm0.05$\\
         149   &       53.091466   &         -27.815463   &          2.0076380   &  $79.85\pm6.56$  &  $10.18\pm0.04$  &  $0.23\pm0.01$\\
         183   &       53.129829   &         -27.813320   &          1.8830679   &  $21.79\pm4.34$  &  $9.63\pm0.02$  &  $0.10\pm0.03$\\
         249   &       53.093552   &         -27.809294   &          2.3463069   &  $63.64\pm20.03$  &  $10.13\pm0.06$  &  $0.33\pm0.04$\\
         250   &       53.128797   &         -27.808940   &          1.8848784   &  $39.40\pm4.68$  &  $10.10\pm0.06$  &  $0.16\pm0.01$\\
         316   &       53.086721   &         -27.806217   &          1.7372309   &  $27.34\pm1.68$  &  $9.59\pm0.04$  &  $0.09\pm0.01$\\
         335   &       53.170742   &         -27.804676   &          1.7632293   &  $29.13\pm3.53$  &  $9.72\pm0.09$  &  $0.17\pm0.01$\\
         365   &       53.115873   &         -27.803383   &          1.6086395   &  $7.97\pm0.50$  &  $9.35\pm0.03$  &  $0.05\pm0.01$\\
         390   &       53.092122   &         -27.801855   &          1.7751231   &  $15.38\pm1.07$  &  $9.52\pm0.12$  &  $0.07\pm0.01$\\
         484   &       53.148588   &         -27.796941   &          1.7651899   &  $21.51\pm2.05$  &  $9.12\pm0.02$  &  $0.06\pm0.01$\\
         487   &       53.151612   &         -27.796397   &          1.7683409   &  $54.20\pm4.34$  &  $10.09\pm0.01$  &  $0.15\pm0.01$\\
         508   &       53.140589   &         -27.795611   &          1.9068955   &  $33.68\pm4.23$  &  $10.18\pm0.07$  &  $0.21\pm0.01$\\
         656   &       53.149308   &         -27.788523   &          1.9074457   &  $167.14\pm15.65$  &  $10.06\pm0.02$  &  $0.34\pm0.01$\\
         675   &       53.161668   &         -27.787455   &          1.8502673   &  $100.70\pm14.07$  &  $10.64\pm0.06$  &  $0.34\pm0.02$\\
         679   &       53.147983   &         -27.787692   &          1.8834080   &  $18.17\pm2.96$  &  $9.67\pm0.01$  &  $0.17\pm0.02$\\
         781   &       53.073845   &         -27.784159   &          1.6064388   &  $15.66\pm0.60$  &  $9.69\pm0.02$  &  $0.06\pm0.01$\\
         858   &       53.155652   &         -27.779271   &          1.8470163   &  $174.62\pm17.96$  &  $10.30\pm0.09$  &  $0.33\pm0.01$\\
         870   &       53.117363   &         -27.780112   &          1.9101266   &  $41.09\pm5.11$  &  $9.92\pm0.06$  &  $0.25\pm0.01$\\
         881   &       53.130519   &         -27.779689   &          2.1330983   &  $27.66\pm6.25$  &  $9.67\pm0.08$  &  $0.23\pm0.03$\\
         894   &       53.149218   &         -27.778804   &          1.8520979   &  $45.91\pm7.37$  &  $9.94\pm0.02$  &  $0.26\pm0.02$\\
         923   &       53.113614   &         -27.777461   &          1.8853886   &  $23.63\pm3.19$  &  $9.72\pm0.03$  &  $0.16\pm0.02$\\
         949   &       53.098763   &         -27.775833   &          2.0772103   &  $56.86\pm26.21$  &  $10.52\pm0.05$  &  $0.39\pm0.06$\\
        1020   &       53.102544   &         -27.772269   &          1.2216150   &  $539.89\pm134.21$  &  $10.39\pm0.13$  &  $0.76\pm0.03$\\
        1050   &       53.074550   &         -27.772420   &          1.5400175   &  $11.58\pm0.80$  &  $9.38\pm0.08$  &  $0.09\pm0.01$\\
        1133   &       53.095309   &         -27.768670   &          1.7253771   &  $29.59\pm3.58$  &  $9.87\pm0.02$  &  $0.22\pm0.01$\\
        1146   &       53.065654   &         -27.767868   &          1.5390371   &  $90.79\pm6.43$  &  $10.45\pm0.07$  &  $0.35\pm0.01$\\
        1224   &       53.073278   &         -27.764309   &          1.8441153   &  $235.02\pm516.93$  &  $10.20\pm0.21$  &  $0.71\pm0.18$\\
        1254   &       53.084079   &         -27.763688   &          1.6113404   &  $13.94\pm0.61$  &  $9.64\pm0.08$  &  $0.06\pm0.01$\\
        1314   &       53.111396   &         -27.761095   &          2.0080381   &  $30.65\pm3.67$  &  $9.85\pm0.06$  &  $0.20\pm0.01$\\
        1372   &       53.090509   &         -27.758247   &          2.0787758   &  $18.40\pm3.08$  &  $9.84\pm0.03$  &  $0.17\pm0.02$\\
        1427   &       53.138121   &         -27.756328   &          1.9204599   &  $18.07\pm2.46$  &  $9.75\pm0.11$  &  $0.25\pm0.02$\\
        1454   &       53.118947   &         -27.755366   &          1.7564771   &  $66.22\pm28.81$  &  $10.08\pm0.04$  &  $0.44\pm0.06$\\
        1464   &       53.117973   &         -27.755219   &          1.7562971   &  $35.53\pm8.15$  &  $10.00\pm0.14$  &  $0.35\pm0.03$\\
        1479   &       53.065364   &         -27.754235   &          2.6741525   &  $90.31\pm23.54$  &  $10.79\pm0.12$  &  $0.26\pm0.03$\\
        1485   &       53.075734   &         -27.754444   &          2.1907669   &  $12.80\pm4.76$  &  $10.22\pm0.06$  &  $0.26\pm0.05$\\
        1486   &       53.078047   &         -27.754012   &          1.8775161   &  $59.98\pm3.71$  &  $9.54\pm0.03$  &  $0.16\pm0.01$\\
        1489   &       53.121496   &         -27.754125   &          2.4325747   &  $38.90\pm6.49$  &  $9.86\pm0.08$  &  $0.17\pm0.02$\\
        1498   &       53.174572   &         -27.753371   &          1.8487168   &  $704.70\pm390.63$  &  $11.06\pm0.07$  &  $0.79\pm0.07$\\
        1624   &       53.075431   &         -27.748652   &          1.7174746   &  $37.50\pm16.11$  &  $10.38\pm0.11$  &  $0.32\pm0.05$\\
        1663   &       53.103062   &         -27.747278   &          2.0267741   &  $25.45\pm3.84$  &  $9.91\pm0.03$  &  $0.19\pm0.02$\\
        1691   &       53.132921   &         -27.745825   &          1.6124008   &  $64.62\pm5.27$  &  $10.39\pm0.01$  &  $0.29\pm0.01$\\
        1748   &       53.084140   &         -27.744149   &          1.8668226   &  $22.51\pm3.81$  &  $9.80\pm0.02$  &  $0.21\pm0.02$\\
        1789   &       53.132642   &         -27.743145   &          1.8841482   &  $51.18\pm7.61$  &  $10.03\pm0.01$  &  $0.34\pm0.02$\\
        1938   &       53.098832   &         -27.736580   &          1.7604884   &  $376.72\pm40.12$  &  $10.94\pm0.02$  &  $0.39\pm0.02$\\
        1980   &       53.062441   &         -27.735547   &          2.6734322   &  $44.92\pm8.03$  &  $10.07\pm0.10$  &  $0.13\pm0.02$\\
        1989   &       53.182840   &         -27.734914   &          2.4317344   &  $181.43\pm14.57$  &  $11.06\pm0.01$  &  $0.28\pm0.01$\\
        2018   &       53.186302   &         -27.733624   &          1.9631536   &  $63.53\pm5.28$  &  $10.02\pm0.03$  &  $0.19\pm0.01$\\
        2032   &       53.188276   &         -27.733722   &          1.9619132   &  $11.45\pm1.92$  &  $10.07\pm0.08$  &  $0.09\pm0.02$\\
        2043   &       53.174451   &         -27.733299   &          2.5777914   &  $139.54\pm21.26$  &  $10.17\pm0.11$  &  $0.36\pm0.02$\\
        2076   &       53.134028   &         -27.732173   &          1.7610586   &  $19.96\pm31.85$  &  $11.09\pm0.16$  &  $0.45\pm0.16$\\
        2099   &       53.131362   &         -27.730782   &          2.1959386   &  $31.26\pm3.57$  &  $9.95\pm0.09$  &  $0.15\pm0.01$\\
        2107   &       53.125361   &         -27.711908   &          1.8844583   &  $46.69\pm7.24$  &  $10.01\pm0.04$  &  $0.23\pm0.02$\\
        2219   &       53.165334   &         -27.718542   &          1.9661846   &  $64.62\pm3.52$  &  $9.99\pm0.02$  &  $0.12\pm0.01$\\
        2252   &       53.079421   &         -27.720914   &          2.4071565   &  $245.75\pm190.56$  &  $10.93\pm0.08$  &  $0.60\pm0.08$\\
        2275   &       53.071987   &         -27.724916   &          1.9072056   &  $35.88\pm7.33$  &  $9.96\pm0.08$  &  $0.27\pm0.02$\\
        2363   &       53.164153   &         -27.709907   &          2.4488600   &  $64.87\pm11.56$  &  $10.53\pm0.15$  &  $0.23\pm0.02$\\
        2381   &       53.161483   &         -27.705100   &          1.4309824   &  $9.23\pm1.41$  &  $10.25\pm0.07$  &  $0.17\pm0.02$\\
        2403   &       53.129007   &         -27.713420   &          1.7654200   &  $45.04\pm10.31$  &  $10.31\pm0.04$  &  $0.34\pm0.03$\\
        2445   &       53.176209   &         -27.712390   &          2.3697045   &  $26.83\pm10.03$  &  $10.28\pm0.08$  &  $0.24\pm0.04$\\
        2450   &       53.181805   &         -27.729937   &          2.3141566   &  $43.41\pm5.77$  &  $10.26\pm0.03$  &  $0.15\pm0.02$\\
        2471   &       53.134824   &         -27.713354   &          2.4313843   &  $80.81\pm9.49$  &  $10.03\pm0.01$  &  $0.16\pm0.01$\\
        2493   &       53.160438   &         -27.707745   &          1.6085995   &  $96.23\pm5.13$  &  $10.43\pm0.06$  &  $0.29\pm0.01$\\
        2526   &       53.157911   &         -27.704309   &          1.8130853   &  $7026.86\pm15755.90$  &  $11.22\pm0.02$  &  $1.07\pm0.20$\\
        2550   &       53.125400   &         -27.703382   &          1.6018674   &  $25.52\pm1.90$  &  $9.99\pm0.05$  &  $0.15\pm0.01$\\
        2562   &       53.138745   &         -27.700470   &          2.4511107   &  $78.26\pm14.27$  &  $10.49\pm0.04$  &  $0.25\pm0.02$\\
        2595   &       53.109119   &         -27.730173   &          2.0773904   &  $14.82\pm5.43$  &  $9.83\pm0.11$  &  $0.21\pm0.05$\\
        2603   &       53.116033   &         -27.718285   &          1.6131810   &  $33.99\pm1.91$  &  $10.04\pm0.04$  &  $0.20\pm0.01$\\
        8005   &       53.088924   &         -27.781955   &          1.9415367   &  $25.33\pm1.25$  &  $9.05\pm0.11$  &  $0.08\pm0.01$\\

\hline

\end{longtable}
\endgroup

$^a$ Global properties of 97 objects with wavelength coverage $2200-2700\AA$ in rest frame. 28 galaxies have [O II] $3727$ detection and therefore
accurate determination of systemic redshift, the rest 69 galaxies have redshift determined from cross correlation.


\clearpage

\begin{table}
\begin{center}
\caption{Absorption Line Properties} \label{tab:abs}

\begin{tabular}{cccccc}
\hline
Abs Line & $f_l$ $^a$ & $P_{fluo}$ from \ion{Fe}{2} $\rightarrow$ $^b$ \ion{Fe}{2}* & $v_{cen}$ $^c$ & $v_{20\%}$ $^c$ & EW  \\ 
             &                  &         &      $ (km/s)$ & $(km/s)$   &  \AA    \\ 
\hline
\ion{Fe}{2} $2344.21$ & $0.114$   &$33.9\%$ & $-126.7\pm9.5$ & $-462.5\pm12.8$ & $2.23\pm0.13$  \\  
\ion{Fe}{2} $2374.46$ & $0.0313$ &$87.8\%$ & $-92.8\pm19.0$   & $-322.0\pm43.5$ & $1.52\pm 0.10$\\ 
\ion{Fe}{2} $2382.77$ & $0.32$     &$0$          & $-122.4\pm18.5$ & $-428.5\pm24.2$ & $2.15\pm0.17$\\
\ion{Fe}{2} $2586.65$ & $0.0691$ &$68.3\%$ & $-105.5\pm16.7$ & $-401.0\pm5.9$ & $1.90\pm0.15$\\
\ion{Fe}{2} $2600.17$ & $0.239$   &$12.6\%$ & $-133.7\pm11.2$ & $-462.2\pm16.5$ & $2.22\pm0.18$\\
\hline
\hline
\ion{Mg}{2} $2796.35$ & $0.6115$ & $0$  & $-193.8\pm 22.4$  &   $-499.2\pm20.3$  & $2.57\pm0.23$\\
\ion{Mg}{2} $2803.53$ & $0.3058$ & $0$ &  $-186.5\pm 25.1$  &   / $^d$  & $2.13\pm0.21$ \\
\hline
\end{tabular}

\end{center}

$^a$ Oscillator Strength \\ 
$^b$ $P_{fluo}(Fe II_{i})=\frac{\sum_{j}A_{ul}(Fe II*_{j})}{\sum_{j}{A_{ul}(Fe II*_j)}+{A_{ul}(Fe II_i)}}$, using atomic data from Morton (2003). \\
$^c$ $v_{cen}$ and $v_{20\%}$ are measured from the median spectrum of the 83 objects covering $2200-2900\AA$ in rest frame.  \\
$^d$  \ion{Mg}{2} $2803.53$ is blended with Mg $2796.35$, $v_{20\%}$ not measurable. \\

\end{table}

\clearpage


\begin{table}
\begin{center}
\caption{Emission Line Properties} \label{tab:ems}

\begin{tabular}{cccccc}
\hline
Ems Line & Associated Abs & $P_{single}$ $^a$ & $\lambda_{range}$ $^b$ & EW  &  $v_{cen}(Fe II*)$ $^e$ \\
             &                  &                           & $\AA$   &  \AA  &  $(km/s)$   \\
\hline
\ion{Fe}{2}* $2365.55$ & \ion{Fe}{2} $2344.21$   &$22.5\%$ & $2358.0-2371.0$ & $-0.45\pm0.10$ & $ -65\pm63$ \\
\ion{Fe}{2}* $2381.49$ & \ion{Fe}{2} $2344.21$   &$11.4\%$ & /$^c$ & / &  ...\\
\ion{Fe}{2}* $2396.36$ & \ion{Fe}{2} $2374.46$ &$87.8\%$ & $2390.0-2404.0$ & $-0.71\pm0.14$ & $-60\pm37$ \\
\ion{Fe}{2}* $2612.65$ & \ion{Fe}{2} $2586.65$ &  $45.4\%$   & $2608.0-2617.0$ & $-0.49\pm0.12$  &  $-69\pm42$ \\
\ion{Fe}{2}* $2626.45$ & \ion{Fe}{2} $2600.17$ &$12.6\%$ & $2620.0-2632.0$ & $-1.04\pm0.21$  &  $-71\pm45$ \\
\ion{Fe}{2}* $2632.11$ & \ion{Fe}{2} $2586.65$ & $22.9\%$ & /$^d$ & / \\
 
\hline
\end{tabular}

\end{center}
EWs are measured from the median spectrum of all 97 objects covering $2200-2700\AA$ in rest frame. \\
$^a$ $P_{single}(Fe II*_{i})=\frac{A_{ul}(Fe II*_{i})}{\sum_{j}{A_{ul}(Fe II*_j)}+{A_{ul}(Fe II_i)}}$. 
Note that if an \ion{Fe}{2} line is associated with only one \ion{Fe}{2}* lines, $P_{single} = P_{fluo}$.  \\
$^b$ Wavelength range for integrating \ion{Fe}{2}* EW. \\
$^c$ Blended with \ion{Fe}{2} $2382.77$.  \\
$^d$ Undetected. \\
$^e$ Flux-weighted velocity centroids. 

\end{table}

\clearpage


\begin{table}

\centering

\caption{Velocities Shifting in Co-added Spectra divided by Global Properties, [\ion{O}{2}] Detected Sample (Upper) and Cross-correlated Sample (Lower)} \label{tab:v_ems}

\resizebox{\textwidth}{!}{

\begin{tabular}{ccccccc}
\hline
Abs Line & $v_{cen}$ (SFR low/high) & $v_{cen}$(sSFR low/high) & $v_{cen}$($M_*$ low/high) & $v_{\%20}$(SFR low/high)  &  $v_{\%20}$(sSFR low/high)  &  $v_{\%20}$($M_*$ low/high) \\
             & $ (km/s)$ & $ (km/s)$ & $ (km/s)$ & $ (km/s)$ &  $ (km/s)$  &  $km/s$  \\
\hline
\ion{Fe}{2} $2344.21$ & $-151\pm26$/$-57\pm50$ & $-158\pm30$/$-84\pm46$ & $-122\pm39$/$-134\pm54$ & $-419\pm26$/$-452\pm27$ & $-450\pm16$/$-426\pm30$ & $-367\pm70$/$-508\pm35$ \\  
\ion{Fe}{2} $2374.46$ & $-102\pm48$/$-82\pm72$ & $-101\pm95$/$-97\pm24$ & $-76\pm103$/$-150\pm30$ & $-323\pm60$/$-332\pm26$ & $-220\pm174$/$-381\pm31$ & $-286\pm14$/$\pm23$ \\
\ion{Fe}{2} $2382.77$ & $-163\pm52$/$-113\pm38$ & $-202\pm57$/$-93\pm51$ & $-129\pm91$/$-148\pm50$ & $-422\pm52$/$-474\pm43$ & $-463\pm26$/$-425\pm27$ & $-368\pm19$/$\pm57$ \\
\ion{Fe}{2} $2586.65$ & $-171\pm71$/$-102\pm23$ & $-141\pm48$/$-104\pm32$ & $-97\pm58$/$-136\pm40$ & $-445\pm30$/$-372\pm20$ & $-312\pm20$/$-449\pm26$ & $-394\pm193$/$-363\pm25$ \\
\ion{Fe}{2} $2600.17$ & $-182\pm32$/$-108\pm48$ & $-181\pm35$/$-113\pm43$ & $-131\pm25$/$-176\pm57$ & $-462\pm35$/$-416\pm25$ & $-374\pm19$/$-464\pm37$ & $-442\pm26$/$\pm20$ \\
\hline
\hline
\ion{Mg}{2} $2796.35$ & $-183\pm51$/$-192\pm34$ & $-232\pm33$/$-144\pm35$ & $-188\pm55$/$-209\pm34$ & $-418\pm44$/$-448\pm28$ & $-433\pm23$/$-425\pm39$ & $-378\pm55$/$-471\pm40$ \\
\hline
\end{tabular}

}

\resizebox{\textwidth}{!}{

\begin{tabular}{ccccccc}
\hline
Abs Line & $v_{cen}$ (SFR low/high) & $v_{cen}$(sSFR low/high) & $v_{cen}$($M_*$ low/high) & $v_{\%20}$(SFR low/high)  &  $v_{\%20}$(sSFR low/high)  &  $v_{\%20}$($M_*$ low/high) \\
             & $ (km/s)$ & $ (km/s)$ & $ (km/s)$ & $ (km/s)$ &  $ (km/s)$  &  $km/s$  \\
\hline
\ion{Fe}{2} $2344.21$ & $-140\pm19$/$-115\pm27$ & $-147\pm16$/$-111\pm22$ & $-122\pm19$/$-139\pm25$ & $-446\pm21$/$-483\pm21$ & $-453\pm12$/$-479\pm33$ & $-411\pm25$/$-521\pm59$ \\  
\ion{Fe}{2} $2374.46$ & $-70\pm36$/$-111\pm31$ & $\-75\pm21$/$-97\pm26$ & $-83\pm26$/$-106\pm33$ & $-266\pm56$/$-360\pm29$ & $-277\pm70$/$\-340pm19$ & $-315\pm40$/$-333\pm34$ \\  
\ion{Fe}{2} $2382.77$ & $-130\pm35$/$-116\pm18$ & $-150\pm16$/$-94\pm33$ & $-103\pm40$/$-139\pm20$ & $-428\pm38$/$-441\pm22$ & $-455\pm26$/-416$\pm28$ & $-358\pm68$/$-470\pm25$ \\  
\ion{Fe}{2} $2586.65$ & $-112\pm23$/$-106\pm24$ & $-127\pm21$/$-95\pm31$ & $-102\pm26$/$-115\pm34$ & $-397\pm49$/$-413\pm28$ & $-377\pm90$/$-434\pm31$ & $-415\pm57$/$-376\pm43$ \\  
\ion{Fe}{2} $2600.17$ & $-127\pm24$/$-138\pm18$ & $-139\pm17$/$-130\pm25$ & $-126\pm18$/$-140\pm19$ & $-418\pm19$/$-496\pm33$ & $-416\pm22$/$-497\pm19$ & $-445\pm14$/$-472\pm28$ \\  
\hline
\hline
\ion{Mg}{2} $2796.35$ & $-224\pm45$/$-184\pm29$ & $-221\pm16$/$-152\pm47$ & $-233\pm36$/$-163\pm25$ & $-421\pm55$/$-583\pm53$ & $-482\pm28$/$-544\pm40$ & $-455\pm47$/$-538\pm74$ \\  
\hline
\end{tabular}

}

\end{table}


\clearpage


\begin{table}

\centering

\caption{Equivalent Widths of \ion{Fe}{2}/\ion{Fe}{2}* lines in Co-added Spectra} \label{tab:ew_bin}

\resizebox{11.5cm}{!}{

\begin{tabular}{ccccc}
\hline
& $EW_1[\AA]$ &  $EW_2[\AA]$ &  $EW_3[\AA]$ & $EW_4[\AA]$  \\
         
$Log(M_*[M_{\bigodot}])$ & 9.56 & 9.96 & 10.31 & 10.92 \\ 
\ion{Fe}{2} 2344 & $-1.76\pm0.17$ & $-2.49\pm0.21$ & $-2.56\pm0.46$ & $-2.52\pm0.30$ \\ 
\ion{Fe}{2} 2374 & $-1.31\pm0.23$ & $-1.52\pm0.16$ & $-1.31\pm0.24$ & $-1.28\pm0.37$ \\ 
\ion{Fe}{2} 2586 & $-1.96\pm0.24$ & $-2.49\pm0.33$ & $-2.37\pm0.53$ & $-1.87\pm0.38$ \\ 
\ion{Fe}{2} 2600 & $-2.14\pm0.28$ & $-2.79\pm0.35$ & $-3.35\pm0.44$ & $-2.90\pm0.36$ \\ 
\hline
\hline
$Log(M_*[M_{\bigodot}])$ & 9.74 & 10.31 \\ 
\ion{Fe}{2}* 2365 & $0.36\pm0.17$ & $0.38\pm0.18$ \\ 
\ion{Fe}{2}* 2396 & $0.73\pm0.18$ & $0.29\pm0.23$ \\ 
\ion{Fe}{2}* 2612 & $0.56\pm0.13$ & $0.24\pm0.21$ \\ 
\ion{Fe}{2}* 2626 & $1.50\pm0.23$ & $0.31\pm0.29$ \\ 
         
\hline
\end{tabular}

}

\resizebox{11.5cm}{!}{

\begin{tabular}{ccccc}
\hline
& $EW_1[\AA]$ &  $EW_2[\AA]$ &  $EW_3[\AA]$ & $EW_4[\AA]$  \\
         
$Log(SFR[M_*/yr])$ & 1.02 & 1.40 & 1.66 & 2.22 \\ 
\ion{Fe}{2} 2344 & $-1.55\pm0.23$ & $-2.21\pm0.16$ & $-2.46\pm0.26$ & $-2.79\pm0.41$ \\ 
\ion{Fe}{2} 2374 & $-0.93\pm0.28$ & $-1.42\pm0.14$ & $-1.58\pm0.12$ & $-1.44\pm0.40$ \\ 
\ion{Fe}{2} 2586 & $-1.81\pm0.23$ & $-2.31\pm0.24$ & $-2.45\pm0.34$ & $-2.76\pm0.40$ \\ 
\ion{Fe}{2} 2600 & $-1.79\pm0.29$ & $-2.63\pm0.28$ & $-3.41\pm0.29$ & $-3.41\pm0.33$ \\ 
\hline
\hline
$Log(SFR[M_*/yr])$ & 1.26 & 1.81 \\ 
\ion{Fe}{2}* 2365 & $0.22\pm0.15$ & $0.49\pm0.14$ \\ 
\ion{Fe}{2}* 2396 & $0.39\pm0.19$ & $0.73\pm0.19$ \\ 
\ion{Fe}{2}* 2612 & $0.47\pm0.13$ & $0.39\pm0.20$ \\ 
\ion{Fe}{2}* 2626 & $1.34\pm0.33$ & $0.55\pm0.20$ \\

\hline
\end{tabular}

}

\resizebox{11.5cm}{!}{

\begin{tabular}{ccccc}
\hline
& $EW_1[\AA]$ &  $EW_2[\AA]$ &  $EW_3[\AA]$ & $EW_4[\AA]$  \\
         
$Log(sSFR[yr^{-1}])$ & -9.08 & -8.58 & -8.34 & -8.02 \\ 
\ion{Fe}{2} 2344 & $-2.47\pm0.42$ & $-2.10\pm0.27$ & $-2.24\pm0.27$ & $-2.41\pm0.34$ \\ 
\ion{Fe}{2} 2374 & $-1.03\pm0.30$ & $-1.54\pm0.16$ & $-1.52\pm0.17$ & $-1.52\pm0.30$ \\ 
\ion{Fe}{2} 2586 & $-2.12\pm0.44$ & $-2.47\pm0.41$ & $-2.03\pm0.18$ & $-2.93\pm0.41$ \\ 
\ion{Fe}{2} 2600 & $-2.29\pm0.51$ & $-2.59\pm0.31$ & $-2.74\pm0.28$ & $-3.45\pm0.32$ \\ 
\hline
\hline
$Log(sSFR[yr^{-1}])$ & -8.67 & -8.28 \\ 
\ion{Fe}{2}* 2365 & $0.49\pm0.19$ & $0.31\pm0.13$ \\ 
\ion{Fe}{2}* 2396 & $0.33\pm0.21$ & $0.71\pm0.16$ \\ 
\ion{Fe}{2}* 2612 & $0.37\pm0.18$ & $0.45\pm0.16$ \\ 
\ion{Fe}{2}* 2626 & $0.91\pm0.30$ & $1.01\pm0.32$ \\

\hline
\end{tabular}

}

\resizebox{11.5cm}{!}{

\begin{tabular}{ccccc}
\hline
& $EW_1[\AA]$ &  $EW_2[\AA]$ &  $EW_3[\AA]$ & $EW_4[\AA]$  \\
         
E(B-V) & 0.09 & 0.17 & 0.25 & 0.39 \\ 
\ion{Fe}{2} 2344 & $-1.70\pm0.16$ & $-2.44\pm0.23$ & $-2.70\pm0.30$ & $-2.51\pm0.36$ \\ 
\ion{Fe}{2} 2374 & $-1.15\pm0.19$ & $-1.63\pm0.21$ & $-1.61\pm0.23$ & $-1.28\pm0.35$ \\ 
\ion{Fe}{2} 2586 & $-1.72\pm0.21$ & $-2.37\pm0.36$ & $-2.68\pm0.33$ & $-2.89\pm0.51$ \\ 
\ion{Fe}{2} 2600 & $-1.90\pm0.31$ & $-2.46\pm0.35$ & $-3.45\pm0.33$ & $-3.76\pm0.30$ \\ 
\hline
\hline
E(B-V) & 0.13 & 0.29 \\ 
\ion{Fe}{2}* 2365 & $0.26\pm0.16$ & $0.47\pm0.16$ \\ 
\ion{Fe}{2}* 2396 & $0.44\pm0.17$ & $0.64\pm0.23$ \\ 
\ion{Fe}{2}* 2612 & $0.60\pm0.13$ & $0.07\pm0.22$ \\ 
\ion{Fe}{2}* 2626 & $1.58\pm0.20$ & $0.14\pm0.29$ \\

\hline
\end{tabular}

}

\end{table}

\clearpage


\begin{table}

\begin{center}

\caption{Mean SFR/$M_*$/Age/E(B-V) of Subsamples $^a$} \label{tab:mean_sub}

\begin{tabular}{cccccc}
\hline
Sample   & $N$   & $M_*$  &  E(B-V) & SFR  & Age    \\
         &       &  $10^9 M_{\odot}$  &  & $M_{\odot}/yr$  & $Myr$    \\
\hline
High $M_*$  &  26   &    32.6  &   0.38   &  227.3   &   850    \\
Low $M_*$  &   20   &    1.9   &   0.17   &   111.8  &     300   \\
High SFR   &   29   &     20.9   &   0.45 &  420.9   &    960   \\
Low SFR    &   36   &     7.1  &   0.15   &   8.7    &      630    \\
High Age   &   23   &    20.3  &   0.26   &  51.3    &    2390    \\
Low Age    &   44   &    11.5  &   0.28   &   210.27  &     30    \\
High E(B-V)  &  37  &    22.8   &   0.44   &  331.4   &    760   \\
Low E(B-V)  &  32   &     5.1   &   0.12    &   12.2    &  950   \\

\hline
\end{tabular}

\end{center}
$^a$ Using exactly the same criteria used by Erb et al. (2012), 
all estimates are based on the Salpeter IMF.

\end{table}

\clearpage

\begin{table}

\begin{center}

\caption{EW(\ion{Fe}{2}) and EW(\ion{Mg}{2}) of subsamples correspoinding to Table~\ref{tab:mean_sub}} \label{tab:ew_sub}

\resizebox{\textwidth}{!}{

\begin{tabular}{cccccccc}
\hline
Sample   & EW(\ion{Fe}{2} $2344$)   & EW(\ion{Fe}{2} $2374$)   &  EW(\ion{Fe}{2} $2382$)  & EW(\ion{Fe}{2} $2586$)  &  EW(\ion{Fe}{2} $2600$)  &  EW(\ion{Mg}{2} $2796$)  &  EW(\ion{Mg}{2} $2803$)   \\
         &   \AA    &  \AA   &  \AA   &  \AA   &   \AA   &   \AA   &   \AA    \\
\hline
High $M_*$  &   $-2.62\pm0.41$  &  $-1.49\pm0.26$  &   $-2.13\pm0.27$   &  $-1.99\pm0.35$   &   $-2.94\pm0.27$   &  $-3.84\pm0.60$  &  $-2.06\pm0.62$   \\
Low $M_*$  &   $-1.69\pm0.28$  &  $-1.17\pm0.28$  &  $-1.26\pm0.39$   &   $-2.08\pm0.35$  &  $-2.34\pm0.27$   &  $-1.11\pm0.39$    &  $-1.51\pm0.39$  \\
High SFR   &   $-2.63\pm0.34$   &  $-1.66\pm0.28$  &  $-2.47\pm0.24$ &  $-2.67\pm0.30$   &  $-3.42\pm0.26$  &  $-3.59\pm0.56$  &  $-2.95\pm0.34$  \\
Low SFR    &   $-1.88\pm0.19$  &  $-1.26\pm0.19$  &   $-1.57\pm0.24$   &   $-2.10\pm0.32$    &  $-2.08\pm0.27$    &  $-1.40\pm0.37$  &  $-1.73\pm0.31$  \\
High Age   &   $-2.62\pm0.39$   &   $-1.65\pm0.31$  &   $-2.55\pm0.39$   &    $-2.63\pm1.06$    &  $-3.07\pm0.55$  &  $-3.54\pm0.61$  &  $-2.96\pm0.64$    \\
Low Age    &   $-2.32\pm0.22$   &   $-1.26\pm0.17$   &   $-1.76\pm0.23$   &   $-2.02\pm0.24$  &   $-2.39\pm0.25$  &  $-2.52\pm0.52$   &   $-1.80\pm0.36$  \\
High E(B-V) &  $-2.77\pm0.32$  &  $-1.52\pm0.22$  &   $-2.29\pm0.33$   &  $-2.71\pm0.39$   &  $-3.46\pm0.32$   &   $-4.01\pm0.89$  &  $-2.60\pm0.68$  \\
Low E(B-V)  &  $-1.85\pm0.20$   &   $-1.40\pm0.22$   &   $-1.58\pm0.31$    &   $-2.06\pm0.23$    &  $-1.93\pm0.28$  &  $-1.59\pm0.31$  &  $-1.75\pm0.32$ \\ 

\hline
\end{tabular}

}

\end{center}

\end{table}

\clearpage

\clearpage

\clearpage

\clearpage

\clearpage

\clearpage

\clearpage

\clearpage

\clearpage

\clearpage

\clearpage

\clearpage

\clearpage

\clearpage

\clearpage

\clearpage 

\clearpage

\clearpage

\clearpage

\clearpage

\clearpage

\clearpage

\clearpage

\clearpage

\clearpage

\clearpage


\end{document}